\documentclass[UTF8,a4paper]{article}
\usepackage{amsmath}
\usepackage{graphicx}
\usepackage{subfigure}
\usepackage{epstopdf}
\usepackage{geometry}
\usepackage{ulem}
\usepackage{indentfirst}
\usepackage{amsfonts,amssymb,dsfont}
\usepackage{setspace}
\usepackage{threeparttable}
\usepackage{amsmath,mathtools,amsthm}
\usepackage{array}
\usepackage{extarrows}
\usepackage{upgreek}

\makeatletter
\renewcommand*\env@matrix[1][\arraystretch]{%
  \edef\arraystretch{#1}%
  \hskip -\arraycolsep
  \let\@ifnextchar\new@ifnextchar
  \array{*\c@MaxMatrixCols c}}
\makeatother
\begin{document}


\title{\bf Electronic properties of the boson mode in a three-point fermion loop and the emergent SYK physics}
\author{Chen-Huan Wu
\thanks{chenhuanwu1@gmail.com}
\\College of Physics and Electronic Engineering, Northwest Normal University, Lanzhou 730070, China}

\maketitle
\vspace{-30pt}
\begin{abstract}
\begin{large} 

We investigate the electronic properties of the boson mode in a three-point fermion loop.
In this framwork, the single-particle excitation and the many-body local (in imaginary time and momentum space) field
effects are investigated in IR or UV limits with the density fluctuation induced by external potential (or bosonic frequency).
The (partly) cancellation effects of the bosonic density-density correlation in a multi-loop particle-hole diagram
with even Green's functions and a one-loop particle-hole diagram with odd Green's functions are studied.
In the limit of vanishing effect of external potential, which is equivalent to UV limit of the fermionic frequency,
the conserving approximation can be applied together with the Luttinger-Ward analysis,
in which case the anomalous contribution to the fermion self-energy 
or the expectation value of many-body interaction term, which is $g^{2}\langle \Delta^{\dag}\Delta\rangle$ ($g$ is the irreducible particle-particle vertex and 
$\Delta$ is the boson field operator), vanishes, and results in a Hartree-Fock type momentum- and frequency-independent fermion self-energy.
The correlator $\langle\Delta^{\dag}\Delta\rangle$ is positive which can be obtained through the local moment sum rule of dynamical susceptibility.
In the long-wavelength limit with low-energy (IR limit of boson mode),
the irreducible particle-particle vertex can be replaced by the bare one,
i.e., the RPA expression, where $\langle\Delta^{\dag}\Delta\rangle=\langle\Delta^{\dag}\rangle\langle\Delta\rangle$
and the electronic compressibility then becomes zero even at finite temperature and with finite chemical potential.
We also verify that the $GG^{0}G^{0}$ approach, where only the first Green's function be dressed, is valid in obtaining the self-consistent relation
(between single fermion property and that of the boson mode)
and the sum rules, even with the bare interaction and beyond the long-wavelength limit.
While the $GGG$ approach with a reducible vertex has been proved be a bad approximation perviously in obtaining sum rules (and it breaks conservation laws),
we found it is useful to obtain the bosonic mass or chemical potential in IR limit.
We also explore the possible emergence of SYK physics in this model
in low-frequency non-Fermi liquid phase,
where the product of irreducible vertices in a three fermion loop can be treated as a Gaussian-distributed
all-to-all random interaction.
The cases with and without pair condensation are studied seperately.
\\

\end{large}

\end{abstract}
\begin{large}

\section{Introduction}

The boson mode always plays a important role in both the quantum dynamics and classical approximations.
The non-Fermi liquid behaviors emerge when the gapless critical bosons coupled with a large number of fermions (or fermi surface),
and unlike the BCS gases, it requries a interaction strength larger than the critical value to drive the system in incoherence non-Fermi liquid phase
towards the pairing state through the zero-temperature quantum-critical point.
Further, in Sachdev-Ye-Kitaev (SYK) model, a stable non-Fermi liquid state can be extended away from the quantum critical point at low temperature,
which is the compressible gapless phases.
For many-body system in condensated theory,
the non-perturbation method is usually used in one-dimensional (or quasi-one dimensional) system
and high-frequency case (with short-time propagator),
where in conserving approximation, the fermion self-energy can be obtained by the functinal derivative of
Luttinger-Ward functional and exhibits a high-frequency asymptotic behavior.
Base on this, we deduce the three-particle self-consistency and the related sum rules.
We found that, the high frequency result of Hartree-Fock self-energy, which is linear in the multiplication of reducible interaction
and the single particle density operator,
is consistent with that in the SYK system\cite{Fu W}, where the conformal relation between fermion self-energy $\Sigma$ and Green's fcuntion $G$
can be seem by the expansion in powers of strong interaction (disorder averaged).
The difference is that for Hartree-Fock self-energy, the single-particle operator needs to be a constant averaged density.
This is because, for fermi liquid (where fermi surface exists),
the Lutinger theorem states that the zero-temperature (also zero frequency) 
behaviors of a region of electrons enclosed by fermi surface depends only on the density 
but not on the interaction.
While in the lower frequency regime,
the bare susceptibility get renormalized by the irreducible interaction which consider the many-body local field 
and thus contains the correlation and exchange effects.
This irreducible interaction can be obtained as the functional derivative of the self-energy with respect to Green's function
in the weak interaction limit and same-time approximation\cite{Allen S}.
And the self-energy containing the renormalization effect of irreducible interaction 
satisfies the self-consistent relation with the multiplication of reducible interaction
and the multi-particle density correlator, which is different from the above Hartree-Fock type one.
Thus the self-consistency obtained by nonperturbative approach in fluctuation dissipation theorem
only valid in weak to intermediate interaction regime, which fit with the Monte Carlo result.
Specially, when dealing with the Fermi gases,
the weak-interaction is also an essential requirement to apply the Kallio-Piilo recipe.

As the nonperturbation approach mainly dealing with the low-dimensional system,
the fluctuation (no matter pairing or particle-hole) can usually not be ignored,
thus considering the effect of particle-particle correlation is very important.
But the renormalization to chemical potential by interaction effect can be ignored at zero-temperature in fermi liquid
due to the Luttinger theorem.
While for the 0-dimensional SYK model\cite{WY,Chowdhury D}, where
the random all-to-all interaction and the disorders are treated perturbatively,
it works with large stiffness (extensive in fermion number $N$) and thus the fluctuations of infinite diagram are suppressed,
but its local criticality may still exhibits finite-dimensional behaviors at low enough temperature\cite{Patel A A}.
Unlike the nonperturbative approach which does not contain the diagrammatic resummation,
for SYK model,
the fluctuation of summation to infinite order in subsets of diagrams can be suppressed by the large-$N$ factor.
That is to say, there are only a few terms left which contribute even in large-$N$ limit.
Thus the gauge symmetry can be preserved as the nonperturbative approach did.
This is why the gapless compressible phase can be preserved by the gauge symmetry and conformal invariance at low-frequency.
We found that in the absence of pair condensation,
the fluctuations around the perturbative conformal solution can be suppressed (or exactly cancelled out)
using the replica process,
to make the saddle-point solution stable.
We prove that in this case the Luttinger-Ward theorem and the self-consistent relations are still valid
just like the nonperturbative case, although there does not exist the fermi surface.
For SYK system with gauge symmetry, the U(1) charge\cite{Gu Y} can be independent of interaction but only relates to the chemical potential,
which can be derived using the Luttinger-Ward analysis\cite{wusyk1,wusyk2}, and
this is also similar to the Luttinger constraint in nonperturbative approach.
When the pair condensation happen, 
we found that the gap equation (or the pairing order parameter) always
introduces a fluctuation of order of $N$ to the saddle-point,
which is a mean-field solution.
In this case, the Luttinger-Ward theorem and the self-consistent relations are no more valid,
and due to this large-N limit,
the phase coherence emerges as long as the pairing order parameter is nonzero.
This is one of the most important result obtained in this paper.
While in high-temperature limit,
the self-energy of SYK model is always consistent with the high frequency asymptotic of the nonperturbative result.

In this paper, we firstly focus on the electronic properties of a boson mode described by three-point fermion loop
for a fermion which excited by external bosonic frequencies to two states with distinct frequencies.
In this model,
two types of interaction are included, the electron-hole interaction and electron-electron interaction,
which are described by different vertices.
In this framwork, the single-particle excitation and the many-body local (in imaginary time and momentum space) field
effects are investigated in IR or UV limits with the density fluctuation induced by external potential (the bosonic frequency).
For the usually particle-hole excitation,
the action of interacting part reads
$S_{{\rm int}}=\int_{q}\phi(q)\psi^{\dag}\psi$, with the order parameter $\phi(q)$ 
(conjugate to the fermion bilinear) describing
the boson field fluctuations and $\psi$ is the fermion field, like the charge density fluctuation or antiferromagnetic spin fluctuation.
In simple particle-hole channel,
the order parameter (boson field) reads $\phi(q)=i\int_{p}g_{b}\psi^{\dag}(p)\psi(p-q)$,
with the bare coupling $g_{b}$.
The boson field operator $\phi$ is
constituted of the fluctuation part contribution and the static (and homogenous) part $\phi(0)$ (related to the saddle point),
i.e., the mean-field part contribution in BCS approximation,
which in $1/N$ expansion reads
$\phi=\frac{1}{\sqrt{N}}(\phi(q)+\phi(0))$,
since each vertex brings a factor of $N^{-1/2}$.
Thus the two-point (density-density) correlation at low temperature 
will exhibits usual fermi-liquid features in large-$N$ limit,
but this is nomore valid for three-point (or more points) correlations.
The large flavor number here plays an essential role in the application of mean-field approach,
but we note that the presence of strong on-site phase fluctuation (like the Hubbard model) or time-dependent symmetry breaking\cite{Winer M}
will invalidate the mean-field approach.
While for the order parameter in particle-particle channel defined by 
anomalous Green's function, it can be reduced to the mean field gap equation in broken-symmetry phase and in
the presence of infinite high-energy cutoff scale (i.e., the interaction effect is erased),
which satisfies the Thouless criterion (for non-self-cosistent $T$-matrix).
Here the gap equation controls the excitation gap and bosonic chemical potential of a single boson,

For the three-point boson mode, with the initial fermion with frequency $i\omega$, 
we define the energies of two excited states as $i\omega+i\Omega_{1}$ and $i\omega+i\Omega_{2}$,
respectively, 
and considering a spatially uniform boson potential here,
thus the action can be written as 
\begin{equation} 
\begin{aligned}
S_{{\rm int}}=\int_{i\omega,i\Omega}\phi(i\Omega)\psi^{\dag}(i\omega+i\Omega_{1})\psi^{\dag}(i\omega+i\Omega_{2})\psi(i\omega),
\end{aligned}
\end{equation}
where the bosonic field $\phi(i\Omega)$ plays a role of normalized wave function.
As we stated above, the fluctuations will leads to disordered Fermi-liquid phase as the temperature is lower that 
the coherence scale $W^{2}/g$ ($W$ is the bandwidth) but higher than other low-energy cutoffs;
While for strong short-range interaction with $g\gg T\gg W^{2}/g$,
or at zero temperature with $g\gg \omega>0$ (in this case the Thouless correlation energy ${\rm max}[\omega,T]$ is frequency
instead of temperature), which can be regarded as root-mean-square random type, 
some results of SYK model can be applied to explore the many-body properties.

At low spatial dimension,
even the weak coupling can significantly affects the behaviors of fermionic Landau quasiparticle,
which is coupled with the massless bosonic critical fluctuation and leads to non-fermi-liquid behaviors.
An example is the one-dimensional Luttinger liquid which is strongly correlated even at weak interaction.
The one-dimensional fermi surface is not continuous and consists of two discrete points
where the gapless modes can exist,
and the momentum distribution function has a power-law singularity near the fermi momentum 
for such a one-dimensional fermi surface,
which also indicates the non-fermi-liquid phase.
Another feature of the one-dimensional Luttinger liquid is the absence of particle-hole excitations 
in low-energy limit,
While for higher spatial dimensions, 
the breakdown of fermi-liquid theory requires strong coupling due to 
the screening effect of conduction electrons and the damping effect from electron-hole pairs.
Note that below critical temperature 
the noninteracting bare bosonic mass term (i.e., the mean field term) will not suppresses the non-fermi-liquid behavior and the
quantum fluctuation, even in the relativistic quantum field theory.
The effect of strong quantum fluctuations induced by particle-hole excitation near the fermi surface 
are not been considered here.
In conserving approximation, where the derivative of Luttinger-Ward functional is used in calculating the self-energy and irreducible vertex,
although the size of Fermi surface ($k_{F}$) is proportional to the total particle number according to the Luttinger's theorem,
a Hatree-Fock type self-energy obtained by high frequency expansion can keeps Fermi surface volume fixed,
thus there would not be much gapless particle-hole excitations near the fermi surface to cause the strong quantum fluctuation.
Some conclusions about IR asymptotic behaviors of SYK model are also applied in this paper,
to deal with the fermions with a marginal fermi-liquid scaling Green's function,
where we found the collective mode cannot be obtained in this case by solving the singularities of dynamical susceptibility.
The effect of strong quantum fluctuations induced by particle-hole excitation near fermi surface 
are not been considered here.

We also explore the emergent SYK physics in the system base on three-point fermion bubble,
in the incoherent non-Fermi liquid phase.
The boson (three-point fermion bubble) becomes critical when it couple to fermions,
and we prove that the static part of boson self-energy $\uppi(0)$ (or the bare bosonic mass $-\uppi(0)$)
will not affect the IR asymptotics behavior of dynamic boson self-energy,
thus the SYK physics-induced incoherence between fermions does not depend on the bare bosonic mass.
Note that when the boson order parameter condenses, the dressed boson mass becomes negative,
and the gap emerges in the many-body spectrum.
This will supresses the non-Fermi liquid phase, and leads to large splitting of eigenvalues of the system Hamiltonian
which signals the emergent off-diagonal long range order.

Lastly, we note that,
in nonperturbative approach,
the most important condition is the same time approximation,
i.e., the propagation time of normal fermion propagator is extremely short,
which corresponds the UV limit in frequency domain,
and this is essential to obtain the constant density operator base on the anticommutation relation of fermion statistic.
And the Luutinger-Ward relation between $G$, $\Sigma$, and the reducible or irreducible vertex
becomes exact only in the same time limit,
with vanishing pairing fluctuation (i.e., close to half-filling).
This is similar to the SYK model, where the self-consistency can be obtained (even in presence of diagrammatic approach
which unlike the nonperturbative approach)
in the limit of vanishing pairing order parameter.
While the SYK model base on the long-time approximation,
which corresponds to the IR limit in frequency domain,
where the conformal saddle point is obtained as $(G_{c},\Sigma_{c})=(G_{\beta=\infty},\Sigma_{\beta=\infty})$.
However, the effect of UV source field to the saddle-point may be treated similar to that of pairing source to the nonperturbative approach,
since the nonnegligible fluctuation $(\delta G,\delta \Sigma)$ can be produced by the UV source field in short-time limit.
In this case, the constant SYK conpling (after disorder average), should be renormalized by the many-body local-field factor.
And the local sum rule can be obtained through the fluctuation-dissipation relation.
Then the product of fermion Green's functions should also be modified by the irreducible coupling.
But this beyonds the scope of this paper, and we will discuss this aspect in another work.

The paper is organized in the following way.
In Sec.2, we describe the physical system we focus on in this paper.
A multi-loop particle-hole diagrams with an odd number of Green's functions is taken into account,
The cancellation effect is partially expected.
In Sec.3, we derive the self-consistent relations and sum rules in IR or UV limit.
The dynamic susceptibility and some important sum rules (about the single fermion spectral weight and boson spectral weight)
related to fluctuation-dissipation theorem are obtained in UV and IR limits.
The $GG_{0}G_{0}$ approach as well as the self-consistency between one- and three-point
quantities are also investigated in this section.
In Sec.4, we discuss the emergence of SYK physics in this model
in the low-frequency non-Fermi liquid phase.
The cases with and without pairing condensation are studied separately,
and we prove that the Luttinger-Ward theory (in particle-particle-hole channel) and the self-consistent relations in non-pertubative approach
(discussed in Sec.3) are valid only in the absence of pair condensation.

\section{Physical system}

\subsection{single fermion Green's function and spectral density}

Consider the finite-temperature effect,
the propagator with Matsubara-frequency  reads
\begin{equation} 
\begin{aligned}
G(p,i\omega)=\int^{\infty}_{-\infty}\frac{d\Omega}{2\pi}
\frac{A(p,\Omega)}{\Omega-i\omega},
\end{aligned}
\end{equation} 
where $A(p,\Omega)=[G^{A}(p,i\omega)-G^{R}(p,i\omega)]/i=[G(p,\omega-i\eta)-G(p,\omega+i\eta)]/i$ 
is the single particle spectral density.
By defining a complex variable $z$ using the retarded analytical continuation,
which is analytical off the real frequency axis but with a single branch cut along the imaginary frequency axis,
the summation over Matsubara frequencies in particle-hole pair propagators
can be replaced by a contour integral over $z$, with the poles brought by the fermi or boson distribution functions
(i.e., the Matsubara frequencies).
While for single-particle Green's function, 
in fermi-liquid phase with weak interaction,
the spectral density can be replaced by the Lorentzian form of delta function,
\begin{equation} 
\begin{aligned}
A(p,\Omega)=\frac{2{\rm Im}\Omega}{(\Omega-{\rm Re}\Omega)^{2}+({\rm Im}\Omega)^{2}},
\end{aligned}
\end{equation}
then the above propagator can be obtained as
\begin{equation} 
\begin{aligned}
G(p,z)=\frac{1}{z-{\rm Re}\Omega_{p}-i{\rm Im}\Omega_{p}},
\end{aligned}
\end{equation} 
This expression corresponds to the integration along the contour only in the upper half complex-frequency plane
as there is a branch cut along the real axis along ${\rm Im}z=0$,
and there does not has any poles within this branch cut since $\Omega_{p}$ is no a real quantity
as long as the interactions exist.
Note that the branch cut correspond to the incoherent continuum of quasiparticle
and the poles correspond to the bounded states (well-defined polaron).

\subsection{Boson mode and related vertices}

The three-point correlator reads
\begin{equation} 
\begin{aligned}
\uppi=-\int^{\beta}_{0}d\tau e^{i\Omega_{2}\tau_{2}}
\int^{\beta}_{0}d\tau e^{i\Omega_{1}\tau_{1}}
\langle\mathcal{T}\rho(q_{2},\tau_{2})\rho(q_{1},\tau_{1})\rho(-q_{1}-q_{2},0)\rangle,
\end{aligned}
\end{equation} 
then by taking the second derivative of the density operator,
we have
\begin{equation} 
\begin{aligned}
\frac{\partial^{2}}{\partial \tau^{2}}\rho(\tau)
=[[\rho(\tau),H],H].
\end{aligned}
\end{equation} 
For three-point correlator,
there are two perturbative terms in the Hamiltonian $H$ and each term contains a density operator at different time.
By taking the Fourier transformation ($F(\partial_{\tau}f(\tau))=i\Omega F(\Omega)$) 
to the equation of motion of propagator
\begin{equation} 
\begin{aligned}
\frac{\partial^{2}}{\partial \tau^{2}}D(\tau_{1},\tau_{2},\tau_{3})=
-\delta(\tau_{1},\tau_{2},\tau_{3})
-\int d\tau_{4}\uppi(\tau_{1},\tau_{4},\tau_{4})D(\tau_{4},\tau_{2},\tau_{3}),
\end{aligned}
\end{equation} 
thus 
\begin{equation} 
\begin{aligned}
(i\Omega)^{2}=-1-\uppi(\Omega)D(\Omega),
\end{aligned}
\end{equation} then
we obtain the bosonic propagator as
\begin{equation} 
\begin{aligned}
D=\frac{1}{-(i\Omega)^{2}-\uppi(i\Omega)},
\end{aligned}
\end{equation} 
where we consider only the states which are close to the fermi surface.
The identity 
\begin{equation} 
\begin{aligned}
\int^{\infty}_{-\infty}dt e^{i\omega t}\int^{\infty}_{-\infty}dt' e^{i\omega t'}\int^{\infty}_{-\infty}dt'' e^{i\omega t''}\delta(t,t',t'')=1
\end{aligned}
\end{equation} 
is used.

As we investigated in Ref.\cite{Wu1}, the particle-hole excitation is essential for the formation of complex bosonic modes, like the bipolaron,
and to leading order the density-density correlation (bosonic self-energy) vanishes for the multi-loop bosonic planar diagram in
particle-hole channel,
as long as the loop number is $\ge 3$. 
Note that there does not exists two loop diagram since
three (or other odd number of) fermionic propagators cannot connected to one single point in a planar diagram).
Thus the only case for nonzero bosonic self-energy (in particle-hole channel) is the one loop diagram.
The above conclusion can be verified in bosonic multi-loop planar diagrams, 
where we choose the one shown in Fig.1, which is a four loop diagram, as an example.
For Fig.1,
the self-energy reads
\begin{equation} 
\begin{aligned}
\uppi(q,\Omega)
=&\int_{a}\int_{b}\Gamma^{2}(q')\\
&\frac{1}{i\nu_{1}-\varepsilon_{k_{1}}}     \frac{1}{i\nu_{1}+i\Omega-\varepsilon_{k_{1}+q}}\\
&\frac{1}{i\nu_{2}-\varepsilon_{k_{2}}}     \frac{1}{i\nu_{2}+i\Omega-\varepsilon_{k_{2}+q}}\\
&\frac{1}{i\nu_{3}-\varepsilon_{k_{3}}}     \frac{1}{i\nu_{3}+i\Omega'-\varepsilon_{k_{3}+q}}\\
=&\int_{a}\int_{b}\Gamma^{2}(q')\\
&\frac{1}{i\nu_{1}-\frac{k^{2}_{1}}{2m}}     \frac{1}{i\nu_{1}+i\Omega-(\frac{k^{2}_{1}}{2m}+\frac{2k_{1}q}{2m}+\frac{q^{2}}{2m} )}\\
&\frac{1}{i\nu_{2}-\frac{k^{2}_{2}}{2m}}     \frac{1}{i\nu_{2}+i\Omega-(\frac{k^{2}_{2}}{2m}+\frac{2k_{2}q}{2m}+\frac{q^{2}}{2m} )}\\
&\frac{1}{i\nu_{3}-\frac{k^{2}_{3}}{2m}}     \frac{1}{i\nu_{3}+i\Omega'-(\frac{k^{2}_{3}}{2m}+\frac{2k_{3}q}{2m}+\frac{q^{2}}{2m} )},
\end{aligned}
\end{equation} 
where we define the $k$-dependent energy terms $a=\frac{k^{2}_{i}}{2m}$, $b=\frac{2k_{i}q}{2m}$.
Since the integral over a product of two Green's functions
provides a nonzero value only when the poles of the two Green's functions are on the different
frequency half-planes, i.e., locates in the opposite sides with respect to the real frequency axis,
we can obtain that, for the integration over $a$,
we must have $|\Omega|>|\nu_{i}|(i=1,2,3)$ to make sure $i\nu_{i}+i\Omega$ have different sign with $i\nu_{i}$, 
thus
for integration over $b$ the poles are $i\nu_{i}+i\Omega-\frac{k^{2}_{i}}{2m}-\frac{q^{2}}{2m}$
which are on the same side of complex frequency plane
(only depends on the sign of $\Omega$).
The result can be extended to arbitrary order of loops.
While for one loop boson diagram, the integration over $b$ has only one pole, which guarantees the finite
value of boson self-energy.
This conclusion is valid until the vertex correction with inelastic scattering, like the phonon scattering, emerges,
since the additional bosonic Green's function would introduces additional poles.
But with such additional poles, the boson self-energy is still finite.
Thus, the planar multiloop particle-hole diagrams with totally even Green's functions will cancel each other and turns out 
to be zero density-density correlation. Such cancellation partly exists even in a three-point loop with odd number of Green's function, as will be shown in 
the following and Appendix.A.
For multi-loop diagram, we found the cancellation effect exists even for the diagram with more than two points,
e.g., the one shown in Fig.1, which has six points (density operator), as long as the total number of Green's function
is even.

Next we discuss the bosonic self-energy in one-loop level.
Firstly, the bosonic self-energy of a three-point fermion loop reads
\begin{equation} 
\begin{aligned}
\uppi(i\Omega_{1},i\Omega_{2})=
\frac{1}{\beta}\sum_{n}
G(i\omega_{n})G(i\omega_{n}+i\Omega_{1})G(i\omega_{n}+i\Omega_{2}).
\end{aligned}
\end{equation} 
Considering the contributions from poles and branch cuts, and using the relation
\begin{equation} 
\begin{aligned}
\frac{1}{\beta}\sum_{n}f(i\omega_{n})
=&\sum_{z_{0}}{\rm Res}[f(z)N_{F}(z)]-\sum_{c}\int^{\infty}_{-\infty}
\frac{d\xi}{2\pi i}N_{F}(\xi)[f(\xi+i\eta)-f(\xi-i\eta)]\\
=&\frac{1}{2\pi i}\int f(z)N_{F}(z)dz-\sum_{c}\int^{\infty}_{-\infty}
\frac{d\xi}{2\pi i}N_{F}(\xi)[f(\xi+i\eta)-f(\xi-i\eta)],
\end{aligned}
\end{equation} 
the $n+1$-point correlation in particle-hole channel can be calculated by using the contour integration in
$z$-plane, with $n+1$ branch cuts.
\begin{equation} 
\begin{aligned}
\uppi(\Omega_{n})
=&\frac{1}{2\pi i}\int_{C}dz N_{F}(z)
G(z)G(z+i\Omega_{1})G(z+i\Omega_{2})\cdot\cdot\cdot G(z+i\Omega_{n})\\
=&\frac{1}{2\pi i}\int^{\infty}_{-\infty}d\xi N_{F}(\xi)
[G(\xi+i\eta)\mathcal{G}-G(\xi-i\eta)\mathcal{G}],\\
\mathcal{G}=&
G(\xi-i\Omega_{1})G(\xi-i\Omega_{2})\cdot\cdot\cdot G(\xi-i\Omega_{n})\\
&+G(\xi+i\Omega_{1})G(\xi-(i\Omega_{2}-i\Omega_{1}))\cdot\cdot\cdot G(\xi-(i\Omega_{n}-i\Omega_{1}))\\
&+\cdot\cdot\cdot\\
&+G(\xi+i\Omega_{n})G(\xi+(i\Omega_{n}-i\Omega_{1}))\cdot\cdot\cdot G(\xi+(i\Omega_{n}-i\Omega_{n-1})),
\end{aligned}
\end{equation} 
where $\xi$ denotes the position of branch cuts.
Thus the three-point density correlation reads
\begin{equation} 
\begin{aligned}
\uppi(i\Omega_{1},i\Omega_{2})
=&\frac{1}{2\pi i}\int^{\infty}_{-\infty}d\xi N_{F}(\xi)
[G(\xi+i\eta)\mathcal{G}-G(\xi-i\eta)\mathcal{G}],\\
\mathcal{G}=
&   G(\xi-i\Omega_{1})G(\xi-i\Omega_{2})\\
&+G(\xi+i\Omega_{1})G(\xi-(i\Omega_{2}-i\Omega_{1}))\\
&+\cdot\cdot\cdot\\
&+G(\xi+i\Omega_{2})G(\xi+i\Omega_{2}-i\Omega_{1}),
\end{aligned}
\end{equation} 
after make some shift in the integration variable,
we obtain
\begin{equation} 
\begin{aligned}
\uppi(i\Omega_{1},i\Omega_{2})
=&\frac{1}{2\pi i}\int^{\infty}_{-\infty}d\xi 
[
N_{F}(\xi+\Omega_{1}+\Omega_{2})
G(\xi+\Omega_{1}+\Omega_{2}+i\eta)G(\xi+\Omega_{2}-i\eta)G(\xi+\Omega_{1}-2i\eta)\\
&+
N_{F}(\xi-\Omega_{1}+\Omega_{2})
G(\xi-\Omega_{1}+\Omega_{2}+i\eta)G(\xi+\Omega_{2}+i\eta)G(\xi-i\eta)\\
&+
N_{F}(\xi+\Omega_{1}-\Omega_{2})
G(\xi+\Omega_{1}-\Omega_{2}+i\eta)G(\xi+\Omega_{1}+2i\eta)G(\xi+i\eta)\\
&-
N_{F}(\xi+\Omega_{1}+\Omega_{2})
G(\xi+\Omega_{1}+\Omega_{2}-i\eta)G(\xi+\Omega_{2}-i\eta)G(\xi+\Omega_{1}-2i\eta)\\
&-
N_{F}(\xi-\Omega_{1}+\Omega_{2})
G(\xi-\Omega_{1}+\Omega_{2}-i\eta)G(\xi+\Omega_{2}+i\eta)G(\xi-i\eta)\\
&-
N_{F}(\xi+\Omega_{1}-\Omega_{2})
G(\xi+\Omega_{1}-\Omega_{2}-i\eta)G(\xi+\Omega_{1}+2i\eta)G(\xi+i\eta)
],
\end{aligned}
\end{equation} 
There are six terms in this expression, but only three of them leads to nonzero results of the integration,
which can be obtained by aid of the procedure introduced in Eq.(3)
as shown in detail in Appendix.A.
Thus the above expression reduced to 
\begin{equation} 
\begin{aligned}
\uppi(i\Omega_{1},i\Omega_{2})
=&\frac{1}{2\pi i}\int^{\infty}_{-\infty}d\xi 
[
N_{F}(\xi+\Omega_{1}+\Omega_{2})
G(\xi+\Omega_{1}+\Omega_{2}+i\eta)G(\xi+\Omega_{2}-i\eta)G(\xi+\Omega_{1}-2i\eta)\\
&-
N_{F}(\xi-\Omega_{1}+\Omega_{2})
G(\xi-\Omega_{1}+\Omega_{2}-i\eta)G(\xi+\Omega_{2}+i\eta)G(\xi-i\eta)\\
&-
N_{F}(\xi+\Omega_{1}-\Omega_{2})
G(\xi+\Omega_{1}-\Omega_{2}-i\eta)G(\xi+\Omega_{1}+2i\eta)G(\xi+i\eta)
],
\end{aligned}
\end{equation} 
or just
\begin{equation} 
\begin{aligned}
\label{13}
\uppi(i\Omega_{1},i\Omega_{2})
=&\frac{1}{2\pi i}\int^{\infty}_{-\infty}d\xi 
[
N_{F}(\xi)
G(\xi+i\eta)G(\xi-i\Omega_{1})G(\xi-i\Omega_{2})\\
&-
N_{F}(\xi)
G(\xi-i\eta)G(\xi+i\Omega_{1})G(\xi-(i\Omega_{2}-i\Omega_{1}))\\
&-
N_{F}(\xi)
G(\xi-i\eta)G(\xi+i\Omega_{2})G(\xi+(i\Omega_{2}-i\Omega_{1}))
].
\end{aligned}
\end{equation} 
Note that the Green's functions $G(\xi\pm i\eta)$ have a negative energy term since they are hole states.
In the limit of $\Omega_{n}\rightarrow 0$,
\begin{equation} 
\begin{aligned}
\uppi
=&-\frac{1}{2\pi i}\int^{\infty}_{-\infty}d\xi 
[
N_{F}(\xi)
G(\xi+i\eta)G^{2}(\xi-i\eta)\\
&-
(N_{F}(\xi)
G(\xi-i\eta)G(\xi+i\eta)G(\xi))^{2}
],
\end{aligned}
\end{equation} 
where we have $G(\xi-i\eta)G(\xi+i\eta)=2\pi\tau
=\frac{-\pi}{{\rm Im}\Sigma}$ in the low-density limit,
with $\tau$ the mean free time (lifetime) of the quasiparticle.

By using the Ward identity, the vertex correction for $n+1$-point correlator loop 
(with one corner with outgoing external frequency and $n$ corners with ingoing external frequency) at vanishing external energy reads
\begin{equation} 
\begin{aligned}
\Gamma_{v}^{(n+1)}
\equiv & \Gamma_{v}(\omega,\omega+\Omega_{1})
         \Gamma_{v}(\omega+\Omega_{1},\omega+\Omega_{2})\cdot\cdot\cdot \Gamma_{v}(\omega+\Omega_{n}) \\
=&\frac{\delta^{n+1}\uppi(\Omega_{n})}
{\delta G(\omega)\delta G(\omega+\Omega_{1})\delta G(\omega+\Omega_{2})+\cdot\cdot\cdot \delta G(\omega+\Omega_{n}) }\\
=&\frac{\delta^{n}\uppi(\Omega_{n})}
{\delta \Omega_{1}\delta (\Omega_{2}-\Omega_{1})\cdot\cdot\cdot\delta(\Omega_{n}-\Omega_{n-1})}\bigg|_{\Omega_{n}=0}.
\end{aligned}
\end{equation} 
as diagrammatically shown in Fig..
Note that using analytical continuation 
(to make the self energy is analytic along the real aixs) 
and the chain rule, we have $\partial/\partial (i\Omega)=\partial/\partial \Omega$.
Thus we obtain the vertex correction for the three-point density correlator as
\begin{equation} 
\begin{aligned}
\Gamma_{v}^{(3)}
=&\frac{\delta^{2}\uppi(\Omega_{n})}
{\delta \Omega_{1}\delta (\Omega_{2}-\Omega_{1})}\bigg|_{\Omega_{n}=0},
\end{aligned}
\end{equation} 
As we defined in Eq.(15) and shown in Fig.2,
we consider only two vertices in a three-point loop.
This is to avoid the double counting of the scattering, in perspective of energy transfer.
Written separately,
the full vertex functions are
\begin{equation} 
\begin{aligned}
i\Omega\Gamma_{1}(i\omega,i\omega+i\Omega)=& G^{-1}(i\omega)-G^{-1}(i\omega+i\Omega),\\
i\Omega\Gamma_{2}(i\omega+i\Omega,i\omega+2i\Omega)=& G^{-1}(i\omega+i\Omega)-G^{-1}(i\omega+2i\Omega),
\end{aligned}
\end{equation} 
thus in the limit of $\Omega\rightarrow 0$,
\begin{equation} 
\begin{aligned}
\Gamma_{1}\Gamma_{2}=&\lim_{\Omega\rightarrow 0} \frac{
(G^{-1}(i\omega)-G^{-1}(i\omega+i\Omega)(G^{-1}(i\omega+i\Omega)-G^{-1}(i\omega+2i\Omega))}
{(i\Omega)^{2}}=(\frac{\partial G^{-1}(i\omega)}{\partial i\omega})^{2},
\end{aligned}
\end{equation} 
i.e., $\Gamma_{1}=\Gamma_{2}$ in this limit.
The Ward identity has also the form of
\begin{equation} 
\begin{aligned}
\Gamma=1\pm
\frac{\delta\Sigma}{\delta \omega},
\end{aligned}
\end{equation} 
in the presence of spin fluctuation\cite{Hertz J A,Katanin A A}.

In the absence of fluctuation at low enough temperature,
the above vertex function reduced to the Thouless criterion,
as the boson self-energy becomes mean-field order parameter and the external frequency is zero.
This leads to $\Gamma^{-1}_{v}\delta\uppi=\Gamma_{v}^{-1}\Delta(0)=0$,
and this relation also appears for systems which
are invariant under rotational or Lorentzian transformations according to the covariance principle\cite{Haussmann R,Jalali-Mola Z}.
Extended to strong-coupling limit with large $n$, by assmuing the ingoing external frequency in each corner has the same value 
($\Omega_{n}=\Omega_{n+1}-\Omega_{n}$),
the vertex can be obtained by the zero-frequency high order slope of the retarded density correlator 
as
\begin{equation} 
\begin{aligned}
\Gamma_{v}^{(n+1)}
=&\frac{\partial^{n}}{\partial \Omega_{n}^{n}}{\rm Re}\uppi(\Omega_{n})\bigg|_{\Omega_{n}=0}\\
=&\frac{\partial^{n}}{\partial \Omega_{n}^{n}}
\int^{\infty}_{-\infty}
\frac{d\Omega'_{n}}{\pi}
\frac{{\rm Im}\uppi(\Omega'_{n})}{\Omega'_{n}-\Omega_{n}}\bigg|_{\Omega_{n}=0}\\
=&
  n!\int^{\infty}_{-\infty}\frac{d\Omega'_{n}}{\pi}\frac{{\rm Im}\uppi(\Omega'_{n})}{(\Omega'_{n})^{n+1}},
\end{aligned}
\end{equation} 
When integral is over the bosonic frequency at branch cuts,
 as long as the $\Omega_{n}$ are all in the same sign,
the $n+1$ branch cuts of $\uppi(z)$ will locate in the same side of real axis,
thus in this case the integration range can replaced by $\int^{\infty}_{0}$ or $\int^{0}_{-\infty}$
for negative and positive $\Omega_{n}$, respectively.

For momentum conservation,
we assume the on-shell condition that
all the external momenta and energies are close to the same fermi surface (while away from the fermi surface (off-shell)
the real part of self-energy is vanishingly small),
then the poles of the density correlation function are independent of the momentum
in contrast to the hydrodynamic pole.
While at higher temperature with thermal excitations, the finite external bosonic momenta may leads to hydrodynamic poles.
In the limit of $(q',\Omega'_{n})\rightarrow 0$ (for particle and hole close to the fermi surface and obey the on-shell action),
we have
\begin{equation} 
\begin{aligned}
\Gamma_{v}^{(3)}
=2!
\frac{1}{\pi}
\frac{3 (832 - 302 \beta \varepsilon_{F} + 7 \beta^{3} \varepsilon_{F}^{3})}
{8 \varepsilon_{F}^{6} \pi},
\end{aligned}
\end{equation} 
and then the vertex function in each corner can be obtained as $\Gamma_{v}=\sqrt{\Gamma_{v}^{(3)}}$.
Since the bosonic spectral density has $A(\Omega)=2{\rm Im}\uppi(\Omega)$,
the above vertex function can also be treated as the
dc conductivity or shear viscosity (of a fluid-like many body system) in
the which can be obtained through the Kubo formula\cite{Jeon S,Mahan G D}.
For example,
$n=0$ ($0!=1$) corresponds to shear viscosity contributed by the two-point correlator with two vertices
with states near fermi surface in fermi-liquid regime.
While the Eq.(20) may indicates the shear viscosity contributed by a larger number of interaction vertices.

The $(n+1)$ point density correlator (also the boson self-energy) can be
asymptotically obtained as
\begin{equation} 
\begin{aligned}
\uppi(\Omega_{n})=\sum^{\infty}_{n=0}\Gamma_{v}^{n+1}\Omega_{n}^{n},
\end{aligned}
\end{equation} 
where $n=0$ correspons to the propagator $G(\xi)$.

Note that the Eq.(8) contains both the contributions from poles and branch cuts
since we extend the variable to the full complex plane.
This extension leads to imaginary discontinuity acoss each cut 
\begin{equation} 
\begin{aligned}
\uppi(\Omega_{n}+i\eta)-\uppi(\Omega_{n}-i\eta)=2i{\rm Im}\uppi(\Omega_{n}).
\end{aligned}
\end{equation} 

\section{Self-consistent relations and sum rules in IR or UV limit}
\subsection{Bosonic self-energy and sum rules}

Next we examine whether three-fermions mode satisfies the self-consistent relation (like the sum rule) similar to the
two-particle self-consistent one\cite{Hertz J A,Kyung B,Katanin A A}.
Firstly,
we rewrite the boson self-energy (Eq.(7)) as (at zero temperature and at half-filling and here we still use $\Omega_{2}=2\Omega_{1}$ for simplicity)
\begin{equation} 
\begin{aligned}
\uppi(i\Omega)
=&\int_{i\omega}G(i\omega)G(i\omega+i\Omega_{1})G(i\omega+i\Omega_{2})\\
=&\int_{i\omega}
\frac{G(i\omega)-G(i\omega+i\Omega_{1})}{G^{-1}(i\omega+i\Omega_{1})-G^{-1}(i\omega)}
G(i\omega+i\Omega_{2})\\
=&\int_{i\omega}
\frac{G(i\omega)G(i\omega+i\Omega_{2})-G(i\omega+i\Omega_{1})G(i\omega+i\Omega_{2})}{i\Omega_{1}}\\
=&\int_{i\omega}
\frac{\frac{G(i\omega)-G(i\omega+i\Omega_{2})}{2i\Omega_{1}}
-\frac{2G(i\omega+i\Omega_{1})-2G(i\omega+i\Omega_{2})}{2i\Omega_{1}}}{i\Omega_{1}}\\
=&\int_{i\omega}
\frac{G(i\omega)-2G(i\omega+i\Omega_{1})+G(i\omega+i\Omega_{2})}{-2\Omega_{1}^{2}}.
\end{aligned}
\end{equation} 
We write the three-fermions dynamic susceptibility as
\begin{equation} 
\begin{aligned}
\chi(i\Omega)=\frac{\uppi(i\Omega)}{1-g^{2}\uppi(i\Omega)},
\end{aligned}
\end{equation} 
whose pole corresponds to the resonant particle-hole pair scattering (Goldstone boson-like)
and can be obtained by solving the relation (using the result of Eq.(70))
\begin{equation} 
\begin{aligned}
1=&g^{2}\uppi(i\Omega)\\
=&g^{2}\int_{i\omega}
\frac{G(i\omega)-2G(i\omega+i\Omega_{1})+G(i\omega+i\Omega_{2})}{-2\Omega_{1}^{2}}\\
=&\frac{g^{2}}{-2\Omega_{1}^{2}}\int_{i\omega}
[G(i\omega)-2G(i\omega+i\Omega_{1})+G(i\omega+i\Omega_{2})]\\
=&\frac{g^{2}}{-2\Omega_{1}^{2}}
[n_{1}-2n_{2}(\Omega_{1})+n_{3}(\Omega_{2})],
\end{aligned}
\end{equation} 
where $n_{1}$ denotes the number density of particle with frequency $i\omega$.
Here the vertex $g^{2}$ is a constant can can also be obtained as a functional derivative $g^{2}=\delta^{3}\uppi/\delta G\delta G'\delta G''$.
Since the Goldstone boson requires $\chi(i\Omega)$ has a pole at zero external frequency limit in which case the system obeys the conservation law,
we obtain $1=g^{2}
\frac{-\delta G(i\omega)}{\delta (i\Omega)}\bigg|_{\Omega=0}$.

The sum rule 
\begin{equation} 
\begin{aligned}
\int_{i\Omega}\chi(i\Omega)\propto n
\end{aligned}
\end{equation} 
 requires the susceptibility to be "exact"\cite{Hertz J A},
i.e., be irreducible.
Here we note that, the irreducible boson self-energy diagram has two kinds,
the first kind is the RPA type (without the ladder expansion; see Fig.2(a)),
while the second one considers the irreducible vertex correction which is the sum of all irreducible electron-hole propagators,
i.e., at least there has one particle-hole scattering running across at least one another rung of the ladder (see Fig.2(c)).
Otherwise, the boson self-energy diagram is reducible, like the diagrams in RPA ladder expansion in Fig.2(b).

By defining the three-point bosonic order parameter (since we only consider the electron-hole scattering within the bubble)
$\Delta$ as 
\begin{equation} 
\begin{aligned}
\Delta^{\dag}=c^{\dag}_{i\omega+i\Omega_{1}}
c^{\dag}_{i\omega+i\Omega_{2}}
c_{i\omega} \equiv c^{\dag}_{1}c^{\dag}_{2}c_{3}.
\end{aligned}
\end{equation} 
Note that the order parametr is approximately $s$-wave type, but we omit the $i$-factor within their expressions
($\Delta^{\dag}=ic^{\dag}_{1}c^{\dag}_{2}c_{3}$)
since the value of product $\Delta\Delta^{\dag}$ will not be affected by this factor.
Then using the anti-commutation relations obtained by identity $x\delta(x)=0$,
$\{c_{i}^{\dag},c_{j}^{\dag}\}=\{c_{i},c_{j}\}=0,
\{c_{i},c_{j}^{\dag}\}=-\{c_{j}^{\dag},c_{i}\}=\delta_{ij}$,
we obtain the sum rule of the susceptibility $\chi(\tau)=-\langle [\Delta(\tau) \Delta^{\dag}(0)]\rangle$ (here the time-ordering is omitted)
\begin{equation} 
\begin{aligned}
\int_{i\Omega}\chi(i\Omega)=\langle\Delta^{\dag}\Delta\rangle
=n_{1}n_{2}(1-n_{3}),
\end{aligned}
\end{equation} 
and we also obtain the sum rule as a consequence of the fluctuation-dissipation theorem
\begin{equation} 
\begin{aligned}
\int^{\infty}_{-\infty}\frac{d\omega}{\pi}
\chi''(i\omega)\equiv &
\int^{\infty}_{-\infty}\frac{d\omega}{\pi}
{\rm Im}\chi(\omega+i\eta)\\
=&-\langle[\Delta,\Delta^{\dag}]\rangle\\
=&( c_{1}^{\dag}c_{2}^{\dag}c_{3}c_{3}^{\dag}c_{2}c_{1}-c_{3}^{\dag}c_{2}c_{1}c_{1}^{\dag}c_{2}^{\dag}c_{3})\\
=&( -c_{1}^{\dag}c_{2}^{\dag}c_{3}c_{1}c_{2}c_{3}^{\dag}+c_{1}c_{2}c_{3}^{\dag}c_{1}^{\dag}c_{2}^{\dag}c_{3})\\
=& n_{1}n_{2}(1-n_{3})-(1-n_{1})(1-n_{2})n_{3},
\end{aligned}
\end{equation}
where $\chi''(i\omega)={\rm Im}\chi^{R}(i\omega)$ (or in imaginary time domain $\chi''(\tau)=-\frac{1}{2}\langle [\Delta(\tau),\Delta^{\dag}(0)]\rangle$
since $\int\frac{d\omega}{2\pi}\langle[\Delta(\tau),\Delta^{\dag}(0)]\rangle=\langle[\Delta(0),\Delta^{\dag}(0)]\rangle$)
is the spectral weigh of the three-point boson mode.
Similarly, we have (through Fourier transform $F(\partial_{\tau} f(\tau))\rightarrow i\omega F(\omega)$),
using the relation $[\Delta,H]=\partial/\partial \tau\Delta$
\begin{equation} 
\begin{aligned}
\label{sumrule}
\int^{\infty}_{-\infty}\frac{d\omega}{\pi}
\omega\chi''(i\omega)
=&-[[\Delta(\tau),H],\Delta^{\dag}(0)]\\
=&[\frac{\partial}{\partial\tau}\Delta(\tau),\Delta^{\dag}(0)]\bigg|_{\tau=0}\\
=&[(\varepsilon_{3}(i\omega)+n_{3}g^{2}-\varepsilon_{2}(i\omega)-n_{2}g^{2}-\varepsilon_{1}(i\omega)-n_{1}g^{2})\Delta(0),\Delta^{\dag}(0)]\bigg|_{\tau=0}\\
=&(\varepsilon_{3}(i\omega)+n_{3}g^{2}-\varepsilon_{2}(i\omega)-n_{2}g^{2}-\varepsilon_{1}(i\omega)-n_{1}g^{2})    [(1-n_{1})(1-n_{2})n_{3}-n_{1}n_{2}(1-n_{3})],\\
\int^{\infty}_{-\infty}\frac{d\omega}{\pi}
\frac{\chi''(i\omega)}{-\omega}
=&\lim_{i\Omega\rightarrow 0}\chi(i\Omega)\\
=&\lim_{\delta\tau\rightarrow\infty}\chi(\tau,\tau')\\
=&\lim_{\delta\tau\rightarrow\infty}\chi(-\delta\tau)\\
=&\int_{\omega}\langle \Delta^{\dag}(\infty)\Delta(0)\rangle_{\omega}\\
=&-\int_{\omega}\langle \Delta(0)\Delta^{\dag}(\infty)\rangle_{\omega}\\
=&n_{1}(i\omega)n_{2}(i\omega)(1-n_{3}(i\omega)),
\end{aligned}
\end{equation} 
The first and second expressions are used in UV and IR limit of the boson mode, respectively.
It can be seen that,
the fluctuation-dissipation theorem relys on the equal-time approximation of the correlation function or commutator,
i.e., the $\tau\rightarrow 0$ limit or $\tau\rightarrow \infty$ limit.
Since the spectral representation of susceptibility can be expanded in IR limit of boson mode as
\begin{equation} 
\begin{aligned}
\chi(i\Omega)=\int\frac{d\omega}{\pi}
\frac{\chi''(\omega)}{i\Omega-\omega}
\approx \int\frac{d\omega}{\pi}
\chi''(\omega)[-\frac{1}{\omega}-\frac{i\Omega}{\omega^{2}}],
\end{aligned}
\end{equation} 
the off-diagonal sum-rule has
\begin{equation} 
\begin{aligned}
\label{38}
-\int\frac{d\omega}{\pi}
\frac{\chi''(\omega)}{\omega^{2}}
=&
\lim_{i\Omega\rightarrow 0}[\frac{\chi(i\Omega)}{i\Omega}-\int\frac{d\omega}{\pi}
\frac{\chi''(\omega)}{-\omega}\frac{1}{i\Omega}]\\
=&\lim_{i\Omega\rightarrow 0}
[\frac{(n_{1}(i\omega)n_{2}(i\omega)(1-n_{3}(i\omega)))+\delta \chi(i\Omega)}{i\Omega}
-\frac{(n_{1}(i\omega)n_{2}(i\omega)(1-n_{3}(i\omega)))+\delta \chi(i\Omega)}{i\Omega}]\\
=&0,
\end{aligned}
\end{equation} 
where $\delta \chi(i\Omega)$ is added as a pertubatively term with small $i\Omega$.
This is because, in IR limit, we have (no matter the coupling is strong or weak)
\begin{equation} 
\begin{aligned}
\lim_{i\Omega\rightarrow 0}\frac{\uppi(i\Omega)}{i\Omega}\approx
\lim_{i\Omega\rightarrow 0}\frac{\chi(i\Omega)}{i\Omega}.
\end{aligned}
\end{equation} 

\subsection{Non-perturbative approach with effective coupling}

When the couplings between 
the particles $n_{1}$ and $n_{2}$ and hole $n_{3}$ are considered,
in which case they are coexist,
the sum rule requires the effective interaction instead of the bare one,
which reads 
\begin{equation} 
\begin{aligned}
g_{eff}^{2}
=& g\frac{\langle n_{1}n_{2}\rangle}{\langle n_{1}\rangle\langle n_{2}\rangle}
g\frac{\langle n_{1}(1-n_{3})\rangle}{\langle n_{1}\rangle\langle (1-n_{3})\rangle}\\
=& g^{2}\frac{\langle n_{1}n_{2}\rangle\langle n_{1}(1-n_{3})\rangle}{\langle n_{1}\rangle^{2}\langle n_{2}\rangle\langle (1-n_{3})\rangle}.
\end{aligned}
\end{equation} 
This is obtained by the inrreducible vertex, which reads $g\frac{\langle n_{1}n_{2}\rangle}{\langle n_{1}\rangle\langle n_{2}\rangle}$
($g\frac{\langle n_{1}(1-n_{2})\rangle}{\langle n_{1}\rangle\langle (1-n_{2})\rangle}$)
for repulsive interaction $g>0$ (attractive interaction $g<0$) between particle density operator (filling) $n_{1}$ and $n_{2}$.
Note that the term $\langle n_{i}n_{j}\rangle$ is free from the Pauli principle here,
and the statistical averages throughtout this paper satisfy,
\begin{equation} 
\begin{aligned}
\label{82}
\langle c_{i}c^{\dag}_{j}\rangle
=&\int^{\infty}_{-\infty}\frac{d\omega}{2\pi}N_{F}(\omega)\rho(\omega)\\
=&\int^{\infty}_{-\infty}\frac{d\omega}{2\pi}N_{F}(\omega)\langle\{c_{i}(\tau),c_{j}^{\dag}(0)\}\rangle_{\omega},\\
\langle \Delta_{i}\Delta^{\dag}_{j}\rangle-\langle \Delta_{i}\rangle\langle \Delta^{\dag}_{j}\rangle
=&-\int^{\infty}_{-\infty}\frac{d\Omega}{\pi}N_{B}(\Omega)\chi''(\Omega)\\
=&\int^{\infty}_{-\infty}\frac{d\Omega}{\pi}N_{B}(\Omega)\frac{1}{2}\langle[\Delta_{i}(\tau),\Delta^{\dag}_{j}(0)]\rangle_{\Omega},\\
\langle n_{i}n_{j}\rangle-\langle n_{i}\rangle\langle n_{j}\rangle
=&\int^{\infty}_{-\infty}\frac{d\Omega}{\pi}N_{B}(\Omega)\frac{1}{2}\langle[c^{\dag}_{i}(\tau)c_{i}(\tau),c^{\dag}_{j}c_{j}]\rangle_{\Omega}.
\end{aligned}
\end{equation} 
Then, since only the particle $n_{1}$ and $n_{2}$ can coexist in the same time, as described by the operator $\Delta^{\dag}$,
the above effective interaction can be divided into two parts
\begin{equation} 
\begin{aligned}
g_{eff}
=& g\frac{\langle n_{1}n_{2}\rangle}{\langle n_{1}\rangle\langle n_{2}\rangle}\\
=&g\frac{1}{1+\frac{g}{\langle n_{1}\rangle\langle n_{2}\rangle}\int_{i\Omega}[\int_{i\omega}G_{2}^{0}G_{3}^{0}]^{2}},\\
g_{eff}
=& g\frac{\langle n_{1}(1-n_{3})\rangle}{\langle n_{1}\rangle\langle (1-n_{3})\rangle}\\
=&g\frac{1}{1+\frac{g}{\langle n_{1}\rangle\langle (1-n_{3})\rangle}\int_{i\Omega}[\int_{i\omega}G_{1}^{0}G_{2}^{0}]^{2}},\\
\end{aligned}
\end{equation}

Further, if we denote
\begin{equation} 
\begin{aligned}
\gamma=& \frac{\langle n_{1}n_{2}\rangle\langle n_{1}(1-n_{3})\rangle}{\langle n_{1}\rangle^{2}\langle n_{2}\rangle\langle (1-n_{3})\rangle},
\end{aligned}
\end{equation} 
we have 
\begin{equation} 
\begin{aligned}
g^{2}\gamma=\int_{i\omega,i\Omega}\Gamma(i\omega,i\omega+i\Omega_{1})\Gamma(i\omega+i\Omega_{1},i\omega+i\Omega_{2}).
\end{aligned}
\end{equation} 
$\gamma=1$ in RPA, but in this case, the diagram is reducible and the Goldstone condition $g\uppi(0)=1$
cannot be meeted until the $\gamma$ is compensately corrected.
In IR limit (for boson mode),
we have
\begin{equation} 
\begin{aligned}
\lim_{i\Omega\rightarrow 0}g_{eff}
=&g(\gamma_{RPA}+\lim_{i\Omega\rightarrow 0}\frac{G_{1}^{-1}-G_{2}^{-1}}{i\Omega})\\
\approx &g(1+\lim_{i\Omega\rightarrow 0}\frac{-\Sigma_{1}+\Sigma_{2}}{i\Omega})\\
= &g(1+\lim_{i\Omega\rightarrow 0}\frac{-n_{1}g^{2}+n_{2}g^{2}+\delta \Sigma(i\Omega)}{i\Omega}),
\end{aligned}
\end{equation} 
where the Hartree-Fock type mean-field self-energy (see Appendix.C) is used and it is important to make sure this self-energy 
enters all the Green's functions of $\uppi(i\Omega)$ to make the
$\chi(i\Omega)$ satisfies the sum rules.
The effective vertex $g_{eff}$ can also be obtained by solving Eq.(81).

Then we have in IR limit (consider the Eq.(\ref{38}))
\begin{equation} 
\begin{aligned}
-\int\frac{d\omega}{\pi}
\frac{\chi''(\omega)}{\omega^{2}}
=&
\lim_{i\Omega\rightarrow 0}[\frac{\chi(i\Omega)}{i\Omega}-\int\frac{d\omega}{\pi}
\frac{\chi''(\omega)}{-\omega}\frac{1}{i\Omega}]\\
=&\lim_{i\Omega\rightarrow 0}[\frac{\uppi(i\Omega)}{1-g_{eff}^{2}\uppi(i\Omega)}\frac{1}{i\Omega}-\int\frac{d\omega}{\pi}
\frac{\chi''(\omega)}{-\omega}\frac{1}{i\Omega}]\\
=&\lim_{i\Omega\rightarrow 0}
[\frac{\uppi(i\Omega)/i\Omega}{(i\Omega)^{-1}
-g^{2}\frac{\langle n_{1}n_{2}(1-n_{3})\rangle}{\langle n_{1}\rangle^{2}\langle n_{2}\rangle\langle (1-n_{3})\rangle}
\frac{\uppi(i\Omega)}{i\Omega}}\frac{1}{i\Omega}-\int\frac{d\omega}{\pi}
\frac{\chi''(\omega)}{-\omega}\frac{1}{i\Omega}]\\
\approx
&\lim_{i\Omega\rightarrow 0}
[\frac{n_{1}(i\omega)n_{2}(i\omega)(1-n_{3}(i\omega))+\delta\chi(i\Omega)}{1-g^{2}
\frac{\langle n_{1}n_{2}\rangle\langle n_{1}(1-n_{3})\rangle}{\langle n_{1}\rangle}}
\frac{1}{i\Omega}
-\frac{-G_{1}(i\omega)G_{2}(i\omega)(1-G_{3}(i\omega))+\delta\chi(i\Omega)}{i\Omega}]\\
\approx
&\lim_{i\Omega\rightarrow 0}
[\frac{1}{(n_{1}(i\omega)n_{2}(i\omega)(1-n_{3}(i\omega)))^{-1}-g^{2}_{eff}}
\frac{1}{i\Omega}
-\frac{n_{1}(i\omega)n_{2}(i\omega)(1-n_{3}(i\omega))}{i\Omega}]\\
=& 0.
\end{aligned}
\end{equation} 
This result is the same as Eq.(\ref{38}) and independent of the value of coupling strenghth,
which reveals that, the off-diagonal sum rule cannot be obtained in IR limit.
Similarly, in UV limit of the boson model, the off-diagonal sum rule is available through the expansion
\begin{equation} 
\begin{aligned}
\chi(i\Omega)=\int\frac{d\omega}{\pi}
\frac{\chi''(\omega)}{i\Omega-\omega}
\approx \int\frac{d\omega}{\pi}
\chi''(\omega)[\frac{1}{i\Omega}+\frac{\omega}{(i\Omega)^{2}}],
\end{aligned}
\end{equation} 
which reads
\begin{equation} 
\begin{aligned}
\int\frac{d\omega}{\pi}
\chi''(\omega)\omega
=&\lim_{i\Omega\rightarrow \infty}
[\chi(i\Omega)(i\Omega)^{2}-\int\frac{d\omega}{\pi}
\chi''(\omega)i\Omega]\\
=&\lim_{i\Omega\rightarrow \infty}
[\frac{\uppi(i\Omega)}{1-g_{eff}^{2}\uppi(i\Omega)}
(i\Omega)^{2}
-[n_{1}n_{2}(1-n_{3})-(1-n_{1})(1-n_{2})n_{3}]i\Omega]\\
=&\lim_{i\Omega\rightarrow \infty}
[\frac{(i\Omega)^{2}[n_{1}n_{2}(1-n_{3})-(1-n_{1})(1-n_{2})n_{3}]}
{i\Omega-g_{eff}^{2}[n_{1}n_{2}(1-n_{3})-(1-n_{1})(1-n_{2})n_{3}]}\\
&-[n_{1}n_{2}(1-n_{3})-(1-n_{1})(1-n_{2})n_{3}]i\Omega]\\
=&g_{eff}^{2}[n_{1}n_{2}(1-n_{3})-(1-n_{1})(1-n_{2})n_{3}]^2.
\end{aligned}
\end{equation} 
This result is equivalent to the first expression of Eq.(\ref{sumrule}).
The relation
\begin{equation} 
\begin{aligned}
\lim_{i\Omega\rightarrow\infty}i\Omega\chi(i\Omega)
=\lim_{i\Omega\rightarrow\infty}i\Omega\uppi(i\Omega)
=\int\frac{d\omega}{\pi}\chi''(i\Omega),
\end{aligned}
\end{equation} 
is used here.
That means the off-diagonal f-sum rule can only be obtained in UV limit of boson mode.

\subsection{$GG_{0}G_{0}$ approximation}

As we stated above, the f-sum rule of $\chi(i\Omega)$ requires each Green's function within $\uppi(i\Omega)$
be dressed by a Hartree-Fock type mean-field self-energy $\Sigma_{MF}=g^{2}n$. 
But what if one of the Green's function is dressed by self-energy beyond mean-field level (and to infinite order of $g$)?
To know this, we redefine the three-point bubble as
\begin{equation} 
\begin{aligned}
\label{ggg}
\uppi(i\Omega)=
\int_{i\omega}G(i\omega)G^{0}(i\omega+i\Omega_{1})G^{0}(i\omega+i\Omega_{2})
\equiv \int_{i\omega}G_{1}G^{0}_{2}G^{0}_{3},
\end{aligned}
\end{equation} 
where $G_{1}$ is, in contrast to $G_{2}$ and $G_{3}$, dressed by the self-energy (in $T$-matrix type approximation)
\begin{equation} 
\begin{aligned}
\Sigma_{1}(i\omega)
=&\int_{i\Omega}
\frac{1}{g^{-2}-\uppi(i\Omega)}G^{0}(i\omega+i\Omega)\\
=&\int_{i\Omega}
(g^{2}+g^{4}\uppi(i\Omega)+g^{6}\uppi^{2}(i\Omega)+\cdot\cdot\cdot)
G^{0}(i\omega+i\Omega)\\
=&\int_{i\Omega}
(g^{2}+g^{4}\uppi(i\Omega)+g^{6}\uppi^{2}(i\Omega)+\cdot\cdot\cdot)
G^{0}(i\omega+i\Omega)\\
=&\int_{i\Omega}
(g^{2}+g^{4}\uppi(i\Omega)+g^{6}\uppi^{2}(i\Omega)+\cdot\cdot\cdot)
G^{0}_{2}\\
=&
g^{2}n^{0}_{2}+\int_{i\Omega}g^{4}\frac{\uppi(i\Omega)}{1+\uppi(i\Omega)g}G_{2}.
\end{aligned}
\end{equation} 
Then single-fermion quantity can be related to a three-fermion quantity self-consistently
as
\begin{equation} 
\begin{aligned}
\label{92}
\int_{i\omega}\Sigma_{1}(i\omega)G_{1}G_{3}
=&\int_{i\Omega}\frac{1}{g^{-2}-\uppi(i\Omega)}\uppi(i\Omega)\\
=&g^{2}\int_{i\Omega}\frac{\uppi(i\Omega)}{1-g^{2}\uppi(i\Omega)}\\
=&g^{2}\int_{i\Omega}\chi(i\Omega)\\
=&g^{2}\langle \Delta^{\dag}\Delta\rangle\\
=&g^{2}n_{1}n_{2}^{0}(1-n_{3}^{0}).
\end{aligned}
\end{equation} 
Here the susceptibility $\chi(i\Omega)$ satisfies the sum rule
just like the one in mean field level did.
But for the spectral weight $\chi''(i\Omega)$,  $\omega\chi''(\omega)$ or $\chi''(\omega)/\omega$,
as can still be obtained by the UV or IR expansion of bosonic mode frequency,
they satisfy the sum rules which are different from the ones in mean-field level,
due to the different order of couplings between mean-field scheme and $G_{1}G^{0}_{2}G^{0}_{3}$ scheme containing $g_{eff}$.
For example,
this can be shown by using the spectral representation of $G_{1}$ in the expression of $\uppi(i\Omega)$,
\begin{equation} 
\begin{aligned}
\uppi(\Omega)
=&\int_{i\omega}G_{1}(i\omega)G^{0}_{2}(i\omega+i\Omega_{1})G^{0}_{3}(i\omega+i\Omega_{2})\\
=&\int_{i\omega}\int\frac{d\Omega}{2\pi}\frac{\rho(\Omega)}{i\omega-\Omega}G^{0}_{2}(i\omega+i\Omega_{1})G^{0}_{3}(i\omega+i\Omega_{2})\\
=&\int\frac{d\Omega}{2\pi}\rho(\Omega)
\int_{i\omega}\frac{1}{i\omega-\Omega}G^{0}_{2}(i\omega+i\Omega_{1})G^{0}_{3}(i\omega+i\Omega_{2}\\
=&\int\frac{d\Omega}{2\pi}\rho(\Omega)
\int_{i\omega}\frac{1}{i\omega-\Omega}\frac{1}{i\omega+i\Omega_{1}-\varepsilon_{F}}
\frac{1}{i\omega+i\Omega_{2}-\varepsilon_{F}}\\
\end{aligned}
\end{equation} 

\subsection{$GGG$}
Using the result in Eq.(\ref{13}),
we can write
\begin{equation} 
\begin{aligned}
\uppi(\Omega)
=&\int\frac{d\Omega'}{2\pi}\rho(\Omega')
\int^{\infty}_{-\infty}d\xi  \frac{1}{2\pi i}
[
G(\xi+i\eta)G(\xi-i\Omega_{1})G(\xi-i\Omega_{2})\\
&-
G(\xi-i\eta)G(\xi+i\Omega_{1})G(\xi-(i\Omega_{2}-i\Omega_{1}))\\
&-
G(\xi-i\eta)G(\xi+i\Omega_{2})G(\xi+(i\Omega_{2}-i\Omega_{1}))
]\\
=&\int\frac{d\Omega'}{2\pi}\rho(\Omega')
\int^{\infty}_{-\infty}d\xi  \frac{1}{2\pi i}
[
\frac{1}{\xi+i\eta-\Omega'}\frac{1}{\xi-i\Omega_{1}-\varepsilon_{F}}\frac{1}{\xi-i\Omega_{2}-\varepsilon_{F}}\\
&-
\frac{1}{\xi-i\eta-\Omega'}\frac{1}{\xi+i\Omega_{1}-\varepsilon_{F}}\frac{1}{\xi-i\Omega_{1}-\varepsilon_{F}}\\
&-
\frac{1}{\xi-i\eta-\Omega'}\frac{1}{\xi+i\Omega_{1}-\varepsilon_{F}}\frac{1}{\xi+2i\Omega_{1}-\varepsilon_{F}}
],
\end{aligned}
\end{equation} 
which turns out to be 
\begin{equation} 
\begin{aligned}
\uppi(\Omega)=
    \begin{cases}
             0, & \Omega\neq -\varepsilon_{F},\\  
             \int\frac{d\Omega'}{2\pi}\rho(\Omega')
\frac{-1}{(i\Omega_{1}+2\varepsilon_{F}+i\eta)(2(i\Omega_{1}+\varepsilon_{F})+i\eta)}, &     \Omega=-\varepsilon_{F},\\
    \end{cases}       
\end{aligned}
\end{equation} 
see also Eq.(46).
By performing the high-frequency expansion of $i\Omega_{1}$,
we obtain (for $\Omega=-\varepsilon_{F}$)
\begin{equation} 
\begin{aligned}
\uppi(\Omega)
=&\int\frac{d\varepsilon}{2\pi}
\frac{-\rho(-\varepsilon)}{2(i\Omega)^{2}}\\
=&\int\frac{d(-\varepsilon)}{2\pi}
\frac{\rho(-\varepsilon)}{2(i\Omega)^{2}},
\end{aligned}
\end{equation} 
thus 
\begin{equation} 
\begin{aligned}
\lim_{i\Omega\rightarrow\infty}(i\Omega)^{2}\uppi(\Omega)
=\frac{1}{2}\langle\{\Delta_{-\varepsilon},\Delta^{\dag}_{-\varepsilon}\}\rangle
=\frac{1}{2},
\end{aligned}
\end{equation} 
or when the fermi distribution is involved,
\begin{equation} 
\begin{aligned}
\lim_{i\Omega\rightarrow\infty}(i\Omega)^{2}\uppi(\Omega)N_{F}(-\varepsilon)
=\frac{1}{2}\langle\Delta_{-\varepsilon},\Delta^{\dag}_{-\varepsilon}\rangle
=\frac{1}{2}n_{-\varepsilon},
\end{aligned}
\end{equation}
Performing the high-frequency expansion to the Eq.(55) after the summation over $i\omega$,
we obtain
\begin{equation} 
\begin{aligned}
\uppi(\Omega)
=&\int\frac{d(-\varepsilon)}{2\pi}\rho(-\varepsilon)
\frac{-1}{(i\Omega_{1}+2\varepsilon+i\eta)(2(i\Omega_{1}+\varepsilon)+i\eta)}\\
\approx &\int\frac{d(-\varepsilon)}{2\pi}\rho(-\varepsilon)
[-\frac{1}{2(i\Omega)^{2}}+ \frac{3 \varepsilon}{2} \frac{1}{(i\Omega)^{3}} - 
 \frac{7}{8} 4 \varepsilon^{2}\frac{1}{(i\Omega)^{4}}  ]\\
=&\int\frac{d(-\varepsilon)}{2\pi}\rho(-\varepsilon)
(-\frac{1}{2(i\Omega)^{2}})
+ \int\frac{d(-\varepsilon)}{2\pi}\varepsilon\rho(-\varepsilon)\frac{3}{2} \frac{1}{(i\Omega)^{3}} 
-\int\frac{d(-\varepsilon)}{2\pi}\varepsilon^{2}\rho(-\varepsilon)
 \frac{7}{8} 4 \frac{1}{(i\Omega)^{4}}  ],
\end{aligned}
\end{equation} 
and the coefficients of this high-frequency expansion can be obtained by using the
relation for single-particle ($n_{3}$ here) spectral function (in contrast with the one shown in Eq.(\ref{sumrule}) 
with ${\rm sgn}[\chi'']=-{\rm sgn}[\rho]=-1$; $\rho(\varepsilon)=-2{\rm Im}G^{R}(\varepsilon)$,
$\rho(\tau)=\langle\{c(\tau),c^{\dag}(0)\}\rangle$)
\begin{equation} 
\begin{aligned}
\int^{\infty}_{-\infty}\frac{d(-\varepsilon)}{2\pi}\rho(-\varepsilon)=&1,\\
\int^{\infty}_{-\infty}\frac{d\omega}{2\pi}\omega\rho(\omega)
=&\langle \{-[c_{3},H],c_{3}^{\dag}\}\rangle\\
=&\{-\frac{\partial}{\partial\tau}c_{3}(\tau),c_{3}^{\dag}\}\\
=&\{-[-(\varepsilon_{3}(i\omega)+\Sigma_{3}^{(1)})c_{3}(0)],c_{3}^{\dag}\}\\
=&\{(\varepsilon_{3}(i\omega)+n_{3}g^{2})c_{3}(0),c_{3}^{\dag}\}\\
=&(\varepsilon_{3}(i\omega)+n_{3}g^{2}),\\
\int^{\infty}_{-\infty}\frac{d\omega}{2\pi}\omega^{2}\rho(\omega)
=&\langle \{-[-[c_{3},H],H],c_{3}^{\dag}\}\rangle\\
=&\{-[(\varepsilon_{3}(i\omega)+n_{3}g^{2}),H],c_{3}^{\dag}\}\\
=&\{-[(\frac{\partial}{\partial\tau}\varepsilon_{3}(i\omega)c_{3}(\tau)+\frac{\partial}{\partial\tau}n_{3}g^{2})c_{3}(\tau)],c_{3}^{\dag}\}\\
=&\{-(-\varepsilon_{3}(i\omega)(\varepsilon_{3}(i\omega)+n_{3}g^{2})c_{3}(\tau)-n_{3}g^{2}(\varepsilon_{3}(i\omega)+n_{3}g^{2})c_{3}(0)),c_{3}^{\dag}\}\\
=&\{\varepsilon_{3}(i\omega)(\varepsilon_{3}(i\omega)+n_{3}g^{2})c_{3}(\tau)+n_{3}g^{2}(\varepsilon_{3}(i\omega)+n_{3}g^{2})c_{3}(0),c_{3}^{\dag}\}\\
=&\{(\varepsilon^{2}_{3}(i\omega)+2n_{3}g^{2}\varepsilon_{3}(i\omega)+n^{2}_{3}g^{4})c_{3}(0),c_{3}^{\dag}\}\\
=&\varepsilon^{2}_{3}(i\omega)+2n_{3}g^{2}\varepsilon_{3}(i\omega)+n^{2}_{3}g^{4}\\
=&(\varepsilon_{3}(i\omega)+n_{3}g^{2})^{2}.
\end{aligned}
\end{equation} 
Thus we find that,
\begin{equation} 
\begin{aligned}
\label{61}
\int^{\infty}_{-\infty}\frac{d\varepsilon}{2\pi}\varepsilon^{2}\rho(\varepsilon)
-[\int^{\infty}_{-\infty}\frac{d\varepsilon}{2\pi}\varepsilon\rho(\varepsilon)]^{2}
=\langle\varepsilon^{2}\rangle-(\langle\varepsilon\rangle)^{2}
=0.
\end{aligned}
\end{equation} 
That is different to the result in the presence of single-site double occupation.
 The time derivation here is obtain in aid of the general definition of thermal Green's function
$G^{-1}(i\omega)=i\omega-(\varepsilon-\mu+\Sigma(i\omega))$ where the $\mu$ is the chemical potential of interacting electron
and setted as zero here.
For the susceptibility dressed by effective coupling,
we have
\begin{equation} 
\begin{aligned}
\chi(i\Omega)
=&\frac{\uppi(i\Omega)}{1-g_{eff}\uppi(i\Omega)}\\
\approx &\int\frac{d(-\varepsilon)}{2\pi}\rho(-\varepsilon)
[-\frac{1}{2(i\Omega)^{2}}+ \frac{3 \varepsilon}{2} \frac{1}{(i\Omega)^{3}} + 
 \frac{1}{8} (-28\varepsilon^{2}-2g_{eff})\frac{1}{(i\Omega)^{4}}  ]\\
=&\int\frac{d(-\varepsilon)}{2\pi}\rho(-\varepsilon)
(-\frac{1}{2(i\Omega)^{2}})
+ \int\frac{d(-\varepsilon)}{2\pi}\varepsilon\rho(-\varepsilon)\frac{3}{2} \frac{1}{(i\Omega)^{3}} 
+\int\frac{d(-\varepsilon)}{2\pi}\rho(-\varepsilon)
 \frac{1}{8} (-28\varepsilon^{2}-2g_{eff}) \frac{1}{(i\Omega)^{4}}  ].
\end{aligned}
\end{equation} 
While the high-frequency expansion of the spectral representation of $\chi(i\Omega)$ reads
\begin{equation} 
\begin{aligned}
\chi(i\Omega)
=&\int\frac{d\omega}{\pi}\frac{\chi''(\omega)}{i\Omega-\omega}
=&\frac{1}{i\Omega}\int\frac{d\omega}{\pi}\chi''(\omega)
+\frac{1}{(i\Omega)^{2}}\int\frac{d\omega}{\pi}\omega\chi''(\omega)
+\frac{1}{(i\Omega)^{3}}\int\frac{d\omega}{\pi}\omega^{2}\chi''(\omega)
+O((i\Omega)^{-4}),
\end{aligned}
\end{equation} 
with the coefficient in third term reads
\begin{equation} 
\begin{aligned}
\int^{\infty}_{-\infty}\frac{d\omega}{\pi}
\omega^{2}\chi''(i\omega)
=&[\frac{\partial}{\partial\tau}
(\varepsilon_{3}(i\omega)+n_{3}g^{2}-\varepsilon_{2}(i\omega)-n_{2}g^{2}-\varepsilon_{1}(i\omega)-n_{1}g^{2})  
c_{3}^{\dag}(\tau)c_{2}(\tau)c_{1}(\tau), c_{1}^{\dag}(\tau)c_{2}^{\dag}(\tau)c_{3}(\tau)]\\
=&
(\varepsilon_{3}(i\omega)+n_{3}g^{2}-\varepsilon_{2}(i\omega)-n_{2}g^{2}-\varepsilon_{1}(i\omega)-n_{1}g^{2}) ^{2}\\
&[(1-n_{1})(1-n_{2})n_{3}-n_{1}n_{2}(1-n_{3}))].
\end{aligned}
\end{equation} 
The interaction-dependence vanishes when the interacting chemical potential is taken into account,
Note that since $\chi''$ is negative unlike the single particle one $\rho$,
this coefficient (Eq.(64)) should also be negative,
thus $[(1-n_{1})(1-n_{2})n_{3}-n_{1}n_{2}(1-n_{3}))]=[\Delta,\Delta^{\dag}]<0$.
Then with $\int^{\infty}_{-\infty}\frac{d\omega}{\pi}
\omega\chi''(i\omega)$ obtained in Eq.(\ref{sumrule}),
we obtain a relation which is different to the Eq.(\ref{61}),
\begin{equation} 
\begin{aligned}
\int^{\infty}_{-\infty}\frac{d\omega}{\pi}
\omega^{2}\chi''(i\omega)
-
(\int^{\infty}_{-\infty}\frac{d\omega}{\pi}
\omega\chi''(i\omega))^{2}
=&\langle\omega^{2}\rangle-(\langle\omega\rangle)^{2}\\
=&(\varepsilon_{3}(i\omega)+n_{3}g^{2}-\varepsilon_{2}(i\omega)-n_{2}g^{2}-\varepsilon_{1}(i\omega)-n_{1}g^{2}) ^{2}\\
&[\Delta,\Delta^{\dag}](1-[\Delta,\Delta^{\dag}])<0.
\end{aligned}
\end{equation}

Next we turn to the imaginary time domain in high-fermion frequency limit,
before that, there are some important relations need to be noted.
Firstly, for in the zero external field limit,
the free fermion Green's function is in diagonal form as
\begin{equation} 
\begin{aligned}
G(\delta \tau)\equiv &G(\tau-\tau') =G(\tau,\tau') \\
G(\tau,\tau^{+})
=& \lim_{\delta\tau\rightarrow 0^{-}}\int\frac{d\omega}{2\pi}N_{F}(\omega)\rho(\omega)\\
=&-\langle c(\tau)c^{\dag}(\tau^{+})\rangle\\
=&\langle c^{\dag}(\tau)c(\tau^{-})\rangle\\
=&\langle n\rangle,\\
G(\tau,\tau^{-})
=&\lim_{\delta\tau\rightarrow 0^{+}}\int\frac{d\omega}{2\pi}N_{F}(\omega)\rho(\omega)\\
=&-\langle c(\tau)c^{\dag}(\tau^{-})\rangle\\
=&\langle c^{\dag}(\tau)c(\tau^{+})\rangle\\
=&\langle -1+n\rangle,
\end{aligned}
\end{equation} 
and the commutation/anticommutation relations
\begin{equation} 
\begin{aligned}
\{c_{i}(\tau),c_{j}^{\dag}(\tau')\}=\delta_{ij}\delta_{\tau\tau'},\\
[\Delta(\tau'),\Delta^{\dag}(\tau)]=\delta_{\tau\tau'}.
\end{aligned}
\end{equation} 
The particle-hole symmetry (or antiperiodicity condition) at half-filling ($n=1/2$) is described by
\begin{equation} 
\begin{aligned}
G(-\delta \tau)=-G(\delta \tau),\\
\Sigma(i\omega)-\Sigma(0)=-\Sigma(-i\omega)+\Sigma(0),
\end{aligned}
\end{equation}
in low-frequency (IR) limit,
or  
\begin{equation} 
\begin{aligned}
G(-\delta \tau)=-G(\delta \tau),\\
\Sigma(i\omega)-\Sigma(\infty)=-\Sigma(-i\omega)+\Sigma(\infty),
\end{aligned}
\end{equation}
in high-frequency (UV) limit, i.e., the non-perturbative treatment.
The imaginary time step is vanishingly small in UV limit of fermions $i\omega\rightarrow\infty$.
Here we assume both the particle and hole are close to the fermi surface.
Then according to Eq.(\ref{92}),
we have the following Ward identity expressionss for the the reducible vertex at zero temperature as
\begin{equation} 
\begin{aligned}
\Gamma(\tau,\tau')=\frac{\delta \Sigma_{1}(\tau',\tau)}{\delta G_{2}^{0}(\tau,\tau')}
=\frac{\delta \Sigma_{1}(\tau',\tau)}{\delta n_{2}},
\end{aligned}
\end{equation} 
this is in contrast with the Hartree-Fock result which ignores the discontinuity at $\delta\tau=0$,
In conserving approximation,
the fermion self-energy as a functional derivative of the Luttinger-Ward functional in UV limit reads
\begin{equation} 
\begin{aligned}
\Sigma_{1}(\tau,\tau')
=\frac{\delta \Phi[G]}{\delta G_{1}(\tau',\tau'')\delta G_{3}^{0}(\tau'',\tau^{\pm})},
\end{aligned}
\end{equation} 
where the Luttinger-Ward functional $\Phi[G]$ of $G$ can be written as
\begin{equation} 
\begin{aligned}
\label{LW}
\delta \Phi[G]
=g^{2}\langle\Delta^{\dag}(\tau)\Delta(\tau^{\pm})\rangle
=-g^{2}\langle\Delta(\tau^{\pm})\Delta^{\dag}(\tau)\rangle,
\end{aligned}
\end{equation} 
thus we have the self-consistent relation in high-frequency limit (see Appendix.C)
\begin{equation} 
\begin{aligned}
\Sigma_{1}(\tau,\tau')\delta G_{1}(\tau',\tau'')\delta G_{3}^{0}(\tau'',\tau^{+})
-
\Sigma_{1}(\tau,\tau')\delta G_{1}(\tau',\tau'')\delta G_{3}^{0}(\tau'',\tau^{-})
=&g^{2} n_{3}=\lim_{i\omega\rightarrow\infty}\Sigma_{1}(i\omega),
\end{aligned}
\end{equation} 
In the presence of finite external field (leads to the anomalous term of the fermion propagator),
 the matrix form of the above Luttinger-Ward functional reads
\begin{equation} 
\begin{aligned}
\delta \Phi[G]=g^{2}
\begin{pmatrix}
\langle\Delta^{\dag}(\tau')\Delta(\tau)\rangle & \langle\Delta(\tau')\Delta(\tau)\rangle\\
\langle\Delta^{\dag}(\tau')\Delta^{\dag}(\tau)\rangle     &  -\langle\Delta^{\dag}(\tau)\Delta(\tau')\rangle
\end{pmatrix},
\end{aligned}
\end{equation} 
where the diagonal elements have
\begin{equation} 
\begin{aligned}
    \langle\Delta^{\dag}(\tau')\Delta(\tau)\rangle 
=&-\langle\Delta(\tau)\Delta^{\dag}(\tau')\rangle\\
=&n_{1}n_{2}(1-n_{3})\\
=& \langle\Delta^{\dag}(\tau)\Delta(\tau)\rangle ,\\
    \langle\Delta^{\dag}(\tau)\Delta(\tau')\rangle 
=&-\langle\Delta(\tau')\Delta^{\dag}(\tau)\rangle\\
=&-(1-n_{1})(1-n_{2})n_{3},
\end{aligned}
\end{equation} 
as can be obtained by using Eq.(66).
The self-consistency between one- and three-point quantities can be seen from the above equations.
Note that at high-frequency limit with $\delta\tau\rightarrow 0$,
there is not a discontinuity in $\chi(\tau=0)$ (since $n_{1}+n_{2}=n_{3}$ is guaranteed by the operator $\Delta^{\dag}$ defined above
and thus $n=n_{1}+n_{2}+(1-n_{3})=1$),
and we have
\begin{equation} 
\begin{aligned}
\chi(\tau,\tau^{+})=\chi(\tau,\tau^{-})=\langle\Delta^{\dag}(\tau^{\pm})\Delta(\tau)\rangle
=-\langle\Delta(\tau)\Delta^{\dag}(\tau^{\pm})\rangle=\langle n_{1}n_{2}(1-n_{3})\rangle,
\end{aligned}
\end{equation} 
unlike the single-particle Green's function which has a jump in high-frequency limit,
\begin{equation} 
\begin{aligned}
G(\tau,\tau^{+})=-G(\tau,\tau^{-})=1/2.
\end{aligned}
\end{equation} 
Note that here the electron-hole symmetry half filling can also be seen
from $2G(\delta\tau^{+})-G(2\delta\tau^{-})=1$, in which case, by using $G(\tau)=-G(-\tau)$,
we can obtain $n_{1}+n_{2}=1/2=(1-n_{3})=n_{3}$.
When close to the half-filling,
it is dominated by the fermi-liquid phase and repulsive interaction,
and thus the Luttinger sum rule $\sum\theta(\mu_{int}-\Sigma^{(1)})$ can also be applied.

If we further replace all the bare Green's function in Eq.(94) by the dressed one
in spectral representation,
we obtain
\begin{equation} 
\begin{aligned}
\uppi(\Omega)
=&\int_{\omega}G_{1}(i\omega)G_{2}(i\omega+i\Omega_{1})G_{3}(i\omega+i\Omega_{2})\\
=&\int_{\omega}
\int\frac{d\varepsilon}{2\pi}\frac{\rho_{1}(\varepsilon)}{i\omega-\varepsilon}
\int\frac{d\varepsilon}{2\pi}\frac{\rho_{2}(\varepsilon)}{i\omega+i\Omega_{1}+\varepsilon}
\int\frac{d\varepsilon}{2\pi}\frac{\rho_{3}(\varepsilon)}{i\omega+2i\Omega_{1}+\varepsilon}\\
=&\int\frac{d^{3}\varepsilon}{(2\pi)^{3}}
\frac{\rho_{1}(\varepsilon) \rho_{2}(\varepsilon) \rho_{3} (\varepsilon)\pi}{\Omega^{2} - 3 i \Omega \varepsilon - 2 \varepsilon^{2}}, 
\end{aligned}
\end{equation} 
this result requires  $2\Omega< \varepsilon$, which means $\varepsilon$ cannot be zero.
Then in IR limit, the bosonic mass of the boson mode is
\begin{equation} 
\begin{aligned}
\uppi(0)=-m_{\phi}^{2}
=-\int\frac{d^{3}\varepsilon}{(2\pi)^{3}}\frac{\rho_{1}(\varepsilon) \rho_{2}(\varepsilon) \rho_{3} (\varepsilon)\pi}{2 \varepsilon^{2}}, 
\end{aligned}
\end{equation} 
and since ${\rm sgn}\rho=-{\rm sgn}\chi''=+$,
$m_{\phi}$ is a real quantity.
While for fermionic self-energy,
the noninteracting chemical potential can be obtained by
\begin{equation} 
\begin{aligned}
\Sigma_{1}(i\omega)
=&\int_{\Omega}
\int\frac{d\varepsilon}{\pi}\frac{\rho_{B}(\varepsilon)}{i\Omega-\varepsilon}
\int\frac{d\varepsilon}{2\pi}\frac{\rho_{F}(\varepsilon)}{i\Omega+i\omega+\varepsilon}\\
=&\int\frac{d^{2}\varepsilon}{2\pi^{2}}
\frac{2 i \rho_{B}(\varepsilon) \rho_{F}(\varepsilon) \pi}{
\sqrt{-4 i \omega\varepsilon + \omega^{2} -  4 \varepsilon^{2}}},
\end{aligned}
\end{equation} 
which becomes, in IR limit (and zero-temperature limit),
\begin{equation} 
\begin{aligned}
 \Sigma_{1}(0)= -\int\frac{d^{3}\varepsilon}{(2\pi)^{3}}\frac{\rho_{B}(\varepsilon) \rho_{F}(\varepsilon) \pi}{\varepsilon}
=-\mu_{0},
\end{aligned}
\end{equation} 
and $\mu_{0}=0$ at half-filling, which corresponds to $\mu_{int}=\Sigma_{i}(\infty)=g^{2}n_{i}$ in UV limit.
Similar to such counteract effect in UV limit with mean-field self-energy,
the boson mass gap can also be counteracted by the interacting chemical potential when away from half-filling (in the non-Fermi liquid phase).
Note that $\rho_{B}(\varepsilon) $ is given by a complicated form shown in Eq.(\ref{bosonspectral}) for IR limit,
but near half-filling,
it can simply be replaced by a delta function form,
\begin{equation} 
\begin{aligned}
\lim_{\Omega\rightarrow 0}\rho_{B}(i\Omega)
\equiv &\lim_{\Omega\rightarrow 0}\chi''(i\Omega)
=-2\pi\delta(i\Omega-0),\\
\lim_{\Omega\rightarrow 0}\rho_{B}^{2}(i\Omega)
=&\frac{-2\pi\delta(i\Omega-0)}{{\rm Im}\Sigma(i\Omega)}.
\end{aligned}
\end{equation} 
Note that here the imaginary part of self-energy is linear in frequency in IR limit.

By using the Hartree-Fock version of Eq.(\ref{LW}) with $\tau^{\pm}$ been replaced by $\tau$,
i.e., the equal-time approximation in IR limit,
the invariant fermion charge density (in IR limit) at half-filling can be obtained by using the current operator $J(\tau)$ 
\begin{equation} 
\begin{aligned}
J(\tau)
=&\sigma(\tau)G(-\tau-\delta\tau)G(\delta\tau)-\sigma(-\tau)G(\tau+\delta\tau)G(-\delta\tau)\\
=&\Sigma(\tau)G(-\tau-\delta\tau)G(\delta\tau)-\Sigma(-\tau)G(\tau+\delta\tau)G(-\delta\tau),
\end{aligned}
\end{equation} 
where $\sigma(\tau)=\Sigma(\tau)-\Sigma(\tau\rightarrow\infty)=\Sigma(\tau)-G(\tau)$
(see Appendix.B) is the UV source field
 which locals in the short time llimit,
and vanishing $\sigma$ in IR limit does not affects the estimation of above net current.
Then we have
\begin{equation} 
\begin{aligned}
n_{f}
=&\frac{-1}{2}\int^{\infty}_{-\infty}d\tau J(\tau)\tau\\
=&\frac{-1}{2}\int^{\infty}_{-\infty}d\tau
(\Sigma(\tau)G(-\tau-\delta\tau)G(\delta\tau)-\Sigma(-\tau)G(\tau+\delta\tau)G(-\delta\tau))\tau\\
=&\int^{\infty}_{-\infty}d\tau\Sigma(\tau)G(-\tau-\delta \tau)G(\delta\tau)\tau,
\end{aligned}
\end{equation} 
Through Fourier transform,
we obtain
\begin{equation} 
\begin{aligned}
n_{f}
=&\frac{1}{2\pi i}P\int^{\infty}_{-\infty}
[\partial_{\omega}G_{1}^{-1}(i\omega)]G_{1}(i\omega)G_{3}(i\omega+i\Omega_{2}),
\end{aligned}
\end{equation} 
which can be rewritten in IR limit of fermion ($i\omega\ll i\Omega$) as
\begin{equation} 
\begin{aligned}
n_{f}
=&\frac{1}{2\pi i}P\int^{\infty}_{-\infty}
[\partial_{\omega}G_{1}^{-1}(i\omega)]G_{1}(i\omega)G_{3}(i\Omega_{2})\\
=&-\frac{1}{2\pi i}P\int^{\infty}_{-\infty}
[\partial_{\omega}G_{1}(i\omega)]G^{-1}_{1}(i\omega)G_{3}(i\Omega_{2})\\
=&-\frac{1}{2\pi i}P\int^{\infty}_{-\infty}
\partial_{\omega}{\rm ln}G_{1}(i\omega)G_{3}(i\Omega_{2})\\
=&-G_{3}(i\Omega_{2})\frac{1}{2\pi i}
(\int^{\infty}_{\eta}
\partial_{\omega}{\rm ln}G_{1}(i\omega)
+\int^{-\eta}_{-\infty}
\partial_{\omega}{\rm ln}G_{1}(i\omega))\\
=&-G_{3}(i\Omega_{2})\frac{1}{2\pi i}
(\int^{\infty}_{\eta}
\partial_{\omega}{\rm ln}G_{1}(i\omega)
-\int^{\infty}_{\eta}
\partial_{\omega}{\rm ln}G_{1}(-i\omega))\\
=&-G_{3}(i\Omega_{2})\frac{1}{2\pi i}
\int^{\infty}_{\eta}
\partial_{\omega}{\rm ln}\frac{G_{1}(i\omega)}{G_{1}(-i\omega)}\\
=&-G_{3}(i\Omega_{2})\frac{1}{2\pi i}
(-i2\delta)\\
=&G_{3}(i\Omega_{2})\frac{\delta}{\pi},
\end{aligned}
\end{equation} 
where $\delta$ is a small angle (related to the phase shift).
This is similar to the fermi-liquid result shown in Eq.(36).
The principal value integral here is to avoid the logarithmic divergence in IR limit.
Note that in IR limit $G(\tau)=\Sigma(\tau) \sim |\tau|^{-1},\ \Sigma(i\omega)=-G^{-1}(i\omega)$,
that is also why UV source field vanishes in this limit.
This will be proved in detail in Appendix.B.

\subsection{Bosonic collective mode}

In IR limit with long-wavelength, the boson mode spectrum can be obtained by using the bare interaction,
however,  the collective excitations cannot be found even in this limit,
by using the IR expression of 
bare boson density-density correlation (as shown in Fig.3 and mentioned in Appendix.B).
There are three ways to make the collective mode exists. 
The first one is to turn back to the Eq.(46) or Eq.(96),
where the zero-temperature boson self-energy reads
\begin{equation} 
\begin{aligned}
\uppi(z)=&\frac{-1}{(z + 2 \varepsilon_{F} + i \eta) (2 (z + \varepsilon_{F}) + i \eta)},
\end{aligned}
\end{equation} 
or for $\varepsilon_{F}=0$,
\begin{equation} 
\begin{aligned}
\uppi(z)=&\frac{i (2 i z + \eta)}{2 z (2 z^2 + 3 i z \eta - \eta^2)}\\
=&\frac{-2  z }{2 z (2 z^2 + 3 i z \eta )},
\end{aligned}
\end{equation} 
whose real part is logarithmically vanishes in UV limit, and the imaginary part is zero
throughout the frequency.
But through the above calculations in Sec.5 (e.g., see Eq.(96)), we know that finite $\varepsilon$ is necessary to obtain well-defined boson mode,
even for the one which is very close to fermi surface (i.e., the boson momentum can be decomposed into two fermion momenta which are very close
to the same fermi surface).
From the above zero-temperature particle-hole boson self-energy, we found that,
for finite $\varepsilon_{F}$, it has $\uppi(\Omega)=\uppi^{*}(-\Omega)$ in the $\eta\rightarrow 0$ limit, but the spectral weight shows two insymmetry local maximum 
away from $\Omega=0$ as shown in Fig.4(b) 
(the particle-hole symmetry corresponds to the symmetry $-\uppi(\Omega)=\uppi^{*}(-\Omega)$).
It has a local minimum in $\Omega=0$ point similar to Fig.3(d),
but that cannot be seen from the spectral at half-filling as shown in Fig.3(c).
While for the RPA susceptibility calculated base on the above boson self-energy, we found
its frequency-dependence increases with the interaction strength.
By adding a frequency-independent Hartree-Fock term to the fermion self-energies (equals to the effect of mean-field shift)
in Eq.(13),
which leads to a result similar in form to Eq.(46), 
\begin{equation} 
\begin{aligned}
\uppi(i\Omega)=
-\frac{1}{(i\Omega+ 2 \varepsilon_{3} + 2 g^{2} n_{3} + i \eta) (2 i\Omega+ 2 \varepsilon_{3} + 2 g^{2} n_{3} + i \eta)}.
\end{aligned}
\end{equation} 
For this expression, we still has the symmetry $\uppi(\Omega)=\uppi^{*}(-\Omega)$.
The plots are shown in Fig.4(c-d).

The second way is by using the collective approximation (or mean-spherical approximation)\cite{Abedinpour S H,Bohm H M},
with the boson self-energy in retarded form
\begin{equation} 
\begin{aligned}
\uppi^{R}(i\Omega)&=\frac{2\varepsilon_{q}}{(\Omega+i\eta)^{2}+\varepsilon_{q}^{2}/S_{0}^{2}},\\
\varepsilon_{q}&=\int^{\infty}_{0}\frac{-1}{\pi}\Omega{\rm Im}\chi^{R}(i\Omega)d\Omega,
\end{aligned}
\end{equation} 
then the susceptibility (density-density response) can be written in a self-consistent form as
\begin{equation} 
\begin{aligned}
\chi^{R}(i\Omega)=\frac{\uppi^{R}(i\Omega)}{1-\frac{\varepsilon_{q}}{2}(\frac{1}{S^{2}}-\frac{1}{S^{2}_{0}})\uppi^{R}(i\Omega)},
\end{aligned}
\end{equation} 
where $S_{0}=\frac{-1}{\pi}\int^{\infty}_{0}d\Omega{\rm Im}\uppi^{R}(i\Omega)$,
$S=\frac{-1}{\pi}\int^{\infty}_{0}d\Omega{\rm Im}\chi^{R}(i\Omega)$ are the intercating and noninteracting static structure factors, respectively.
Obviously, this set of equation requires finite value of boson momentum,
while in long-wavelength limit where $q,\Omega\rightarrow 0$,
the static structure factors are not well-defined, e.g., $S_{0}=\frac{-1}{2 \pi \eta}{\rm ln}\frac{16}{9}\rightarrow\infty$ in this limit.
It is easy to see that, the boson self-energy in collective approximation is a Kronecker delta function for $q=0$: $\uppi(i\Omega)=\delta(i\Omega)$,
which satisfy the condition $\chi(q=0,i\Omega\neq 0)=0$, which is however violated in our model with three point excitations even for the undressed loop.

\section{The emergence of SYK model formed by bosons of three-point fermion loop}

In the presence of three-point fermion loop,
we assume the disordered complex Gaussian random coupling between fermions,
then the expectation $\langle\Delta^{\dag}\Delta\rangle$,
which is an essential quantity as we study in detail in above,
can be related to the SYK Hamiltonian.
For the case of $\langle\Delta_{\alpha}^{\dag}\Delta_{\beta}\rangle=\langle c^{\dag}_{i}c^{\dag}_{j}c_{k}c_{l}^{\dag}c_{m}c_{n}\rangle$,
the six fermion indices are independent with each other,
it is thus a SYK$_{6}$ model,
described by 
\begin{equation} 
\begin{aligned}
H_{SYK_{6}}=\frac{g_{ijklmn}}{2^{2}N^{5/2}}\sum^{N}_{ijklmn}c^{\dag}_{i}c^{\dag}_{j}c_{k}c_{l}^{\dag}c_{m}c_{n}.
\end{aligned}
\end{equation}
While for the case of $\langle\Delta_{\alpha}^{\dag}\Delta_{\alpha'}\rangle=\langle c^{\dag}_{i}c^{\dag}_{j}c_{k}c_{k'}^{\dag}c_{j'}c_{i'}\rangle$,
the six fermion indices are not completely independent with each other,
and we can only found two set of independent indices, $ijk$ and $k'j'i'$,
it is thus a SYK$_{3}\times$SYK$_{3}$ model,
described by 
\begin{equation} 
\begin{aligned}
\label{SYKHa}
H_{SYK_{3}\times SYK_{3}}=\frac{g_{ijk;k'j'i'}}{2^{5/2}N^{2}}\sum^{N}_{ijk;k'j'i'}c^{\dag}_{i}c^{\dag}_{j}c_{k}c_{k'}^{\dag}c_{j'}c_{i'}.
\end{aligned}
\end{equation}
The prefactors of the above two Hamiltonians can be obtained follow the rule found in Refs.\cite{wusyk1,wusyk2}.
We only focus on the latter one, i.e., the SYK$_{3}\times$SYK$_{3}$ model, in this paper,
where the SYK coupling satisfies the Gaussian distribution $\overline{g_{ijk;k'j'i'}}=0$,
and after disorder average in Gaussian unitary ensemble (GUE), we have
\begin{equation} 
\begin{aligned}
\frac{g_{ijk;k'j'i'}}{2^{5/2}N^{2}}\mathcal{O}_{ijk}^{\dag}\mathcal{O}_{k'j'i'}\rightarrow 
\frac{g^{2}2\mathcal{O}_{ijk}\mathcal{O}_{ijk}^{\dag}2\mathcal{O}^{\dag}_{k'j'i'}\mathcal{O}_{k'j'i'}}
{2^{5}N^{4}}
=\frac{g^{2}\mathcal{O}_{ijk}\mathcal{O}_{ijk}^{\dag}\mathcal{O}^{\dag}_{k'j'i'}\mathcal{O}_{k'j'i'}}
{2^{3}N^{4}},
\end{aligned}
\end{equation}
i.e., the variance is $\sigma^{2}=\frac{1}{2^{5}N^{4}}\overline{g_{ijk;k'j'i'}^{2}}=\frac{g^{2}}{2^{3}N^{4}}$.

The SYK non-fermi liquid appears in the presence of fermion incoherence near the
quantum critical point at zero temperature, and prevent the formation of pairs in Cooper channel.
Note that if the SYK coupling is short-range type (or on-site type) and strong enough,
the boson modes in interacting system can be gapless even without turn to the quantum critical point,
which is the so-called Bose metal in an incoherence critical metal phase.
However, when the pair condensation happen,
i.e., 
\begin{equation} 
\begin{aligned}
\langle\Delta_{\alpha}^{\dag}\Delta_{\alpha'}\rangle=&
\langle c^{\dag}_{i}c^{\dag}_{j}c_{k}c_{k'}^{\dag}c_{j'}c_{i'}\rangle\\
=&\langle c^{\dag}_{i}c^{\dag}_{j}(\delta_{k,k'}-c_{k'}^{\dag}c_{k})c_{j'}c_{i'}\rangle\\
\rightarrow &
\langle c^{\dag}_{i}c^{\dag}_{j}\rangle \langle c_{j'}c_{i'}\rangle\delta_{k,k'}-
\langle c^{\dag}_{i}c^{\dag}_{j}c^{\dag}_{k'}\rangle \langle c_{k}c_{j'}c_{i'}\rangle,
\end{aligned}
\end{equation}
the emergent off-diagonal long-range order and eigenvalue splitting (which leads to discrete spectrum),
and finite many-body spectral gap, will suppress the SYK non-fermi liquid phase,
and lead to the phase transition to disordered fermi liquid phase.
For $k\neq k'$, the Hamiltonian reads $H=-\frac{g_{ijk;k'j'i'}}{2^{5/2}N^{2}}\sum_{ijk;k'j'i'}\langle
c^{\dag}_{i}c^{\dag}_{j}c^{\dag}_{k'}\rangle \langle c_{k}c_{j'}c_{i'}\rangle$.

\subsection{Replica procedure}

Firstly, the Euclidean time path integral reads
\begin{equation} 
\begin{aligned}
Z=\int D[c^{\dag},c] e^{-S},
\end{aligned}
\end{equation}
where the action reads
\begin{equation} 
\begin{aligned}
S=&S_{0}+S_{int},\\
S_{0}=&\int d\tau d\tau' c^{\dag}(\tau)[\partial_{\tau}-\mu]\delta(\tau-\tau')c(\tau'),\\
S_{int}=&\int d\tau \frac{g_{ijk;k'j'i'}}{2^{5/2}N^{2}}\sum^{N}_{ijk;k'j'i'}c^{\dag}_{i}c^{\dag}_{j}c_{k}c_{k'}^{\dag}c_{j'}c_{i'}.
\end{aligned}
\end{equation}
After disorder average we have
\begin{equation} 
\begin{aligned}
Z=&\int D[c^{\dag},c] e^{-S_{0}}
\int D[c^{\dag},c]\int D[g_{ijk;k'j'i'},g_{ijk;k'j'i'}^{*}]\\
&e^{\frac{-1}{\sigma^{2}}\frac{1}{{2^{5}N^{4}}}|g_{ijk;k'j'i'}^{2}|}
e^{-(\frac{g_{ijk;k'j'i'}}{2^{5/2}N^{2}}\sum^{N}_{ijk;k'j'i'}c^{\dag}_{i}c^{\dag}_{j}c_{k}c_{k'}^{\dag}c_{j'}c_{i'}+H.c.)}\\
=&\int D[c^{\dag},c] e^{-S_{0}}
\int D[c^{\dag},c]\int D[g_{ijk;k'j'i'},g_{ijk;k'j'i'}^{*}]\\
&e^{\frac{-1}{\sigma^{2}}\frac{1}{{2^{5}N^{4}}}\overline|g_{ijk;k'j'i'}^{2}|}\\
&e^{-(\frac{g_{ijk;k'j'i'}}{2^{5/2}N^{2}}\sum^{N}_{ijk;k'j'i'}c^{\dag}_{i}c^{\dag}_{j}c_{k}c_{k'}^{\dag}c_{j'}c_{i'}
+\frac{g_{ijk;k'j'i'}}{2^{5/2}N^{2}}\sum^{N}_{ijk;k'j'i'}c^{\dag}_{i'}c^{\dag}_{j'}c_{k'}c_{k}^{\dag}c_{j}c_{i})}.
\end{aligned}
\end{equation}
Since 
\begin{equation} 
\begin{aligned}
&\int D[c^{\dag},c]\int D[g_{ijk;k'j'i'},g_{ijk;k'j'i'}^{*}]
e^{\frac{-1}{\sigma^{2}}\frac{1}{{2^{5}N^{4}}}\overline|g_{ijk;k'j'i'}^{2}|}\\
&e^{-(\frac{g_{ijk;k'j'i'}}{2^{5/2}N^{2}}\sum^{N}_{ijk;k'j'i'}c^{\dag}_{i}c^{\dag}_{j}c_{k}c_{k'}^{\dag}c_{j'}c_{i'}
+\frac{g_{ijk;k'j'i'}}{2^{5/2}N^{2}}\sum^{N}_{ijk;k'j'i'}c^{\dag}_{i'}c^{\dag}_{j'}c_{k'}c_{k}^{\dag}c_{j}c_{i})}\\
=&\int D[c^{\dag},c]\int D[g_{ijk;k'j'i'},g_{ijk;k'j'i'}^{*}]\\
&e^{\frac{-1}{\sigma^{2}}
(\frac{g_{ijk;k'j'i'}}{2^{5/2}N^{2}}+\sigma^{2}c^{\dag}_{i'}c^{\dag}_{j'}c_{k'}c_{k}^{\dag}c_{j}c_{i})
(\frac{g^{*}_{ijk;k'j'i'}}{2^{5/2}N^{2}}+\sigma^{2}c^{\dag}_{i}c^{\dag}_{j}c_{k}c_{k'}^{\dag}c_{j'}c_{i'})
}\\
&e^{\sigma^{2}(c^{\dag}_{i'}c^{\dag}_{j'}c_{k'}c_{k}^{\dag}c_{j}c_{i})(c^{\dag}_{i}c^{\dag}_{j}c_{k}c_{k'}^{\dag}c_{j'}c_{i'})}\\
=&e^{\sigma^{2}(c^{\dag}_{i'}c^{\dag}_{j'}c_{k'}c_{k}^{\dag}c_{j}c_{i})(c^{\dag}_{i}c^{\dag}_{j}c_{k}c_{k'}^{\dag}c_{j'}c_{i'})},
\end{aligned}
\end{equation}
where $D[c^{\dag},c]=\Pi_{\alpha=i,j,k,k',j',i'}dc^{\dag}_{\alpha}dc_{\alpha}$,
the action becomes
\begin{equation} 
\begin{aligned}
S_{int}
=&-\sigma^{2}(c^{\dag}_{i'}c^{\dag}_{j'}c_{k'}c_{k}^{\dag}c_{j}c_{i})(c^{\dag}_{i}c^{\dag}_{j}c_{k}c_{k'}^{\dag}c_{j'}c_{i'})\\
=&-\frac{g^{2}}{2^{3}N^{4}}(c^{\dag}_{i'}c^{\dag}_{j'}c_{k'}c_{k}^{\dag}c_{j}c_{i})(c^{\dag}_{i}c^{\dag}_{j}c_{k}c_{k'}^{\dag}c_{j'}c_{i'}).
\end{aligned}
\end{equation}
Using the Lagrange multiplier field
\begin{equation} 
\begin{aligned}
G(\tau,\tau')=\sum_{i}^{N}c^{\dag}_{i}(\tau')c_{i}(\tau),
\end{aligned}
\end{equation}
and the identity
\begin{equation} 
\begin{aligned}
1=&\int DG\delta(G(\tau',\tau)-\frac{1}{N}\sum_{i}c^{\dag}_{i}(\tau)c_{i}(\tau'))\\
=&\int DG\delta(NG(\tau',\tau)-\sum_{i}c^{\dag}_{i}(\tau)c_{i}(\tau'))\\
=&\int DGD\Sigma e^{N\Sigma(\tau,\tau')G(\tau',\tau)-\Sigma(\tau,\tau')\sum_{i}c^{\dag}_{i}(\tau)c_{i}(\tau')},
\end{aligned}
\end{equation}
we obtain in large-N limit (mean-field treatment)
\begin{equation} 
\begin{aligned}
S
=&\sum_{i}\int d\tau d\tau' c_{i}^{\dag}(\tau)[\partial_{\tau}-\mu+\Sigma(\tau,\tau')]\delta(\tau-\tau')c_{i}(\tau')-\frac{g^{2}N^{2}}{2^{3}}g^{3}(\tau',\tau)(-g(\tau,\tau'))^{3}
+N\Sigma(\tau,\tau')G(\tau',\tau)\\
=&-N\int d\tau d\tau' {\rm Tr}{\rm ln}(\partial_{\tau}-\mu+\Sigma(\tau,\tau'))-\frac{g^{2}N^{2}}{2^{3}}g^{3}(\tau',\tau)(-g(\tau,\tau'))^{3}
+N\Sigma(\tau,\tau')G(\tau',\tau).
\end{aligned}
\end{equation}
Then the saddle-point equations at half-filling (with particle-hole symmetry) are
\begin{equation} 
\begin{aligned}
\frac{\partial S(\tau,\tau')}{\partial \Sigma(\tau',\tau)}=NG(\tau,\tau')-\frac{N}{\partial_{\tau}+\Sigma(\tau,\tau')-\mu}=0,\\
\frac{\partial S(\tau,\tau')}{\partial G(\tau',\tau)}=N\Sigma(\tau,\tau')-\frac{g^{2}N^{2}}{2^{3}}6G^{5}(\tau',\tau)=0,
\end{aligned}
\end{equation}
thus
\begin{equation} 
\begin{aligned}
G(\tau,\tau')=&\frac{1}{\partial_{\tau}+\Sigma(\tau,\tau')},\\
\Sigma(\tau,\tau')=&g^{2}N\frac{3}{4}G^{5}(\tau',\tau)=-g^{2}N\frac{3}{4}G^{5}(\tau,\tau'),
\end{aligned}
\end{equation}
that is, in frequency space,
\begin{equation} 
\begin{aligned}
G(i\omega)=&\frac{1}{-i\omega+\Sigma(i\omega)},\\
\Sigma(i\omega)=&-g^{2}N\frac{3}{4}G^{5}(i\omega).
\end{aligned}
\end{equation}

\subsection{Density calculation of bosons in SYK model base on Luttinger-Ward analysis}

Next we calculate the boson number densities by using the boson Green's function 
(instead of the order parameter propagator) base on the Luttinger-Ward procedure,
\begin{equation} 
\begin{aligned}
n_{b}
=&-\frac{1}{\beta}\sum_{n'}G_{B}(i\Omega_{n'})e^{i\Omega_{n'}\eta}\\
=&-\int^{i\infty}_{-i\infty}\frac{dz}{2\pi i}
G_{B}(z)e^{z\eta}N_{b}(z),
\end{aligned}
\end{equation}
where $N_{b}(z)=1/(e^{\beta z}-1)$, $G_{B}(i\Omega_{n'})=i\Omega_{n'}-g^{3}\uppi(\Omega_{n'})$ ($g$ denotes the interaction vertex). 
Using the identity
$\partial_{z}(G_{B}^{-1}+g^{3}\uppi)=1$,
we have
\begin{equation} 
\begin{aligned}
n_{b}
=&-\int^{i\infty}_{-i\infty}\frac{dz}{2\pi i}
G_{B}(z)e^{z\eta}\partial_{z}(G_{B}^{-1}(z)+g^{3}\uppi(z))
N_{b}(z)\\
=&-\int^{i\infty}_{-i\infty}\frac{dz}{2\pi i}
G_{B}(z)e^{z\eta}\partial_{z}G_{B}^{-1}(z)
N_{b}(z)
-\int^{i\infty}_{-i\infty}\frac{dz}{2\pi i}
G_{B}(z)e^{z\eta}\partial_{z}g^{3}\uppi(z)
N_{b}(z)
\equiv I_{1}+I_{2}.
\end{aligned}
\end{equation} 
The integral in the first term can be rewritten as
\begin{equation} 
\begin{aligned}
I_{1}= &
-\int^{i\infty}_{-i\infty}\frac{dz}{2\pi i}
G_{B}(z)e^{z\eta}\partial_{z}G_{B}^{-1}(z)
N_{b}(z)\\
=&
\int^{i\infty}_{-i\infty}\frac{dz}{2\pi i}
G_{B}^{-1}(z)e^{z\eta}\partial_{z}G_{B}(z)
N_{b}(z)\\
=&
\int^{i\infty}_{-i\infty}\frac{dz}{2\pi i}
\partial_{z}{\rm ln}G_{B}(z)e^{z\eta}
N_{b}(z).
\end{aligned}
\end{equation} 
Note that the second step is valid even when one of the dressed boson propagator is replaced by the bosonic self-energy,
i.e.,
\begin{equation} 
\begin{aligned}
\int\frac{dz}{2\pi i}D(z)\partial_{z}\uppi(z)=-\int\frac{dz}{2\pi i}\uppi(z)\partial_{z}D(z),
\end{aligned}
\end{equation} 
for SYK model or Luttinger-Ward model\cite{Luttinger J M}.
For $I_{1}$, by taking the principal value integral along the contour,
$P\int^{i\infty}_{-i\infty}=\int^{i\infty}_{i\eta}+\int^{-i\eta}_{-i\infty}=\int^{\infty+i\eta}_{i\eta}-\int^{\infty-i\eta}_{-i\eta}
=\int^{-\infty+i\eta}_{i\eta}-\int^{-\infty-i\eta}_{-i\eta}$,
and defining the phase shift
\begin{equation} 
\begin{aligned}
\delta=\frac{i}{2} {\rm ln}\frac{G(z+i\eta)}{G(z-i\eta)}
=\frac{1}{-i} {\rm ln}\frac{G(z+i\eta)}{G(z)}
=\frac{1}{i} {\rm ln}\frac{G(z-i\eta)}{G(z)}
=-{\rm Arg}G(z+i\eta),
\end{aligned}
\end{equation} 
we arrive at
\begin{equation} 
\begin{aligned}
I_{1}
=&P\int^{i\infty}_{-i\infty}\frac{dz}{2\pi i}
\partial_{z}{\rm ln}G_{B}(z)e^{z\eta}
N_{B}(z).
\end{aligned}
\end{equation} 
At zero temperature limit, $N_{b}(z)\rightarrow -1$,
thus 
\begin{equation} 
\begin{aligned}
I_{1}
=&-P\int^{i\infty}_{-i\infty}\frac{dz}{2\pi i}
\partial_{z}{\rm ln}G_{B}(z)e^{z\eta}\\
=&-(\int^{\infty+i\eta}_{0+i\eta}-\int^{\infty-i\eta}_{0-i\eta})\frac{dz}{2\pi i}
\partial_{z}{\rm ln}G_{B}(z)e^{z\eta}\\
=&-\int^{\infty}_{0}\frac{dz}{2\pi i}\partial_{z}{\rm ln}\frac{G_{B}(z+i\eta)}{G_{B}(z-i\eta)}e^{z\eta}\\
=&\frac{1}{\pi}({\rm Arg}G_{B}(\infty+i\eta)-{\rm Arg}G_{B}(i\eta))\\
=&\frac{1}{\pi}(0-(-\frac{\pi}{2}-\theta))\\
=&\frac{1}{2}+\frac{\theta}{\pi}.
\end{aligned}
\end{equation} 
This fermi-liquid result can also be obtained by using the relation between
Feynman Green's function $G^{F}_{B}(i\Omega)$ and the retarded Green's function $G^{R}_{B}(i\Omega)\equiv G_{B}(\Omega+i\eta)$,
\begin{equation} 
\begin{aligned}
\label{37}
I_{1}'
=&-P\int^{\infty}_{-\infty}\frac{d\Omega}{2\pi i}
\partial_{\Omega}{\rm ln}G_{B}^{F}(\Omega)e^{i\Omega\eta}\\
=&\frac{1}{2}-\frac{\theta}{\pi},
\end{aligned}
\end{equation} 
\begin{equation} 
\begin{aligned}
\label{38}
-P\int^{\infty}_{-\infty}\frac{d\Omega}{2\pi i}
\partial_{\Omega}{\rm ln}G_{B}(\Omega+i\eta)e^{i\Omega\eta}
-(-I'_{1})=&
P\int^{\infty}_{-\infty}\frac{d\Omega}{2\pi i}
\partial_{\Omega}{\rm ln}G_{B}(\Omega+i\eta)e^{i\Omega\eta}
-
P\int^{\infty}_{-\infty}\frac{d\Omega}{2\pi i}
\partial_{\Omega}{\rm ln}G_{B}^{F}(\Omega)e^{i\Omega\eta}\\
=&\frac{{\rm Arg}G_{B}(-\eta+i\eta)-{\rm Arg}G_{B}(-\infty+i\eta)}{\pi}\\
=&\frac{\frac{-3\pi}{4}-\theta-(-\pi)}{\pi}\\
=&\frac{1}{4}-\frac{\theta}{\pi},
\end{aligned}
\end{equation} 
where the Feynman Green's function and the retarded Green's function are related by the relation
$\frac{1}{x+i\eta}=P(\frac{1}{x})-i\pi\delta(x)$,
and the first term in the left hand side can be obtained as
\begin{equation} 
\begin{aligned}
\label{39}
P\int^{\infty}_{-\infty}\frac{d\Omega}{2\pi i}
\partial_{\Omega}{\rm ln}G_{B}(\Omega+i\eta)e^{\Omega\eta}
=&P\int^{i\infty}_{-i\infty}\frac{dz}{2\pi i}
\partial_{z}{\rm ln}G_{B}(z+i\eta)e^{iz\eta}\\
=&(\int^{\infty+i\eta}_{0+i\eta}-\int^{\infty-i\eta}_{0-i\eta})\frac{dz}{2\pi i}
\partial_{z}{\rm ln}G_{B}(z+i\eta)e^{iz\eta}\\
=&\int^{\infty}_{0}\frac{dz}{2\pi i}
\partial_{z}{\rm ln}\frac{G_{B}(z+2i\eta)}{G_{B}(z)}e^{iz\eta}\\
=&\int^{\infty}_{0}\frac{dz}{2\pi i}
\partial_{z}(-i\delta)e^{iz\eta}\\
=&\int^{\infty}_{0}\frac{dz}{2\pi i}
\partial_{z}(i{\rm Arg}G_{B}(z+2i\eta))e^{iz\eta}\\
=&\int^{\infty}_{0}\frac{dz}{2\pi}
\partial_{z}{\rm Arg}G_{B}(z+2i\eta)e^{iz\eta}\\
=&\frac{{\rm Arg}(\infty+2i\eta)-{\rm Arg}(0+2i\eta)}{2\pi}\\
=&\frac{0-(-\frac{\pi}{2}-\theta)}{2\pi}\\
=&\frac{1}{4}+\frac{\theta}{2\pi}\approx \frac{1}{4},
\end{aligned}
\end{equation} 
Substituting Eq.(\ref{39}) back into Eq.(\ref{38}), we obtain the result in Eq.(\ref{37}).
Note that
in the integral path of thrid line, we donot add an infinitesimal shift in real axis since there is not IR divergent
for the integrand.

Eq.(\ref{38}) can also be cast into the integral over the complex frequency $z$ as
\begin{equation} 
\begin{aligned}
P\int^{i\infty}_{-i\infty}\frac{dz}{2\pi i}
\partial_{z}{\rm ln}G^{R}_{B}(z)e^{z\eta}
-P\int^{\infty}_{-\infty}\frac{dz}{2\pi i}
\partial_{z}{\rm ln}G_{B}^{F}(z)e^{z\eta},
\end{aligned}
\end{equation} 
where the Feynman and retarded propagator can be simply written as
\begin{equation} 
\begin{aligned}
G^{F}_{B}(z)=&\frac{1}{z^{2}+i\eta},\\
G^{R}_{B}(z)=&\frac{1}{z(z+i\eta)}.
\end{aligned}
\end{equation} 
Each propagator has two poles, $z_{+}z_{-}=z^{2}$ and $z_{+}\neq z_{-}$.
The poles of Feynman propagator can be easily solved as
$z^{2}=-i\eta$, $z_{\pm}=(-i\eta)/z_{\mp}$, i.e., the two poles locate in the imaginary axis
(one in the upper half plane and the another in the lower half plane); 
while the poles of retarded propagator cannot be exactly solved which requires $z_{+}z_{-}=-\eta^{2}$ and $\sqrt{z_{+}z_{-}}=-i\eta$
be satisfied in the same time,
thus the poles can only be approximatly written as $z_{\pm}=(-\eta^{2})/z_{\mp}-i\eta$,
i.e., the poles locate only in the lower half plane.
Then by choosing the integral contour in upper half-plane,
the integral over $z$ of $G_{B}^{F}(z)$ (contribution from residues at poles) vanishes, i.e.,
\begin{equation} 
\begin{aligned}
\sum_{z_{0}}{\rm Res}_{{\rm z}>0}G_{B}^{R}=0.
\end{aligned}
\end{equation} 
Thus Eq.(\ref{37}) reduces to 
\begin{equation} 
\begin{aligned}
P\int^{i\infty}_{-i\infty}\frac{dz}{2\pi i}
\partial_{z}{\rm ln}G_{B}^{F}(z)e^{z\eta}
=&(\int^{\infty+i\eta}_{0+i\eta}-\int^{\infty-i\eta}_{0-i\eta})\frac{dz}{2\pi i}
\partial_{z}{\rm ln}G_{B}^{F}(z)e^{z\eta}\\
=&\int^{\infty}_{0}\frac{dz}{2\pi i}\partial_{z}{\rm ln}\frac{G^{F}_{B}(z+i\eta)}{G^{F}_{B}(z-i\eta)}e^{z\eta}\\
=&\frac{1}{2}-\frac{\theta}{\pi},
\end{aligned}
\end{equation} 
which is the same with the above result.
Note that this integral $G_{B}^{F}(z)$ is free from IR divergence (unlike the $G_{B}^{R}(z)$), and we can safely let $z$ be zero during the integration,
and thus there will also not be a branch cut along the real axis.
While for the retarded one $G_{B}^{R}(z)$, although there is a branch cut along real axis ($z$ cannot be zero),
but since there is only one integration contour, the contribution from branch cut vanishes,
i.e., $\int\frac{dz}{2\pi i}[G^{R}_{B}(z+i\eta)-G^{R}_{B}(z-i\eta)]=\int\frac{dz}{2\pi i}2i{\rm Im}G^{R}_{B}(z)=0$.
The contour setted in above is closed at $|z|=\infty$ since the integrands are convergent,
\begin{equation} 
\begin{aligned}
\int^{\infty}_{-\infty}dz G_{B}^{F}=&f(\eta),\\
{\rm P}\int^{\infty}_{-\infty}dz G^{R}_{B}=&\pi/\eta,
\end{aligned}
\end{equation} 
where $f$ is an analytic function of $\eta$.

$I_{1}$ term correspond to the fermi liquid result with the summation of states under fermi surface being ignored here.
After the analytic continuation, the integral $I_{2}$ at zero temperature limit reads
\begin{equation} 
\begin{aligned}
I_{2}
=&\int^{i\infty}_{-i\infty}\frac{dz}{2\pi i}
G_{B}(z)e^{z\eta}\partial_{z}g^{3}\uppi(z)\\
=&\int^{\infty}_{-\infty}\frac{dz}{2\pi}
{\rm Im}
[G_{B}(\Omega)e^{\Omega\eta}\partial_{\Omega}g^{3}\uppi(\Omega)].
\end{aligned}
\end{equation} 
After calculation
the result turns out to be zero, which is obtained by using
\begin{equation} 
\begin{aligned}
\uppi(z)
=&\int^{\infty}_{-\infty}\frac{d\xi}{2\pi i}
[G(\xi+i\eta)G(\xi-z)G(\xi-2z)\\
&-G(\xi-i\eta)G(\xi+z)G(\xi-z)-G(\xi-i\eta)G(\xi+z)G(\xi+2z)]\\
=&\frac{-1}{(z + 2 \varepsilon_{F} + i \eta) (2 (z + \varepsilon_{F}) + i \eta))},\\
I_{2}
=&\int^{i\infty}_{-i\infty}\frac{dz}{2\pi i}
\frac{e^{z\eta}\partial_{z}g^{3}\uppi(z)}{z-g^{3}\uppi(z)}
\approx 0.
\end{aligned}
\end{equation} 
This is obtained by taking the limit $\eta\rightarrow 0$ before integration.
This is valid for large $z$,
but invaid for IR limit.
To obtain the more precise result,
we use the expression of self-energy in IR limit,
i.e., for a long-time behavior,
which reads $\uppi(\tau)=G_{B}(\tau)^{1}(-G_{B}(-\tau)^{1-1})=G_{B}(\tau)$ for two-fermions interaction (i.e., $q=2$ mode\cite{Gu Y}),
then we have
\begin{equation} 
\begin{aligned}
\uppi(z)
=&-\int_{\Omega_{1}>0\ \cup\ \Omega_{1}<0}d\Omega_{1}
    \frac{1}{\pi}\frac{\rho(\Omega_{1})}{\Omega_{1}-z-i\eta{\rm sgn}\Omega_{1}}\\
=&P\int d\Omega_{1}
    \frac{1}{\pi}\frac{\rho(\Omega_{1})}{z-\Omega_{1}+i\eta{\rm sgn}\Omega_{1}},\\
\partial_{z}\uppi(z)
=&\frac{1}{\pi} P\int d\Omega_{1}\frac{-\rho(\Omega_{1})}{(z-\Omega_{1}+i\eta{\rm sgn}\Omega_{1})^{2}},\\
I_{2}
=&P\int^{i\infty}_{-i\infty}\frac{dz}{2\pi i}
G^{F}_{B}(z)\partial_{z}\uppi(z)e^{z\eta}\\
=&P\int^{i\infty}_{-i\infty}\frac{dz}{2\pi i}
[P\int^{\infty}_{-\infty} d\Omega_{0}\frac{\rho(\Omega_{0})}{\pi[z-(\Omega_{0}-i\eta{\rm sgn}\Omega_{0})]}
\frac{1}{\pi}P\int^{\infty}_{-\infty} d\Omega_{1}\frac{-\rho(\Omega_{1})}{(z-\Omega_{1}+i\eta{\rm sgn}\Omega_{1})^{2}}
],
\end{aligned}
\end{equation} 

At zero temperature in which case $N_{F}=\theta(\varepsilon_{F}-\xi)$,
if we use the free form of boson and fermion Green's function, then
we have
\begin{equation} 
\begin{aligned}
\label{47}
\uppi(\Omega)
=&
\frac{-1}{2 (\Omega + \varepsilon_{F} + i \eta) (\Omega + 2 \varepsilon_{F} + 2 i \eta)},\\
I_{2}
=&-\int^{\infty}_{-\infty}\frac{dz}{2\pi}
{\rm Im}[G_{B}(\Omega)e^{\Omega\eta}\partial_{\Omega}g^{3}\uppi(\Omega)]=0,
\end{aligned}
\end{equation} 
i.e., the $I_{2}$ vanishes as long as we let the convergent factor $e^{\Omega\eta}\rightarrow 1$ before integration.
This is obviously not the non-fermi-liquid result.
Similar to Ref.\cite{Altshuler B L},
we can use a full derivative form to deal with the above integral,
\begin{equation} 
\begin{aligned}
I_{2}
=&-\int^{\infty}_{-\infty}\frac{d\Omega}{2\pi}
{\rm Im}
\partial_{\Omega}Y(\Omega)
\frac{e^{\Omega\eta}}{e^{\beta \Omega}-1},\\
\delta Y=&\int\frac{d\Omega}{2\pi}
\uppi(\Omega)\delta G_{B}(\Omega^{1/3}).
\end{aligned}
\end{equation} 
Nonzero $I_{2}$ requires a finite nonanalytic term in function $Y$,
which is
\begin{equation} 
\begin{aligned}
Y_{NA}\sim &
{\rm ln}\frac{G_{0}(\omega)G_{0}(\omega+\Omega_{1})G_{0}(\omega+\Omega_{2})}
{G(\omega)G(\omega+\Omega_{1})G(\omega+\Omega_{2})}\\
=&{\rm ln}\frac{\uppi_{nsc}(\Omega)}
{\uppi_{sc}(\Omega)},
\end{aligned}
\end{equation} 
where $\uppi_{nsc}(\Omega)$ and $\uppi_{sc}(\Omega)$ correspond to the non-self-consistent and self-consistent self-energies.
In IR asymptotic, $G_{B}^{-1}(\Omega)=-\uppi(\Omega)$,
thus a finite and analytic self-energy guarantees the absence of singularities or zeros of $G_{B}(\Omega)$.
Only for the case that $\uppi_{nsc}(\Omega)$ can be continuously turned to $\uppi_{sc}(\Omega)$ in the frequency space
(even to the IR limit),
the integral $I_{2}$ can be reduced to the values of $Y(\Omega)$ to UV and IR limits (in the low-temperature limit)
which cancels out by the integration domain ($I_{2}=0$).
Note that the self-energy $\uppi(\Omega)$ can be rewritten as the functional derivative
\begin{equation} 
\begin{aligned}
\uppi^{nsc/sc}(\Omega)
=-T\frac{\delta \Phi^{nsc/sc}[G_{B}]}{\delta G_{B}}
=-T\frac{\delta \Phi^{nsc/sc}[G_{B}]}{\delta (i\Omega-\uppi^{nsc/sc}(\Omega))^{-1}},
\end{aligned}
\end{equation} 
or in IR limit as $\uppi^{nsc/sc}(\Omega)=\delta \Phi^{nsc/sc}[G_{B}]
\delta\uppi^{nsc/sc}(\Omega)$,
where $\Phi^{nsc/sc}[G_{B}]$ is the infinite perturbation series of three-density Feynman diagrams
consistent of the bare/dressed fermion propagators.
The $\Phi^{nsc/sc}[G_{B}]$ can also be referred to the Luttinger-Ward functional in fermi-liquid theory,
which can be expressed by the product of foward and backward propagators in time domain,
$\Phi^{nsc/sc}[G_{B}(\tau)]=\int d\tau G^{2}_{B}(\tau)G^{2}_{B}(-\tau)$.
Nonzero $Y_{NA}$ reveals that $\Phi^{nsc}[G_{B}]/\Phi^{sc}[G_{B}]$ cannot be perturbatively expanded order by order.
Note that the discussion here is valid when the $G_{B}$ is replaced by the boson field propagator $D$.

To deal with Luttinger-Ward integral $I_{2}$ at zero temperature in non-fermi-liquid picture, we use the IR asymptotic (i.e., $\tau\rightarrow\infty$
in the imaginary time domain) of the fermion and boson dressed self-energies.
For $q=2$ mode (the scaling dimension of the field is $1/2$),
$G_{B}(\tau)\equiv G_{B}(\tau_{1},\tau_{1}+\tau)=\uppi(\tau)=|\tau|^{-1}$,
then we have the following self-consistent relations
\begin{equation} 
\begin{aligned}
\uppi(i\Omega)=&-\int^{\infty}_{-\infty}\frac{d\omega}{2\pi}
\Sigma^{-1}(i\omega)\Sigma^{-1}(i\omega+i\Omega_{1})\Sigma^{-1}(i\omega+i\Omega_{2}),\\
\Sigma(i\omega)=&-\int^{\infty}_{-\infty}\frac{d\Omega}{2\pi}
\uppi^{-1}(i\Omega)\Sigma^{-1}(i\omega-i\Omega_{1}),\\
\Sigma(\pm i\omega)=&\pm ie^{\mp i\theta}\omega_{NFL}^{1-\alpha}\omega^{\alpha}.
\end{aligned}
\end{equation} 
Note that the non-fermi-liquid energy scale $\omega_{NFL}$ ($\gg\Omega,\omega$ in IR limit) is fixed and indispensable for $q=2$ mode
(the bosonic version of SYK$_{2}$ model).
Since in IR limit the linear frequency term can be treated as a small perturbation,
the fermion Green's function reads
\begin{equation} 
\begin{aligned}
G_{F}(\pm i\omega)
=\pm ie^{\pm i\theta}\omega_{NFL}^{-1+\alpha}\omega^{-\alpha},
\end{aligned}
\end{equation} 
with the factor $0\le \alpha<\frac{1}{2}$.
Note that $ie^{i\theta}\approx (i+{\rm tan}\theta)$ in IR limit.
The $G_{F}$ is logarithmically divergent in $\omega\rightarrow 0$ limit.

Then through calculation we have
\begin{equation} 
\begin{aligned}
\label{54}
\uppi( i\Omega)=&
-(1/(4 \Gamma[\alpha] \Gamma[(1 + \alpha)/2]))
 i e^{3 i \theta} \Omega^(1 - 3 \alpha) \sqrt{\pi}
   \omega_{NFL}^{-3 +3 \alpha} \\
&(2 \sqrt{\pi}
      {\rm csc}[2 \alpha \pi] \Gamma[-\frac{1}{2} + (3 \alpha)/2] + 
    2^\alpha {\rm csc}[\alpha \pi] \Gamma[-\frac{1}{2} + \alpha] \Gamma[(1 + \alpha)/
      2] {}_{2}F_{1}[1 - \alpha, \alpha, 2 - 2 \alpha, 1/2])\\
&	  -(1/(2 \Gamma[\alpha]))
 i (-1)^{-\alpha} e^{-3 i \theta} (-(1/\Omega))^{\alpha} \Omega^{1 - 2 \alpha}
   \omega_{NFL}^{-3 +3 \alpha} \\
&(\sqrt{\pi}
      {\rm csc}[\alpha \pi] \Gamma[-\frac{1}{2} + \alpha] {}_{2}F_{1}[
      1 - \alpha, \alpha, 2 - 2 \alpha, 2] + 
    2 \Gamma[1 - 2 \alpha] \Gamma[-1 + 3 \alpha] {}_{2}F_{1}[
      \alpha, -1 + 3 \alpha, 2 \alpha, 2])\\
\propto &
-\Omega^{1 - 3 \alpha}  \omega_{NFL}^{-3 +3 \alpha}
-\Omega^{1 - 2 \alpha}   \omega_{NFL}^{-3 +3 \alpha},
\end{aligned}
\end{equation} 
thus 
\begin{equation} 
\begin{aligned}
G_{B}(i\Omega)\propto  -[g^{3}(
-\Omega^{1 - 3 \alpha}   \omega_{NFL}^{-3 +3 \alpha}
-\Omega^{1 - 2 \alpha}   \omega_{NFL}^{-3 +3 \alpha})]^{-1}.
\end{aligned}
\end{equation} 
The obtained boson mode self-energies in IR limit are shown in Fig.3,
for the cases of half-filling ($\theta=0,\ \alpha=0.16$; Fig.3(a)) and away from the half filling ($\theta=\pi/4,\ \alpha\rightarrow 0.5$; Fig.3(b)).
While at full filling which corresponds to $\theta=\pi/2$,
the boson self-energy reduces to $-\uppi(0)=m_{\phi}^{2}$ which is almost a constant.

Since the variable transformation in analytical continuation $i\omega\rightarrow \omega+i\eta$
is equivalent to $\omega\rightarrow -i\omega+i\eta\approx -i\omega$,
the spectral density reads
\begin{equation} 
\begin{aligned}
\label{bosonspectral}
\rho(\Omega)=&{\rm Im}G_{B}^{R}
={\rm Im}[-g^{3}\uppi^{R}(\Omega)]^{-1}\\
\approx &
-((2 g^{3} (\Omega^{2})^{-\frac{1}{2} + \frac{3 \alpha}{2}}
      \omega_{NFL}^{4} (\omega_{NFL}^{2})^{-\frac{1}{2} - \frac{3 \alpha}{2}}
      \Gamma[\alpha]\\
& ((\frac{1}{\Omega^{2}})^{\alpha/2} (\Omega^{2})^{\alpha/2}         {\rm cos}[\alpha \pi + 3 \theta - 
          \alpha {\rm Arg}[-(i/\Omega)] + (-1 + 2 \alpha) {\rm Arg}[-i \Omega] - 
          3 (-1 + \alpha) {\rm Arg}[\omega_{NFL}]] \\
		  &+ 
       2 \sqrt{\pi}
         {\rm cos}[3 \theta + (1 - 3 \alpha) {\rm Arg}[-i \Omega] + 
          3 (-1 + \alpha) {\rm Arg}[\omega_{NFL}]] \Gamma[\alpha]))
		  /((\frac{1}{\Omega^{2}})^{\alpha} (\Omega^{2})^{\alpha} \\
		  &+ 
     4 (\frac{1}{\Omega^{2}})^{\alpha/2} (\Omega^{2})^{\alpha/2} \sqrt{\pi}
       {\rm cos}[\alpha \pi + 6 \theta - \alpha {\rm Arg}[-(i/\Omega)] - 
        \alpha {\rm Arg}[-i \Omega]] \Gamma[\alpha] + 4 \pi \Gamma[\alpha]^{2}))\\
=&
-((2 g^{3} (\Omega^{2})^{-\frac{1}{2} + \frac{3 \alpha}{2}}
      \omega_{NFL}^{4} (\omega_{NFL}^{2})^{-\frac{1}{2} - \frac{3 \alpha}{2}}
      \Gamma[\alpha] \\
&     (   {\rm cos}[\alpha \pi + 3 \theta - 
          \alpha {\rm Arg}[-(i/\Omega)] + (-1 + 2 \alpha) {\rm Arg}[-i \Omega] - 
          3 (-1 + \alpha) {\rm Arg}[\omega_{NFL}]] \\
		  &+ 
       2 \sqrt{\pi}
         {\rm cos}[3 \theta + (1 - 3 \alpha) {\rm Arg}[-i \Omega] + 
          3 (-1 + \alpha) {\rm Arg}[\omega_{NFL}]] \Gamma[\alpha]))\\
&		  /(1+      4  \sqrt{\pi}
       {\rm cos}[\alpha \pi + 6 \theta - \alpha {\rm Arg}[-(i/\Omega)] - 
        \alpha {\rm Arg}[-i \Omega]] \Gamma[\alpha] + 4 \pi \Gamma[\alpha]^{2})).
\end{aligned}
\end{equation} 
By assuming the variation $\delta {\rm Re}\Omega$
is small, the spectral function can be cast into the form $\rho(\Omega)=f_{0}(\Omega)f_{\pm}$,
where $f_{0}(\Omega)$ only depends on the value of $\Omega$ and independent of the sign of $\Omega$,
while $f_{\pm}$ only depends on the sign of $\Omega$ and independent of the value of $\Omega$,
\begin{equation} 
\begin{aligned}
\label{57}
f_{0}(\Omega)=& 
2 g^{3} (\Omega^{2})^{-\frac{1}{2} + \frac{3 \alpha}{2}}
      \omega_{NFL}^{4} (\omega_{NFL}^{2})^{-\frac{1}{2} - \frac{3 \alpha}{2}}
      \Gamma[\alpha],\\
f_{\pm}= &-
 (   {\rm cos}[\alpha \pi + 3 \theta - 
          \alpha {\rm Arg}[-(i/(\pm \Omega))] + (-1 + 2 \alpha) {\rm Arg}[-i (\pm \Omega)] - 
          3 (-1 + \alpha) {\rm Arg}[\omega_{NFL}]] \\
		  &+ 
       2 \sqrt{\pi}
         {\rm cos}[3 \theta + (1 - 3 \alpha) {\rm Arg}[-i (\pm \Omega)] + 
          3 (-1 + \alpha) {\rm Arg}[\omega_{NFL}]] \Gamma[\alpha])\\
&		  /( 1+     4  \sqrt{\pi}
       {\rm cos}[\alpha \pi + 6 \theta - \alpha {\rm Arg}[-(i/(\pm \Omega))] - 
        \alpha {\rm Arg}[-i (\pm \Omega)]] \Gamma[\alpha] + 4 \pi \Gamma[\alpha]^{2}).
\end{aligned}
\end{equation} 
There are two important terms within $f_{\pm}$,
\begin{equation} 
\begin{aligned}
{\rm Arg}[-(i/\Omega)]=&
{\rm Arg}[-\frac{y}{x^{2}+y^{2}}-i\frac{x}{x^{2}+y^{2}}]={\rm atan}[\frac{x}{y}]-\pi
={\rm Arg}[-(i/(-\Omega))]-\pi,\\
{\rm Arg}[-i \Omega] =& 
{\rm Arg}[y-ix] ={\rm atan}[\frac{-x}{y}]={\rm Arg}[-i (-\Omega)]+\pi,
\end{aligned}
\end{equation} 
Note that we rewrite the frequency variable as $\Omega=x+iy$ with $x,y>0$ here,
and ${\rm Re}\Omega\equiv x >0$ is required to make the above expression of boson self-energy valid,
while ${\rm Im}\Omega\equiv y$ could be positive or negative, but we only discuss the $y>0$ case here and $y<0$ case is similar.
Since the variation $\delta {\rm Re}\Omega=\delta x$ is small,
the contribution of ${\rm atan}[\frac{\pm x}{y}]$ to the angle of triangle function can be ignored,
Then $f_{\pm}$ is independent of the value of $\Omega$.

Then we can rewrite the Eq.(\ref{47}) as
\begin{equation} 
\begin{aligned}
I_{2}
=&P\int^{i\infty}_{-i\infty}\frac{dz}{2\pi i}
[\int^{\infty}_{-\infty} d\Omega_{0}\frac{\rho(\Omega_{0})}{\pi[z-(\Omega_{0}-i\eta{\rm sgn}\Omega_{0})]}
\frac{1}{\pi}\int^{\infty}_{-\infty} d\Omega_{1}\frac{-\rho(\Omega_{1})}{(z-\Omega_{1}+i\eta{\rm sgn}\Omega_{1})^{2}}
]\\
=&P\int^{-i\infty}_{i\infty}\frac{d(-z)}{2\pi i}
[\int^{-\infty}_{\infty} d(-\Omega_{0})\frac{\rho(-\Omega_{0})}{\pi[-z-(-\Omega_{0}+i\eta{\rm sgn}\Omega_{0})]}
\frac{1}{\pi}\int^{-\infty}_{\infty} d(-\Omega_{1})\frac{-\rho(-\Omega_{1})}{(-z+\Omega_{1}-i\eta{\rm sgn}\Omega_{1})^{2}}
]\\
=&-P\int^{i\infty}_{-i\infty}\frac{dz}{2\pi i}
[\int^{\infty}_{-\infty} d\Omega_{0}\frac{\rho(-\Omega_{0})}{\pi[z-\Omega_{0}+i\eta{\rm sgn}\Omega_{0})]}
\frac{1}{\pi}\int^{\infty}_{-\infty} d\Omega_{1}\frac{-\rho(-\Omega_{1})}{(z-\Omega_{1}+i\eta{\rm sgn}\Omega_{1})^{2}}
],
\end{aligned}
\end{equation} 
or 
\begin{equation} 
\begin{aligned}
I_{2}
=&P\int^{-i\infty}_{i\infty}\frac{d(-z)}{2\pi i}
[\int^{\infty}_{-\infty} d\Omega_{0}\frac{\rho(\Omega_{0})}{\pi[-z-(\Omega_{0}-i\eta{\rm sgn}\Omega_{0})]}
\frac{1}{\pi}\int^{-\infty}_{\infty} d(-\Omega_{1})\frac{-\rho(-\Omega_{1})}{(-z+\Omega_{1}-i\eta{\rm sgn}\Omega_{1})^{2}}
]\\
=&P\int^{i\infty}_{-i\infty}\frac{dz}{2\pi i}
[\int^{\infty}_{-\infty} d\Omega_{0}\frac{\rho(\Omega_{0})}{\pi[z+(\Omega_{0}-i\eta{\rm sgn}\Omega_{0})]}
\frac{1}{\pi}\int^{\infty}_{-\infty} d\Omega_{1}\frac{-\rho(-\Omega_{1})}{(z-\Omega_{1}+i\eta{\rm sgn}\Omega_{1})^{2}}
],
\end{aligned}
\end{equation} 
or
\begin{equation} 
\begin{aligned}
I_{2}
=&P\int^{i\infty}_{-i\infty}\frac{dz}{2\pi i}
[\int^{-\infty}_{\infty} d(-\Omega_{0})\frac{\rho(-\Omega_{0})}{\pi[z-(-\Omega_{0}+i\eta{\rm sgn}\Omega_{0})]}
\frac{1}{\pi}\int^{\infty}_{-\infty} d\Omega_{1}\frac{-\rho(\Omega_{1})}{(z-\Omega_{1}+i\eta{\rm sgn}\Omega_{1})^{2}}
]\\
=&P\int^{i\infty}_{-i\infty}\frac{dz}{2\pi i}
[\int^{\infty}_{-\infty} d\Omega_{0}\frac{\rho(-\Omega_{0})}{\pi[z+\Omega_{0}-i\eta{\rm sgn}\Omega_{0})]}
\frac{1}{\pi}\int^{\infty}_{-\infty} d\Omega_{1}\frac{-\rho(\Omega_{1})}{(z-\Omega_{1}+i\eta{\rm sgn}\Omega_{1})^{2}}
].
\end{aligned}
\end{equation} 
Using these three expressions of $I_{2}$,
we have
\begin{equation} 
\begin{aligned}
\label{i2}
I_{2}
=&\frac{1}{4}[
P\int^{i\infty}_{-i\infty}\frac{dz}{2\pi i}
[\int^{\infty}_{-\infty} d\Omega_{0}\frac{\rho(\Omega_{0})}{\pi[z-(\Omega_{0}-i\eta{\rm sgn}\Omega_{0})]}
\frac{1}{\pi}\int^{\infty}_{-\infty} d\Omega_{1}\frac{-\rho(\Omega_{1})}{(z-\Omega_{1}+i\eta{\rm sgn}\Omega_{1})^{2}}]
\\
&-P\int^{i\infty}_{-i\infty}\frac{dz}{2\pi i}
[\int^{\infty}_{-\infty} d\Omega_{0}\frac{\rho(-\Omega_{0})}{\pi[z-\Omega_{0}+i\eta{\rm sgn}\Omega_{0})]}
\frac{1}{\pi}\int^{\infty}_{-\infty} d\Omega_{1}\frac{-\rho(-\Omega_{1})}{(z-\Omega_{1}+i\eta{\rm sgn}\Omega_{1})^{2}}
]\\
&+\frac{1}{4}
[P\int^{i\infty}_{-i\infty}\frac{dz}{2\pi i}
[\int^{\infty}_{-\infty} d\Omega_{0}\frac{\rho(\Omega_{0})}{\pi[z+(\Omega_{0}-i\eta{\rm sgn}\Omega_{0})]}
\frac{1}{\pi}\int^{\infty}_{-\infty} d\Omega_{1}\frac{-\rho(-\Omega_{1})}{(z-\Omega_{1}+i\eta{\rm sgn}\Omega_{1})^{2}}]
+\\
&P\int^{i\infty}_{-i\infty}\frac{dz}{2\pi i}
[\int^{\infty}_{-\infty} d\Omega_{0}\frac{\rho(-\Omega_{0})}{\pi[z+\Omega_{0}-i\eta{\rm sgn}\Omega_{0})]}
\frac{1}{\pi}\int^{\infty}_{-\infty} d\Omega_{1}\frac{-\rho(\Omega_{1})}{(z-\Omega_{1}+i\eta{\rm sgn}\Omega_{1})^{2}}
]\\
=&\frac{1}{4}[P\int^{i\infty}_{-i\infty}\frac{dz}{2\pi i}
[P\int^{\infty}_{-\infty} d\Omega_{0}\frac{1}{\pi}\int^{\infty}_{-\infty} \frac{d\Omega_{1}}{\pi}
\frac{-\rho(\Omega_{1})\rho(\Omega_{0})+\rho(-\Omega_{0})\rho(-\Omega_{1})}{\pi[z-(\Omega_{0}-i\eta{\rm sgn}\Omega_{0})]}
\frac{1}{(z-\Omega_{1}+i\eta{\rm sgn}\Omega_{1})^{2}}
]]\\
&+\frac{1}{4}[
P\int^{i\infty}_{-i\infty}\frac{dz}{2\pi i}
[P\int^{\infty}_{-\infty} d\Omega_{0}P\int^{\infty}_{-\infty} \frac{d\Omega_{1}}{\pi}
\frac{-\rho(\Omega_{0})\rho(-\Omega_{1})-\rho(-\Omega_{0})\rho(\Omega_{1})}{\pi[z+(\Omega_{0}-i\eta{\rm sgn}\Omega_{0})]}
\frac{1}{(z-\Omega_{1}+i\eta{\rm sgn}\Omega_{1})^{2}}
]]\\
=&\frac{1}{4}[P\int^{i\infty}_{-i\infty}\frac{dz}{2\pi i}
[P\int^{\infty}_{-\infty} d\Omega_{0}\frac{1}{\pi}\int^{\infty}_{-\infty} \frac{d\Omega_{1}}{\pi}
\frac{-f_{0}(\Omega_{1})f_{0}(\Omega_{0})(f_{+}f_{-}-f_{-}f_{+})}{\pi[z-(\Omega_{0}-i\eta{\rm sgn}\Omega_{0})]}
\frac{1}{(z-\Omega_{1}+i\eta{\rm sgn}\Omega_{1})^{2}}
]]\\
&+\frac{1}{4}[
P\int^{i\infty}_{-i\infty}\frac{dz}{2\pi i}
[\int^{\infty}_{-\infty} d\Omega_{0}\int^{\infty}_{-\infty} \frac{d\Omega_{1}}{\pi}
\frac{-f_{0}(\Omega_{0})f_{0}(\Omega_{1})[f_{+}f_{-}+f_{-}f_{+}]}{\pi[z+(\Omega_{0}-i\eta{\rm sgn}\Omega_{0})]}
\frac{1}{(z-\Omega_{1}+i\eta{\rm sgn}\Omega_{1})^{2}}
]]\\
=&\frac{1}{4}[
P\int^{i\infty}_{-i\infty}\frac{dz}{2\pi i}
[\int^{\infty}_{-\infty} d\Omega_{0}\int^{\infty}_{-\infty} \frac{d\Omega_{1}}{\pi}
\frac{-f_{0}(\Omega_{0})f_{0}(\Omega_{1})[f_{+}f_{-}+f_{-}f_{+}]}{\pi[z+(\Omega_{0}-i\eta{\rm sgn}\Omega_{0})]}
\frac{1}{(z-\Omega_{1}+i\eta{\rm sgn}\Omega_{1})^{2}}.
\end{aligned}
\end{equation} 
We found that the integrals over frequencies within the above expression are free from IR divergence
base on the above IR asymptotics of Green's function (and self-energy),
and thus we can replace the principal value integration by an integration along real axis,
\begin{equation} 
\begin{aligned}
&\int^{\infty}_{-\infty} \frac{d\Omega_{0}}{\pi}
\frac{-f_{0}(\Omega_{0})}{z-(\Omega_{0}-i\eta{\rm sgn}\Omega_{0})}
=\\
&
\frac{1}{\pi}
2 g^{3} \pi (\omega_{NFL}^{2})^{-\frac{3}{2} (-1 + \alpha)} ((-i \eta - z)^{-1 + 3 \alpha}
- (-i \eta + z)^{-1 +3 \alpha}) {\rm csc}[3 \alpha \pi] \Gamma[\alpha],\\
&\int^{\infty}_{-\infty} \frac{d\Omega_{1}}{\pi}
\frac{f_{0}(\Omega_{1})}{(z-\Omega_{1}+i\eta{\rm sgn}\Omega_{1})^{2}}
=\\
&
\frac{1}{\pi}
2 (-1 + 3 \alpha) g^{3} \pi (\omega_{NFL}^{2})^{-\frac{3}{2} (-1 + \alpha)} 
(\frac{(-i \eta - z)^{3 \alpha}}{(\eta - i z)^{2}} - (\frac{1}{-i \eta + z})^{2 - 3 \alpha}) 
{\rm csc}[3 \alpha \pi] \Gamma[\alpha].
\end{aligned}
\end{equation} 
Since the variable $\Omega$ has positive real part and nonzero imaginary part,
we can simply replace the term $\Omega-i\eta{\rm sgn}[\Omega]$ by $\Omega$,
then we have
\begin{equation} 
\begin{aligned}
\label{i22}
&\int^{\infty}_{-\infty} \frac{d\Omega_{0}}{\pi}
\frac{-f_{0}(\Omega_{0})}{z-\Omega_{0}}
=\\
&
-((2 g^{3} \pi (\omega_{NFL}^{2})^{-\frac{3}{2} (-1 + \alpha)} ((-z)^{3 \alpha} 
+ z^{3 \alpha}) {\rm csc}[3 \alpha \pi] \Gamma[\alpha])/\pi z),\\
&\int^{\infty}_{-\infty} \frac{d\Omega_{1}}{\pi}
\frac{f_{0}(\Omega_{1})}{(z-\Omega_{1})^{2}}
=\\
&
	-((2 (-1 + 3 \alpha) g^{3} \pi (\omega_{NFL}^{2})^{-\frac{3}{2} (-1 + \alpha)}
	((-z)^{3 \alpha} + z^{3 \alpha}) {\rm csc}[3 \alpha \pi] \Gamma[\alpha])/\pi z^{2}).
\end{aligned}
\end{equation} 
Thus the final result of $I_{2}$ can be obtained by substituting the Eqs.(\ref{57},\ref{i22}) into Eq.(\ref{i2}). 
Note that this result requires $1/3<\alpha<1/2$,
which corresponds to $1/6>\sqrt{M/2\pi N}>0$ for non-Fermi liquid in Yukawa-SYK model\cite{Wang Y}
where $M$ and $N$ denote the flavor of fermions and bosons, respectively.
The $z$ integration is also IR finite now and the principal value integration is not needed.
The final expression of $I_{2}$ can then be obtained by substituting the Eq.(\ref{57})
into the above expression.

\subsection{Boson self-energy at zero frequency}
Next we examine whether a
finite boson self-energy at zero frequency will plays a role of tuning parameter of fermionic quantum critical behavior
(since it affects directly the dispersion of fermion excitation)
within the boson field propagator.
Firstly we note that at half-filling in coherent states, 
$\Sigma(0)=-\mu=0$ and $\theta=0$.
Although there should be Fermi liquid state due to the low fermion density at half-filling
and thus obey the Luttinger's theorem,
the fermi surface discontinuity is hard to be seen from the curve of energy/momentum distribution function\cite{Singh R R P}.
This is partly due to the weak energy/momentum-dependence of the quasiparticles at half-filling.
Also,
the charge fluctuation is most dominating in the half-filling,
otherwise the pair fluctuation (Cooper channel) is dominant when away from half-filling
(i.e., enters the non-fermi-liquid regime, until it goes to the non-fermi-liquid end-point with band filling equals one).
The gap term of charge fluctuation can be written as
\begin{equation} 
\begin{aligned}
\uppi(0)
=&\int^{\infty}_{-\infty}\frac{d\omega}{2\pi}(-\Sigma(i\omega))^{-3}\\
=&-\int^{\infty}_{0}\frac{d\omega}{2\pi}
i e^{3 i \theta} \omega^{-3 \alpha} \omega_{NFL}^{-3 + 3 \alpha}
+\int^{0}_{-\infty}\frac{d\omega}{2\pi}
i e^{-3 i \theta} \omega^{-3 \alpha} \omega_{NFL}^{-3 + 3 \alpha}.
\end{aligned}
\end{equation} 
The analytic solution of this expression requires either an UV or IR cutoff,
which are
\begin{equation} 
\begin{aligned}
2\pi\uppi(0)
=&-((i e^{3 i \theta} \Lambda_{UV}^{1 - 3 \alpha} \omega_{NFL}^{-3 + 3 \alpha})/(
 1 - 3 \alpha))+
 (i e^{-3 i \theta} (-\Lambda_{UV})^{1 - 3 \alpha} \omega_{NFL}^{-3 + 3 \alpha})/(-1 + 
 3 \alpha),\\
2\pi\uppi(0)
=& -((i e^{3 i \theta} \Lambda_{IR}^{1 - 3 \alpha} \omega_{NFL}^{-3 + 3 \alpha})/(-1 + 
  3 \alpha))+
  (i e^{-3 i \theta} (-\Lambda_{IR})^{-3 \alpha} \Lambda_{IR} \omega_{NFL}^{-3 + 3 \alpha})/(-1 + 
 3 \alpha),
\end{aligned}
\end{equation} 
respectively,
where the first solution requires $0<\alpha<1/3$ and the second one requires $1/3<\alpha<1/2$.
At half-filling, we have
\begin{equation} 
\begin{aligned}
2\pi\uppi(0)
=&-((i  \Lambda_{UV}^{1 - 3 \alpha} \omega_{NFL}^{-3 + 3 \alpha})/(
 1 - 3 \alpha))+
 (i  (-\Lambda_{UV})^{1 - 3 \alpha} \omega_{NFL}^{-3 + 3 \alpha})/(-1 + 
 3 \alpha),\\
2\pi\uppi(0)
=& -((i  \Lambda_{IR}^{1 - 3 \alpha} \omega_{NFL}^{-3 + 3 \alpha})/(-1 + 
  3 \alpha))+
  (i  (-\Lambda_{IR})^{-3 \alpha} \Lambda_{IR} \omega_{NFL}^{-3 + 3 \alpha})/(-1 + 
 3 \alpha),
\end{aligned}
\end{equation} 
and the particle-hole symmetry guaruatees that there does not exists a linear-in-frequency term in the denominator of boson field propagator.
The second line of Eq.(52) then beomes
\begin{equation} 
\begin{aligned}
\Sigma(i\omega)=&\int^{\infty}_{-\infty}\frac{d\Omega}{2\pi}
(-\uppi^{-1}(i\Omega)+\uppi^{-1}(0))\Sigma^{-1}(i\omega-i\Omega_{1}),
\end{aligned}
\end{equation} 
which implies that the dynamic part of boson self-energy $-\uppi^{-1}(i\Omega)+\uppi^{-1}(0)$ is the sum of all one-loop
fermion self-energy diagrams.
Comparing Eq.(\ref{54}) and the above results, we can see that for small $\alpha$ (i.e., large fermion flavor number),
the static part of boson self-energy (bosonic mass term) $\uppi(0)$ will not affect the IR asymptotics behavior of dynamic 
boson self-energy, in the presence of UV cutoff.
Thus we can savely use $\uppi(i\Omega)$ instead of $\uppi(i\Omega)-\uppi(0))$ throughout this paper.

\subsection{Low-rank SYK}

When the pair condensation happen, the above pairing order parameter
can be defined by $\Delta_{\phi}(\tau)=|\Delta_{\phi}|e^{i\phi(\tau)}$,
with the long-range (real space) condensate phase fluctuation $\phi(\tau)$.
The phase $\phi(\tau)$ is not fixed by the saddle point approximation.

To study the $g_{ijk;k'j'i'}$ in the matrix form,
we rewrite it as
\begin{equation} 
\begin{aligned}
\label{gmatrix}
\frac{1}{2^{5/2}N^{2}}g_{ijk;k'j'i'}=\frac{1}{2}\sum^{R}_{n}\lambda_{n}\psi^{(n)}_{\alpha}\psi_{\alpha'}^{(n)},
\end{aligned}
\end{equation} 
where $\psi^{(n)}_{\alpha}$ and $\psi_{\alpha'}^{(n)}$ correspond to $\Delta_{\alpha}^{\dag}$ and $\Delta_{\alpha'}$, respectively.
This is similar to the definition of low-rank SYK coupling\cite{Kim J}.
Similar to the discussion we presented in Ref.\cite{wusyk1,wusyk2},
for SYK$_{3}\times$SYK$_{3}$ model,
since the sets of indices $ijk$ and $k'j'i'$ are not completely indepedent with each other,
but correlated by some certain mechanism, 
e.g., the mapping between $c_{i}$ to $c_{i'}$ is the same with that between $c_{j}$ to $c_{j'}$,
there are only four degrees of freedom although there are six indices.
Thus the two-dimensional matrix $g_{ijk;k'j'i'}$ is a $N^{2}\times N^{2}$ matrix.
Then the SYK Hamiltonian Eq.(\ref{SYKHa}) can be rewritten as
\begin{equation} 
\begin{aligned}
H=\frac{1}{2}\sum^{R}_{n}\sum_{ijk;k'j'i'}^{N}\lambda_{n}\psi^{(n)}_{\alpha}\psi_{\alpha'}^{(n)}
c^{\dag}_{i}c^{\dag}_{j}c_{k}c_{k'}^{\dag}c_{j'}c_{i'}.
\end{aligned}
\end{equation} 
Here we note the following mean and variance 
\begin{equation} 
\begin{aligned}
\label{variance}
&\overline{\psi_{\alpha}^{(n)}}=\overline{\psi_{\alpha'}^{(n')}}=0,\\
&\frac{1}{2}\overline{\sum_{n}\lambda_{n}\psi_{\alpha}^{(n)}\psi_{\alpha'}^{(n')}}=\frac{g}{2^{3/2}N^{2}}\delta_{n,n'}.
\end{aligned}
\end{equation} 
Through Hubbard-Stratonovich transformation,
\begin{equation} 
\begin{aligned}
e^{-\frac{1}{2^{5/2}N^{2}}g_{ijk;k'j'i'}\sum^{N}_{\alpha,\alpha'}\Delta_{\alpha}^{\dag}\Delta_{\alpha'}}
=&e^{-\frac{1}{2}\sum^{R}_{n}\lambda_{n}\psi^{(n)}_{\alpha}\psi_{\alpha'}^{(n)}\sum^{N}_{\alpha,\alpha'}\Delta_{\alpha}^{\dag}\Delta_{\alpha'}}\\
=&\int D[\Delta_{\phi'},\Delta_{\phi}^{\dag}]
e^{\sum_{n}[-2\lambda^{-1}\Delta_{\phi}^{\dag (n)}\Delta^{(n)}_{\phi'}-i\Delta_{\phi}^{\dag (n)}\psi_{\alpha}^{(n)}\Delta_{\alpha}^{\dag}
 -i\Delta_{\phi'(n)}\psi_{\alpha'}^{(n)}\Delta_{\alpha'}]},
\end{aligned}
\end{equation} 
the partition function reads
(follow the above procedure)
\begin{equation} 
\begin{aligned}
Z=&\int D[c^{\dag},c]
e^{-\int d\tau d\tau'[c^{\dag}(\tau)(\partial_{\tau}\delta(\tau-\tau')+\Sigma(\tau-\tau'))c(\tau')
+\sum_{n}2\lambda_{n}^{-1}\Delta_{\phi}^{\dag (n)}\Delta_{\phi' (n)}]}\\
&\int D[\psi_{\alpha},\psi_{\alpha}^{\dag}]e^{-\frac{|\psi_{\alpha}|^{2}}{\sigma^{2}}}\\
&\int D[\psi_{\alpha'},\psi_{\alpha'}^{\dag}]e^{-\frac{|\psi_{\alpha'}|^{2}}{\sigma^{2}}}\\
&\int D[\psi_{\alpha'},\psi_{\alpha'}^{\dag}] D[\psi_{\alpha},\psi_{\alpha}^{\dag}]
e^{-\int d\tau d\tau' \sum_{n}(i\Delta_{\phi}^{\dag (n)}\psi_{\alpha}^{(n)}\Delta_{\alpha}^{\dag}
+i\Delta_{\phi' (n)}\psi_{\alpha'}^{(n)}\Delta_{\alpha'}
)},
\end{aligned}
\end{equation} 
where $\Delta_{\phi}$ is the condensated boson operator, while $\Delta_{\alpha}$ is the uncondensated one, as defined in the begining of this section.
Here $\sigma^{2}=\lambda_{n}\overline{\psi_{\alpha}^{2}}=\frac{g}{2^{3/2}N^{2}}$.
Using the relation
\begin{equation} 
\begin{aligned}
&\int d\psi_{\alpha}d\psi_{\alpha'}
e^{-\frac{1}{\sigma^{2}}
(\psi_{\alpha}^{(n)}+i\Delta_{\phi}^{(n)}\Delta_{\alpha}\sigma^{2})
(\psi_{\alpha}^{\dag(n)}+i\Delta^{\dag (n)}_{\phi}\Delta_{\alpha}^{\dag}\sigma^{2})}
e^{-\sigma^{2}\Delta_{\phi}^{(n)}\Delta_{\phi}^{\dag (n)}\Delta_{\alpha}\Delta_{\alpha}^{\dag}}\\
&=e^{-\sigma^{2}\Delta_{\phi}^{(n)}\Delta_{\phi}^{\dag (n)}\Delta_{\alpha}\Delta_{\alpha}^{\dag}},
\end{aligned}
\end{equation} 
the action can be obtained as
\begin{equation} 
\begin{aligned}
S=&\int d\tau d\tau'[c^{\dag}(\tau)(\partial_{\tau}\delta(\tau-\tau')+\Sigma(\tau,\tau'))c(\tau')
+\int d\tau d\tau' 2\sum_{n}\lambda_{n}^{-1}\Delta_{\phi}^{\dag (n)}(\tau)\Delta^{(n)}_{\phi'}(\tau)]\\
&+\int d\tau d\tau'[\frac{gN}{2^{3/2}}\Delta_{\phi}^{(n)}\Delta_{\phi}^{\dag (n)}G^{2}(\tau,\tau')G(\tau',\tau)
+\frac{gN}{2^{3/2}}\Delta_{\phi'}^{(n)}\Delta_{\phi'}^{\dag (n)}G^{2}(\tau,\tau')G(\tau',\tau)]\\
&-\int d\tau d\tau'[N\Sigma(\tau,\tau')G(\tau',\tau)]\\
=&-N{\rm Tr}{\rm ln}[-i\omega+\Sigma(i\omega)]+\int d\tau d\tau'2\sum_{n}\lambda_{n}^{-1}\Delta_{\phi}^{\dag (n)}\Delta^{(n)}_{\phi'}\\
&+\int d\tau d\tau'[\frac{gN}{2^{3/2}}\Delta_{\phi}^{(n)}\Delta_{\phi}^{\dag (n)}G^{2}(\tau,\tau')G(\tau',\tau)
+\frac{gN}{2^{3/2}}\Delta_{\phi'}^{(n)}\Delta_{\phi'}^{\dag (n)}G^{2}(\tau,\tau')G(\tau',\tau)]\\
&-\int d\tau d\tau'[N\Sigma(\tau,\tau')G(\tau',\tau)].
\end{aligned}
\end{equation} 
Through saddle-point equation we have (at saddle point we have $\Delta_{\phi}=\Delta_{\phi'}$)
\begin{equation} 
\begin{aligned}
\frac{\partial S}{\partial G}=&-N\Sigma(i\omega)+\frac{gN}{2^{1/2}}\Delta_{\phi}^{(n)}\Delta_{\phi}^{\dag (n)}G^{2}(i\omega)=0,\\
\frac{\partial S}{\partial \Sigma}=&-NG(i\omega)-\frac{N}{-i\omega+\Sigma(i\omega))}=0,\\
\frac{\partial S}{\partial \Delta^{\dag}}=&2\sum_{n}\lambda_{n}^{-1}\Delta_{\phi}-\sum_{n}\frac{2g}{2^{3/2}N^{2}}\Delta_{\phi}G^{3}(\tau,\tau')=0,
\end{aligned}
\end{equation} 
thus 
\begin{equation} 
\begin{aligned}
\label{gapequation1}
\Sigma(i\omega)=&\frac{g}{2^{1/2}}\Delta_{\phi}^{(n)}\Delta_{\phi}^{\dag (n)}G^{2}(i\omega),\\
G(i\omega)=&\frac{-1}{-i\omega+\Sigma(i\omega)},\\
\frac{2\Delta_{\phi}}{g}=&\sum_{n}\frac{\lambda_{n}N}{2^{1/2}}\Delta_{\phi}G^{3}(\tau,\tau').
\end{aligned}
\end{equation} 
At saddle point, $\Delta_{\phi}=\Delta_{\phi'}$, and $\Delta_{\phi}$ is independent of time and site.
This subsection discuss the low-rank SYK ($R\sim N^{2}$) without the pair condensation,
while for $R\gg N^{2}$, it becomes the standard SYK model.

Note that for the above case,
it also obeys the Luttinger-Ward theorem in particle-particle-hole channel,
where the irreducible vertice $g$-dependent part of the action $S[G]$ can be treated as Luttinger-Ward functional.
The $S[G]$ satisfies
\begin{equation} 
\begin{aligned}
\frac{\partial S[G]}{\partial G(\tau',\tau)}=\Sigma(\tau,\tau'),
\end{aligned}
\end{equation} 
and the irreducible vertices has
\begin{equation} 
\begin{aligned}
g\sim \frac{\partial^{3} S[G]}{\partial G^{3}(\tau',\tau)}=\frac{\partial^{2} \Sigma}{\partial G^{2}(\tau',\tau)}.
\end{aligned}
\end{equation} 
The self-consistent relation as we discussed in Sec.3 is still valid here
\begin{equation} 
\begin{aligned}
\Sigma(\tau,\tau')G(\tau',\tau)=\frac{g}{2^{1/2}}\Delta_{\phi}^{(n)}\Delta_{\phi}^{\dag (n)}(1-n)n^{2}.
\end{aligned}
\end{equation} 
The only difference to the self-consistent relation presented in Sec.3 is the existence of translational invariance 
in saddle point approximation.

\subsection{Low-rank SYK with weak pair condensation}

Next we discuss the case when the SYK non Fermi liquid phase is supressed by the large many-body spectrum gap
generated by weakly condensated pairing order operator,
in which case the largest eigenvalue split is realized in a Gaussian orthogonal ensemble (GOE).
By "weak", we mean that the dynamical anomalous self-energy is vanishingly small,
when compares to the static anomalous self-energy $\Delta_{\phi}$.
In this case, $R=N^{2}$, and we have the relation
\begin{equation} 
\begin{aligned}
\overline{\lambda^{2}_{n}}=\frac{1}{2^{5}N^{4}}\overline{g_{ijk;k'j'i'}^{2}}
=\frac{1}{4}\sum_{n}\overline{(\lambda_{n}\psi_{\alpha})^{2}}\overline{(\lambda_{n}\psi_{\alpha'})^{2}}
=\frac{g^{2}}{2^{3}N^{4}}+\frac{\delta_{ijk',kj'i'}}{N^{4}}
\sim O(N^{-4})+\delta_{ijk',kj'i'}O(N^{-4}),
\end{aligned}
\end{equation} 
thus $\sum_{n}\lambda_{n}^{2}=\frac{g^{2}}{2^{3}N^{2}}$.
Then, since the $N^{2}\times N^{2}$ matrix $g_{ijk;k'j'i'}$ is not a positive define matrix,
we have the largest eigenvalues $\lambda_{max}=\pm \frac{g}{2^{5/2}N}$, 
and there are $\frac{N^{2}-2}{2}$ eigenvalues $\lambda=\frac{1}{N\sqrt{N^{2}-2}}$
and $\frac{N^{2}-2}{2}$ eigenvalues $\lambda=\frac{-1}{N\sqrt{N^{2}-2}}$.

Thus the Eq.(\ref{gmatrix}) can be rewritten as
\begin{equation} 
\begin{aligned}
\frac{1}{2^{5/2}N^{2}}g_{ijk;k'j'i'}=\frac{1}{2}[\frac{g}{2^{3/2}N}\psi_{\alpha}^{(1)}\psi_{\alpha'}^{(1)}-\frac{g}{2^{3/2}N}\psi_{\alpha}^{(2)}\psi_{\alpha'}^{(2)}],
\end{aligned}
\end{equation} 
and the SYK Hamiltonian reads
\begin{equation} 
\begin{aligned}
H=\frac{1}{2}[\frac{g}{2^{3/2}N}\psi_{\alpha}^{(1)}\psi_{\alpha'}^{(1)}-\frac{g}{2^{3/2}N}\psi_{\alpha}^{(2)}\psi_{\alpha'}^{(2)}
+\frac{\sqrt{N^{2}-2}}{2N}\psi^{(3)}_{\alpha}\psi_{\alpha'}^{(3)}-\frac{\sqrt{N^{2}-2}}{2N}\psi^{(4)}_{\alpha}\psi_{\alpha'}^{(4)}]
\sum^{N}_{ijk',kj'i'}\langle c^{\dag}_{i}c^{\dag}_{j}c^{\dag}_{k'}\rangle \langle c_{k}c_{j'}c_{i'}\rangle.
\end{aligned}
\end{equation} 
Here we assume that $\psi^{(n)}_{\alpha}\psi_{\alpha'}^{(n)}=\psi^{(m)}_{\alpha}\psi_{\alpha'}^{(m)}$ if and only if $\lambda_{n}=\lambda_{m}$.

We have the Hubbard-Stratonovich transformations
\begin{equation} 
\begin{aligned}
&e^{-\frac{1}{2}\frac{g}{2^{3/2}N}\psi_{\alpha}^{(1)}\psi_{\alpha'}^{(1)}
\sum^{N}_{ijk',kj'i'}\langle c^{\dag}_{i}c^{\dag}_{j}c^{\dag}_{k'}\rangle \langle c_{k}c_{j'}c_{i'}\rangle}\\
&=e^{-\frac{g}{2^{5/2}N}\psi_{\alpha}^{(1)}\psi_{\alpha'}^{(1)}
\sum^{N}_{ijk',kj'i'}\langle c^{\dag}_{i}c^{\dag}_{j}c^{\dag}_{k'}\rangle \langle c_{k}c_{j'}c_{i'}\rangle}\\
&=\int D[\Delta_{\phi}^{(1)},\Delta_{\phi}^{\dag (1)}]e^{-g^{-1}2^{5/2}N\Delta_{\phi}^{(1)}\Delta_{\phi}^{\dag (1)}
-\sum_{ijk'}i\Delta_{\phi}^{(1)}\psi_{\alpha}^{(1)}\langle c^{\dag}_{i}c^{\dag}_{j}c^{\dag}_{k'}\rangle
-\sum_{kj'i'}i\Delta_{\phi}^{\dag (1)}\psi_{\alpha'}^{(1)}\langle c_{k}c_{j'}c_{i'}\rangle},\\
&e^{-\frac{1}{2}\frac{\sqrt{N^{2}-2}}{2N}\psi^{(3)}_{\alpha}\psi_{\alpha'}^{(3)}
\sum^{N}_{ijk',kj'i'}\langle c^{\dag}_{i}c^{\dag}_{j}c^{\dag}_{k'}\rangle \langle c_{k}c_{j'}c_{i'}\rangle}\\
&=e^{-\frac{\sqrt{N^{2}-2}}{4N}\psi^{(3)}_{\alpha}\psi_{\alpha'}^{(3)}
\sum^{N}_{ijk',kj'i'}\langle c^{\dag}_{i}c^{\dag}_{j}c^{\dag}_{k'}\rangle \langle c_{k}c_{j'}c_{i'}\rangle}\\
&=\int D[\Delta_{\phi}^{(3)},\Delta_{\phi}^{\dag (3)}]e^{-\frac{4N}{\sqrt{N^{2}-2}}  \Delta_{\phi}^{(1)}\Delta_{\phi}^{\dag (1)}
-\sum_{ijk'}i\Delta_{\phi}^{(3)}\psi_{\alpha}^{(3)}\langle c^{\dag}_{i}c^{\dag}_{j}c^{\dag}_{k'}\rangle
-\sum_{kj'i'}i\Delta_{\phi}^{\dag (3)}\psi_{\alpha'}^{(3)}\langle c_{k}c_{j'}c_{i'}\rangle}.
\end{aligned}
\end{equation}
Next we only focus on the first term: the largest eigenvalue with $\psi^{(1)}_{\alpha}\psi^{(1)}_{\alpha'}$. 
Note that unlike the Eq.(\ref{variance}),
when the superscript $n$ of wave functions $\psi^{(n)}_{\alpha}$ $\psi^{(n)}_{\alpha'}$ are fixed,
these wave functions are identified as constants.
Thus they are omitted in the following calculations.
This is the most important premise for the availability of many-body spectrum's gap equation.

Note that here the boson field (pairing order parameter) $\Delta_{\phi}=\langle c^{\dag}_{k'}c_{j}c_{i}\rangle$ plays the role of anomalous component of the
boson propagator, insteads of the single fermion propgator, thus the action should written as
\begin{equation} 
\begin{aligned}
S=&\sum_{ijk}^{N}\int d\tau d\tau'[c^{\dag}_{i}(\tau)c_{j}^{\dag}(\tau)(\partial_{\tau}\delta(\tau-\tau')+\Pi(\tau,\tau'))c_{k}(\tau')]
+\sum^{N}_{ijk',kj'i'}\int  d\tau [i\Delta_{\phi}^{(1)}(\tau)\langle c^{\dag}_{i}c^{\dag}_{j}c^{\dag}_{k'}\rangle\\
&+i\Delta_{\phi}^{\dag (1)}(\tau)\langle c_{k}c_{j'}c_{i'}\rangle]
+g^{-1}2^{5/2}N\int d\tau \Delta_{\phi}^{(1)}(\tau)\Delta_{\phi}^{\dag (1)}(\tau)
-N^{3}\Pi(\tau,\tau')D(\tau',\tau)\\
=&N^{3}\int d\tau d\tau'[c^{\dag}_{i}(\tau)c_{j}^{\dag}(\tau)(\partial_{\tau}\delta(\tau-\tau')+\Pi(\tau,\tau'))c_{k}(\tau')]
+N^{3}\int  d\tau [i\Delta_{\phi}^{(1)}(\tau)\langle c^{\dag}_{i}c^{\dag}_{j}c^{\dag}_{k'}\rangle\\
&+i\Delta_{\phi}^{\dag (1)}(\tau)\langle c_{k}c_{j'}c_{i'}\rangle]
+g^{-1}2^{5/2}N\int d\tau \Delta_{\phi}^{(1)}(\tau)\Delta_{\phi}^{\dag (1)}(\tau)
-N^{3}\Pi(\tau,\tau')D(\tau',\tau)\\
=&-N^{3}\sum_{\omega}{\rm ln}{\rm Det}
\begin{pmatrix}
-i\omega+\Pi(i\omega) & i\Delta_{\phi}^{(1)}\\
-i\Delta_{\phi}^{\dag (1)} & -i\omega-\Pi(-i\omega) 
\end{pmatrix}\\
&+g^{-1}2^{5/2}N\int^{\beta}_{0}d\tau  \Delta_{\phi}^{(1)}(\tau)\Delta_{\phi}^{\dag (1)}(\tau)
-N^{3}\Pi(\tau,\tau')D(\tau',\tau)\\
=&-N^{3}\sum_{\omega}{\rm ln}
[(-i\omega+\Pi(i\omega))(-i\omega-\Pi(-i\omega))-\Delta_{\phi}^{(1)}\Delta_{\phi}^{\dag (1)} ]\\
&+g^{-1}2^{5/2}N\beta  \Delta_{\phi}^{(1)}\Delta_{\phi}^{\dag (1)}
-N^{3}\Pi(\tau,\tau')D(\tau',\tau),
\end{aligned}
\end{equation}
where we define the long-range (in real space) propagator with only the normal component
$D(\tau,\tau')=\frac{1}{N^{3}}\sum^{N}_{ijk'}c^{\dag}_{i}(\tau')c^{\dag}_{j}(\tau')c_{k'}(\tau)$.
Here the anomalous propagator does not exist in the presence of small $\Delta_{\phi}$ since in this case
$\Delta_{\phi}$ is a BCS term which is single-time-dependent and thus cannot forms
a progation process.
Through saddle point equation we have
\begin{equation} 
\begin{aligned}
\frac{\partial S}{\partial \Pi(-i\omega)}=&N^{3}D(i\omega)-
N^{3}\frac{-(-i\omega+\Pi(i\omega))}{(-i\omega+\Pi(i\omega))(-i\omega-\Pi(-i\omega))-\Delta_{\phi}^{(1)}\Delta_{\phi}^{\dag (1)} }=0,\\
\frac{\partial S}{\partial D(-i\omega)}=&N^{3}\Pi(i\omega)=0,\\
\frac{\partial S}{\partial \Delta^{\dag (1)}_{\phi}}=&-N^{3}\sum_{\omega}\frac{-\Delta_{\phi}}
{(-i\omega+\Sigma(i\omega))(-i\omega-\Sigma(-i\omega))-\Delta_{\phi}^{(1)}\Delta_{\phi}^{\dag (1)}}-g^{-1}2^{5/2}N\beta \Delta_{\phi}=0,
\end{aligned}
\end{equation} 
thus at saddle point we have
 \begin{equation} 
\begin{aligned}
D(i\omega)=&
\frac{-(-i\omega+\Pi(i\omega))}{(-i\omega+\Pi(i\omega))(-i\omega-\Pi(-i\omega))-\Delta_{\phi}^{(1)}\Delta_{\phi}^{\dag (1)} },\\
\Pi(i\omega)=&0,
\end{aligned}
\end{equation} 
and we obtain the gap equation as
\begin{equation} 
\begin{aligned}
g^{-1}2^{5/2} \Delta_{\phi}=TN^{2}\sum_{\omega}\frac{\Delta_{\phi}}
{(-i\omega+\Pi(i\omega))(-i\omega-\Pi(-i\omega))-\Delta_{\phi}^{(1)}\Delta_{\phi}^{\dag (1)}},
\end{aligned}
\end{equation} 
and the condensation happen in the singularity point, where
\begin{equation} 
\begin{aligned}
(-i\omega+\Pi(i\omega))(-i\omega-\Pi(-i\omega))=\Delta_{\phi}^{(1)}\Delta_{\phi}^{\dag (1)}.
\end{aligned}
\end{equation} 
It is obvious that the above gap equation has a rather different form compares to the one in the standard SYK phase (Eq.(\ref{gapequation1})).
Since in saddle point the pairing order parameter is site- and time-independent,
the Luttinger-Ward theorem and the self-consistent relation,
which are valid in the absence of pair condensation as shown in the above subsection,
are no more valid here,
as long as the dynamical anomalous self-energy is vanishingly small when compares to the static one.
This is one of the most important result of this paper.

Note that the $H_{SYK}$ can be reformed to be positive-define so that all eigenvalues are $\ge 0$
through the procedure of Refs.\cite{Pikulin D I,Lantagne-Hurtubise}.

\subsection{Subextensive rank}

For subextensive rank $R=2\ll N^{2}$ the SYK non-Fermi liquid phase is completely supressed by
the many-body spectrum gap.
In this case, it is possible to assume all $N^{2}$-components vectors $\psi_{\alpha}$ and $\psi_{\alpha'}$ are mutually orthogonal
as long as $R<N^{2}/2$,
i.e., vectors $\psi_{\alpha}^{(1)}$, $\psi_{\alpha}^{(2)}$, $\psi_{\alpha'}^{(1)}$, and $\psi_{\alpha'}^{(2)}$ are mutually orthogonal,
Then there are $2R=4$ eigenvectors with eigenvalues $\pm \frac{g}{4RN^{2}}=\frac{g}{8N^{2}}$,
and $N^{2}-2R$ eigenvectors with zero eigenvalues.

In regime of subextensive rank, 
$\psi_{\alpha}^{(n)}\psi_{\alpha'}^{(n)}$ is approximately independent of the superscript $n$
and thus we have $g_{ijk;k'j'i'}\rightarrow 0$,
i.e., the free fermion limit.
Then when the temperature is lower than superconductivity critical temperature,
the phase coherence is generated by any nonzero $\Delta_{\phi}$, and the dynamical anomalous self-energy is nonzero,
which leads to the free fermion Green's function
\begin{equation} 
\begin{aligned}
G(i\omega)=\frac{2i}{\omega+{\rm sgn}[\omega](\omega^{2}+4\Pi_{A}/D_{A})},
\end{aligned}
\end{equation}
where $\Pi_{A}$ and $D_{A}$ are the anomalous self-energy and anomalous Green's function, respectively
While when the temperature is higher than the critical one,
the anomalous self-energy is zero,
and free fermion Green's function reduces to $G(i\omega)=i/\omega$, which is consistent with the high-frequency result.
Note that the critical temperature can be identified by solving the gap equation in the $\Delta_{\phi}\rightarrow 0$ limit.

Then the action becomes
\begin{equation} 
\begin{aligned}
S=&\sum_{ijk}^{N}\int d\tau d\tau'[c^{\dag}_{i}(\tau)c_{j}^{\dag}(\tau)(\partial_{\tau}\delta(\tau-\tau')+\Pi(\tau,\tau'))c_{k'}(\tau')\\
&+\Pi_{A}c^{\dag}_{i}(\tau)c_{j}^{\dag}(\tau)c^{\dag}_{k'}(\tau')+\Pi^{\dag}_{A}c_{k}(\tau)c_{j'}(\tau)c_{i'}(\tau')]
+\sum^{N}_{ijk',kj'i'}\int  d\tau [i\Delta_{\phi}^{(1)}(\tau)\langle c^{\dag}_{i}c^{\dag}_{j}c^{\dag}_{k'}\rangle\\
&+i\Delta_{\phi}^{\dag (1)}(\tau)\langle c_{k}c_{j'}c_{i'}\rangle]
+g^{-1}2^{5/2}N\int d\tau \Delta_{\phi}^{(1)}(\tau)\Delta_{\phi}^{\dag (1)}(\tau)\\
&-N^{3}\Pi_{A}(\tau,\tau')D_{A}(\tau',\tau)
-N^{3}\Pi(\tau,\tau')D(\tau',\tau)\\
=&-N^{3}\sum_{\omega}{\rm ln}
[(-i\omega+\Pi(i\omega))(-i\omega-\Pi(-i\omega))-(\Pi_{A}(i\omega)+i\Delta_{\phi}^{(1)})(\Pi_{A}^{\dag}(i\omega)-i\Delta_{\phi}^{\dag (1)} )]\\
&+g^{-1}2^{5/2}N\beta  \Delta_{\phi}^{(1)}\Delta_{\phi}^{\dag (1)}
-N^{3}\Pi_{A}(\tau,\tau')D_{A}(\tau',\tau)
-N^{3}\Pi(\tau,\tau')D(\tau',\tau),
\end{aligned}
\end{equation}
thus leads to the anomalous part of Boson Green's function
\begin{equation} 
\begin{aligned}
&D_{A}=\frac{-(\Pi_{A}^{\dag}(i\omega)-i\Delta_{\phi}^{\dag (1)} )}
{(-i\omega+\Pi(i\omega))(-i\omega-\Pi(-i\omega))-(\Pi_{A}(i\omega)+i\Delta_{\phi}^{(1)})(\Pi_{A}^{\dag}(i\omega)-i\Delta_{\phi}^{\dag (1)} )},\\
&g^{-1}2^{5/2} \Delta_{\phi}=TN^{2}\sum_{\omega}
\frac{i(\Pi_{A}^{\dag}(i\omega)+i\Delta_{\phi}^{ (1)} )}
{(-i\omega+\Pi(i\omega))(-i\omega-\Pi(-i\omega))-(\Pi_{A}(i\omega)+i\Delta_{\phi}^{(1)})(\Pi_{A}^{\dag}(i\omega)-i\Delta_{\phi}^{\dag (1)} )}.
\end{aligned}
\end{equation} 
But since $g_{ijk;k'j'i'}\rightarrow 0$, we still have $\Pi_{A}=0$ at saddle-point.
Note that unlike the normal self-energy,
anomalous self-energy has $\Pi_{A}(\tau,\tau')=\Pi_{A}(\tau',\tau)$.

\section{Conclusion}

The self-consistent relations and sum rules in IR or UV limit are investigated in Sec.3.
Where the zeroth, first, and second moment are obtained through the fluctuation-dissipation theorem,
for the dynamical susceptibility which related to the three-point (density-density-density response function) boson mode.
The zeroth moment also provides the static structure factor.
Unless in the $GG^{0}G^{0}$ approximation as discussed in Sec.5 and Appendix.B,
the self-consistent relation and the sum rules requires 
the intercation entering the susceptibility through ladder form should be irreducible vertices (see Fig.2),
by considering the many-body local-field effect,
which means except the RPA-type bare interactions (i.e., the
particle-hole interaction or the pseudo-potential), the pair fluctuation (correlation) of exchange should be incorprated (Fig.2(c)).
Once the vertex is irreducible, even the noninteracting boson self-energy entering susceptibility leads to sum rule.

As can be seen from Eq.(\ref{82}),
the irreducible vertex reduced to the bare RPA one in the long-wavelength limit (corresponds to IR limit),
where $\langle n_{i}n_{j}\rangle-\langle n_{i}\rangle \langle n_{j}\rangle=0$ due to the vanishing imaginary part of susceptibility.
This is relates to the zero compressibility at finite temperature for systems between half-filling and full filling,
which is, $\kappa=dn/d\mu\propto (\langle n_{i}n_{j}\rangle-\langle n_{i}\rangle \langle n_{j}\rangle)=0$.
That can also easily be seen from the Eq.(126), where boson self-energy $\uppi(\Omega)$ vanish in $\Omega\rightarrow 0$ limit.
Note that the compressible non-Fermi liquid also supports the gapless SYK modes in the presence of small chemical potential,
which exhibits instability to the pairing state when it is gapped out by the condensated boson order parameter.

It is important to note that, no matter for the $G^{0}G^{0}G^{0}$ (Eq.(12); which is Lindhard-type) or $GG^{0}G^{0}$ (\ref{ggg}) approximations,
the interactions are in the $s$-wave channel (static particle-hole interaction or the so-called pseudo-potential),
which corresponds to the boson formed by single-particle excitation.
However, to go beyond the RPA and consider the many-body local-field effect with short-range interactions
(in which case the $T$-matrix can be viewed as momentum-independent),
we define the order parameter $\Delta^{\dag}$, which is approximately $s$-wave type as we assuming the particles
 $n_{1}$ and $n_{2}$ and hole $n_{3}$ coexist in a same imaginary time (see Fig.2(g);
and we have $n_{1}+n_{2}=n_{3}$ in half-filling),
then the exchange or pair correlation effects can be introduced (see Fig.2(h)), through the effective irreducible interaction $g_{eff}$ (see Eq.(40,42)),
into the $s$-wave susceptibility $\chi(\Omega)$.
Note that although there are three-vertices in each three-point fermion loop,
we consider only two of them, i.e., the interaction between $n_{1}$ and $n_{2}$ and that between $n_{1}$ and $1-n_{3}$, as shown in Fig.2.
Then the conservation law in many-body theorem can be used to obtain the self-consisitent relations and the local moment sum rule,
through the Luttinger-Ward analysis.
Except the local moment sum rules, the self-consistent relation can also be obtained
in $GG^{0}G^{0}$ approximation,
as shown in Eq.(52),
where single particle quantity can be seen that is equivalent to the three-particle quantiity (the susceptibility).

By treating the effective coupling,
i.e., the product of irreducible vertices,
as a SYK-type coupling,
we also discuss in detail the emergent SYK physics in this paper.
For the Gaussian distributed SYK coupling, we obtain the variance in the SYK$_{3}\times$SYK$_{3}$ model,
which plays an important role in detecting the dynamic of SYK model quantitatively.
We prove that the Luttinger-Ward theory (in particle-particle-hole
channel) and the self-consistent relations in non-pertubative approach (discussed in Sec.3) are
valid only in the absence of pair condensation.
While in the presence of pair condensation, the SYK non-Fermi liquid phase is suppressed,
and the many-body spectrum is gapped out by the condensated pairing order parameter,
and the phase transition from the chaotic non-Fermi liquid phase to Fermi liquid or insulator will happen
(depending on whether the Fermi surface is well-defined).
Experimentally,
this would potentially realized in the compressible quantum Hall state under artificial external field\cite{Halperin B I},
or the SYK-Kondo model which contains both the coherent and incoherent electrons\cite{Erdmenger J}.

\section{Appendix.A}
There are six terms within the bracket of Eq.(16).
The first term can be rewritten as, by transfrom the integral over frequencies into that over momenta
\begin{equation} 
\begin{aligned}
&\frac{1}{2\pi i}\int^{\infty}_{-\infty}d\xi 
[N_{F}(\xi+\Omega_{1}+\Omega_{2})
G(\xi+\Omega_{1}+\Omega_{2}+i\eta)
G(\xi+\Omega_{2}-i\eta)
G(\xi+\Omega_{1}-i\eta)\\
=&
\int^{\infty}_{-\infty}dk
[N_{F}(\xi+\Omega_{1}+\Omega_{2})
\frac{1}{\xi+\Omega_{1}+\Omega_{2}+i\eta-\frac{(k+q_{1}+q_{2})^{2}}{2m}}
\frac{1}{\xi+\Omega_{2}-i\eta-\frac{(k+q_{2})^{2}}{2m}}
\frac{1}{\xi+\Omega_{1}-i\eta-\frac{(k+q_{1})^{2}}{2m}}\\
=&
\int^{\infty}_{-\infty}da\int^{\infty}_{-\infty}db\int^{\infty}_{-\infty}dc
[N_{F}(\xi+\Omega_{1}+\Omega_{2})
\frac{1}{\xi+\Omega_{1}+\Omega_{2}+i\eta-(a+b+c+\frac{q_{1}^{2}+2q_{1}q_{2}+q_{2}^{2}}{2m})}\\
&
\frac{1}{\xi+\Omega_{2}-i\eta-(a+c+\frac{q_{2}^{2}}{2m})}
\frac{1}{\xi+\Omega_{1}-i\eta-(a+b+\frac{q_{1}^{2}}{2m})}\\
\end{aligned}
\end{equation} 
where we define
$a=\frac{k^{2}}{2m},\ b=\frac{2kq_{1}}{2m},\ c=\frac{2kq_{2}}{2m}$.
Since the integration over $c$ is nozero as the related poles locate in different sides of real aixs in complex plane,
due to the mixing of retarded (second and third propagators) and advanced propagators (first propagators).
Similarly we can obtain the integration in the fourth term of Eq.(16) vanishes.
While for the second term of Eq.(16),
which can be rewritten as
\begin{equation} 
\begin{aligned}
&\frac{1}{2\pi i}\int^{\infty}_{-\infty}d\xi 
[N_{F}(\xi-\Omega_{1}+\Omega_{2})
G(\xi-\Omega_{1}+\Omega_{2}+i\eta)
G(\xi+\Omega_{2}+i\eta)
G(\xi-i\eta)\\
=&
\int^{\infty}_{-\infty}dk
[N_{F}(\xi-\Omega_{1}+\Omega_{2})
\frac{1}{\xi-\Omega_{1}+\Omega_{2}+i\eta-\frac{(k-q_{1}+q_{2})^{2}}{2m}}
\frac{1}{\xi+\Omega_{2}+i\eta-\frac{(k+q_{2})^{2}}{2m}}
\frac{1}{\xi-i\eta-\frac{k^{2}}{2m}}\\
=&
\int^{\infty}_{-\infty}da\int^{\infty}_{-\infty}db\int^{\infty}_{-\infty}dc
[N_{F}(\xi-\Omega_{1}+\Omega_{2})
\frac{1}{\xi+\Omega_{1}+\Omega_{2}+i\eta-(a-b+c+\frac{q_{1}^{2}-2q_{1}q_{2}+q_{2}^{2}}{2m})}\\
&
\frac{1}{\xi+\Omega_{2}-i\eta-(a+c+\frac{q_{2}^{2}}{2m})}
\frac{1}{\xi+\Omega_{1}-i\eta-(\frac{a^{2}}{2m})},
\end{aligned}
\end{equation} 
since the integral over $c$ has the related poles $\xi-i\Omega_{1}^{A}+i\Omega_{2}^{R}$ and $\xi+i\Omega_{R}$
where $i\Omega^{A/R}=\Omega\mp i\Omega$
(and we have $\Omega_{2}\ge \Omega_{1}$ which leads to the same result even we replace the pole
$\xi-i\Omega_{1}^{A}+i\Omega_{2}^{R}$ with pole $\xi+i\Omega_{1}^{A}+i\Omega_{2}^{R}$),
thus these two poles always in the same side of the real axis in complex plane (depends on the sign of $\Omega_{2}$ only).
Similarly we can obtain the integration in the third term of Eq.(11) vanishes, and the fifth and sixth terms are nonzero.

\section{Appendix.B: Relation of $G(\tau)=\Sigma(\tau) \sim |\tau|^{-1}$ in SYK model in IR limit and in conserving approximation}

In this paper, some conclusions of the SYK$_{2}$ model are used,
which can be described as $H=i\sum_{i=1,2}gc_{i}^{\dag}c_{i}$,
where the $i$ factor here is important to keep the particle-hole symmetry $H^{*}=H$ as the $g$ is real,
i.e., $g$ is the particle-hole amplitude.
Note that usually the $i$ factor is incorporated into the coupling $g$\cite{Lau P H C}
and thus $g$ is complex and of Gaussian type (with zero mean value), unless for the Majorana fermions\cite{Lau P H C,Garcia-Garcia A M}.
Also, the conformal symmetry properties of SYK$_{2}$ model is used in IR limit
in dealing with the relation between $\Sigma$ and $G$.
The single fermion self-energy in IR limit can be obtained through the equal-time Luttinger-Ward functional $\delta\Phi$,
according to Eq.(94) and Eq.(113) based on the $GG^{0}G^{0}$ approximation
\begin{equation} 
\begin{aligned}
\Sigma_{1}(\tau)
=&\frac{\delta\Phi}{\delta G_{1}(-\tau-\delta\tau)\delta G_{3}(\delta\tau)}\\
=&\frac{\int d\tau g^{2}\langle\Delta^{\dag}(\tau)\Delta(\tau)\rangle}{\delta G_{1}(-\tau-\delta\tau)\delta G^{0}_{3}(\delta\tau)}\\
=&\frac{\int d\tau g^{2}\langle\Delta(-\tau)\Delta(\tau)\rangle}{\delta G_{1}(-\tau-\delta\tau)\delta G^{0}_{3}(\delta\tau)}\\
\approx & g^{2}G_{1}(\tau).
\end{aligned}
\end{equation} 
For fermion spectral function, we have
\begin{equation} 
\begin{aligned}
\int^{\infty}_{-\infty}\frac{d\varepsilon}{2\pi}\rho(\varepsilon)=1,\\
\int^{\infty}_{0}\frac{d\varepsilon}{2\pi}\rho(\varepsilon)e^{-\varepsilon\tau}=G(\tau),\\
\int^{0}_{-\infty}\frac{d\varepsilon}{2\pi}\rho(\varepsilon)e^{-\varepsilon\tau}=1-G(\tau),
\end{aligned}
\end{equation} 
where the last two equations can be obtained through spectral decomposition.
Since $\rho(\varepsilon)=-2{\rm Im}G(\varepsilon+i\eta)$,
using the ansatz
\begin{equation} 
\begin{aligned}
G_{F}(\pm i\omega)
=\pm ie^{\pm i\theta}\omega_{NFL}^{-1+\alpha}\omega^{-\alpha},
\end{aligned}
\end{equation} 
we have
\begin{equation} 
\begin{aligned}
\rho(\pm\omega)
=&-2{\rm Im}
ie^{\pm i\theta}\omega_{NFL}^{-1+\alpha}(\mp i\omega)^{-\alpha}\\
=&-2
(
\omega^{-\alpha} \omega_{NFL}^{ -1 + \alpha}
   {\rm cos}[\theta] {\rm cos}[
   \alpha {\rm Arg}[\mp i \omega]] {\rm cos}[(-1 + \alpha) {\rm Arg}[
     \omega_{NFL}]] 
	 \\
	 +& \omega^{-\alpha} \omega_{NFL}^{ -1 + \alpha}
   {\rm cos}[(-1 + \alpha) {\rm Arg}[\omega_{NFL}]] {\rm sin}[\theta] {\rm sin}[
   \alpha {\rm Arg}[\mp i \omega]] 
   \\
   -& \omega^{-\alpha} \omega_{NFL}^{ -1 + \alpha}
   {\rm cos}[\alpha {\rm Arg}[\mp i \omega]] {\rm sin}[
   \theta] {\rm sin}[(-1 + \alpha) {\rm Arg}[\omega_{NFL}]] 
   \\
   +& \omega^{-\alpha} \omega_{NFL}^{ -1 + \alpha}
   {\rm cos}[\theta] {\rm sin}[\alpha {\rm Arg}[\mp i \omega]] {\rm sin}[(-1 + \alpha) {\rm Arg}[\omega_{NFL}]]
),
\end{aligned}
\end{equation} 
thus for $\tau>0$
\begin{equation} 
\begin{aligned}
G(\tau)=&
\int^{\infty}_{0}\frac{d\varepsilon}{2\pi}\rho(\varepsilon)e^{-\varepsilon\tau}\\
=&\frac{-1}{\pi}
[
\tau^{-1 + \alpha} \omega_{NFL}^{-1 + \alpha}
  {\rm cos}[(\alpha \pi)/2] {\rm cos}[\theta] {\rm cos}[(-1 + \alpha) {\rm Arg}[\omega_{NFL}]] \Gamma[
  1 - \alpha]
  \\
  &-\tau^{-1 + \alpha} \omega_{NFL}^{-1 + \alpha}
  {\rm cos}[(-1 + \alpha) {\rm Arg}[\omega_{NFL}]] \Gamma[1 - \alpha] {\rm sin}[(\alpha \pi)/
  2] {\rm sin}[\theta]\\
  &-
  \tau^{-1 + \alpha} \omega_{NFL}^{-1 + \alpha}
  {\rm cos}[(\alpha \pi)/2] \Gamma[1 - \alpha] {\rm sin}[
  \theta] {\rm sin}[(-1 + \alpha) {\rm Arg}[\omega_{NFL}]]
  \\
  &-\tau^{-1 + \alpha} \omega_{NFL}^{-1 + \alpha}
  {\rm cos}[\theta] \Gamma[1 - \alpha] {\rm sin}[(\alpha \pi)/
  2] {\rm sin}[(-1 + \alpha) {\rm Arg}[\omega_{NFL}]]
].
\end{aligned}
\end{equation} 
and for $\tau<0$
\begin{equation} 
\begin{aligned}
G(\tau)=&
\int^{\infty}_{0}\frac{d\varepsilon}{2\pi}\rho(-\varepsilon)e^{\varepsilon\tau}\\
=&\frac{-1}{\pi}
[
\tau^{-1 + \alpha} \omega_{NFL}^{-1 + \alpha}
  {\rm cos}[(\alpha \pi)/2] {\rm cos}[\theta] {\rm cos}[(-1 + \alpha) {\rm Arg}[\omega_{NFL}]] \Gamma[
  1 - \alpha]
  \\
  +&\tau^{-1 + \alpha} \omega_{NFL}^{-1 + \alpha}
  {\rm cos}[(-1 + \alpha) {\rm Arg}[\omega_{NFL}]] \Gamma[1 - \alpha] {\rm sin}[(\alpha \pi)/
  2] {\rm sin}[\theta]
  \\
  -&\tau^{-1 + \alpha} \omega_{NFL}^{-1 + \alpha}
  {\rm cos}[(\alpha \pi)/2] \Gamma[1 - \alpha] {\rm sin}[
  \theta] {\rm sin}[(-1 + \alpha) {\rm Arg}[\omega_{NFL}]]
  \\
  +&\tau^{-1 + \alpha} \omega_{NFL}^{-1 + \alpha}
  {\rm cos}[\theta] \Gamma[1 - \alpha] {\rm sin}[(\alpha \pi)/
  2] {\rm sin}[(-1 + \alpha) {\rm Arg}[\omega_{NFL}]]
].
\end{aligned}
\end{equation} 

Thus in the $\alpha\rightarrow 0$ limit (e.g., in half-filling), we have $G(\tau)\sim |\tau|^{-1}$ ($\tau\gg 1$).
Eq.(135) can also be verified through the relation
\begin{equation} 
\begin{aligned}
\Sigma_{1}(\pm \omega)
=&-g^{2}G^{-1}(\omega)\\
=&-g^{2}[\int^{\infty}_{0}d\tau G(\tau)e^{-\omega\tau}]^{-1}\\
=&g^{2}(\frac{1}{\pi} (\pm\omega)^{-\alpha} \omega_{NFL}^{ -1 + \alpha}
   {\rm cos}[(\alpha \pi)/2 + \theta + (-1 + \alpha) {\rm Arg}[\omega_{NFL}]] \Gamma[
   1 - \alpha] \Gamma[\alpha])^{-1}\\
=&g^{2}\frac{\pi \omega^{\alpha} \omega_{NFL}^{1 - \alpha}
   {\rm sec}[(\alpha \pi)/2 + \theta + (-1 + \alpha) {\rm Arg}[\omega_{NFL}]]}{
 \Gamma[1 - \alpha] \Gamma[\alpha]}.
\end{aligned}
\end{equation} 
Now the relation of $G(\tau)=\Sigma(\tau) \sim |\tau|^{-1}$ for three-point boson mode is proved.

Unlike the $q=4$ (four point) mode, for SYK$_{2}$ fermions,
the fermion Green's function imaginary time domain follows the marginal fermi-liquid scaling, 
$G(\tau)\sim |\tau|^{-1}$, i.e., linear in frequency, which locates in neither the fluctuation-dominated regime or the interaction-dominated regime.
We note that, through a phase field $\phi(\tau)$,
the three-point boson mode (correlator) can also be expressed as\cite{Altland A,Pavlov A I}
\begin{equation} 
\begin{aligned}
D(\tau',\tau)
=\langle\Delta^{\dag}(\tau')\Delta(\tau)\rangle\\
=\langle e^{-i\phi(\tau')}e^{i\phi(\tau)}\rangle_{\phi}\\
=e^{-m_{\phi}\frac{|\tau'-\tau|}{2}},
\end{aligned}
\end{equation} 
which becomes $D(\tau',\tau)=1$ and $D(\tau',\tau)=0$ in UV and IR limit, respectively.
Here $m_{\phi}$ is the bosonic mass of the boson dispersion which can easily be obtain in frequency domain.

The saddle-point equations of Green's functions and self-energyies obtained in IR limit, 
are equivalent to the Schwinger-Dyson equations in diagrammatic approach,
where the time-derivative term is irrelavant in the low-energy and zero-temperature limit,
Since the resulting boson self-energies are proportional to the frequency in IR limit,
it can not be used in the expression of susceptibility to find the instability,
which corresponds to the singularities in dynamic regime.

\section{Appendix.C: Hartree-Fock type self-energy in UV limit}

The Hartree-Fock type self-energy is related to the high-frequency (UV) limit of the fermions, and is momentum and frequency-independent.
It can be obtained by using the Luttinger-Ward functional derivative in equal time limit
\begin{equation} 
\begin{aligned}
\Sigma^{HF}_{1}(\tau,\tau')
=&\frac{\delta \Phi[G]}{\delta G_{1}(\tau',\tau'')\delta G_{3}^{0}(\tau'',\tau)},
\end{aligned}
\end{equation} 
thus we have
\begin{equation} 
\begin{aligned}
\lim_{i\omega\rightarrow\infty} \Sigma^{HF}_{1}(i\omega)
=&\lim_{i\omega\rightarrow\infty} g^{2}G_{2}(i\omega+i\Omega_{1})\\
=&g^{2}G_{1}(i\omega)\\
=&g^{2}[\int\frac{d\varepsilon}{2\pi}\frac{\rho(\varepsilon)}{i\omega}
+\int\frac{d\varepsilon}{2\pi}\frac{\varepsilon\rho(\varepsilon)}{(i\omega)^{2}}]\\
=& g^{2}[\frac{1}{i\omega}
+\frac{\varepsilon_{3}(\varepsilon)+n_{3}g^{2}}{(i\omega)^{2}}]\\
=&\lim_{i\omega\rightarrow\infty} g^{2}\frac{1}{i\omega-\varepsilon_{3}(\varepsilon)-\Sigma_{1}}\\
=& g^{2}[\frac{1}{i\omega}
+\frac{\varepsilon_{3}(\varepsilon)+\Sigma_{1}}{(i\omega)^{2}}],
\end{aligned}
\end{equation} 
where we can see that $\lim_{i\omega\rightarrow\infty} \Sigma^{HF}_{1}(i\omega)=n_{3}g^{2}$.
Eq.(101) and Eq.(104) are used here.
Here the Luttinger-Ward functional $\delta \Phi[G]=g^{2}\langle\Delta^{\dag}(\tau)\Delta(\tau)\rangle$ is also related to the expectation value of interaction term.
The simplest $T$-matrix type approximation corresponds to the equal-time approximation in UV limit,
in which case the irreducible vertices is just the bare one, in contrast to the Eq.(71).
In this case, the pair-order induced phase and amplitude fluctuations around the saddle-point vanishes, 
then the self-energy is of the Hartree-Fock type which can be absorbed into he interacting chemical potential,
and the Luttinger-Ward functional $\delta\phi[G]$ becomes diagonal,
i.e., the anomalous components ($\Delta^{\dag}(\tau')\Delta^{\dag}(\tau)$,$\Delta(\tau')\Delta(\tau)$) vanish,
that is consistent with the high-frequency (UV limit) asymopitic behavior of fermion self-energy.
Note that the interaction here is of the charge or spin fluctuation type instead of the nematic type which is frequency-dependent.

\end{large}
\renewcommand\refname{References}

\clearpage
\begin{figure}

\centering
\begin{subfigure}
  \centering
  \includegraphics[width=0.4\linewidth]{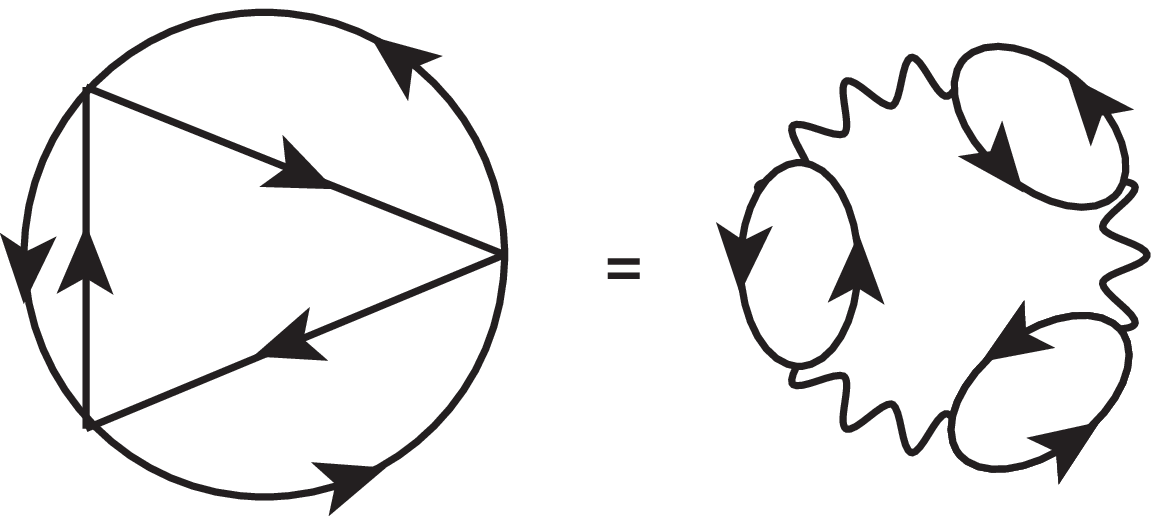}
\end{subfigure}
\caption{The four-loop planar particle-hole diagram.
In this figure,
the solid straight line is the fermion propagator and the wavy line is the boson propagator.
There are totally six density operators in this diagram.}
\end{figure}

\clearpage
\begin{figure}
\centering
\begin{subfigure}
  \centering
  \includegraphics[width=0.4\linewidth]{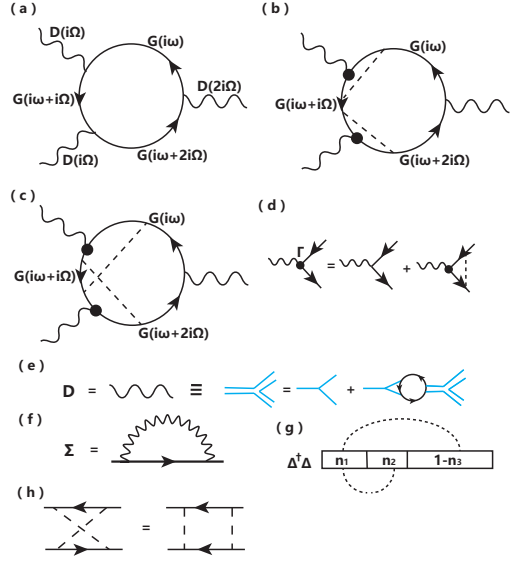}
\end{subfigure}
\caption{The boson mode described by the three-point fermion loop.
(a) corresponds to the first diagram of the RPA ladder expansion which is shown in (b).
(c) contains the pair correlation or short-range exchange.
The diagrams (a) and (c) are irreducible while (b) is reducible.
(d) shows the vertex function. (e) is the boson propagator which is needed in the calculation of fermion self-energy (f).
(g) shows the interaction term whose expectation value is $g^{2}\langle\Delta^{\dag}\Delta$.
The half-filling case corresponds to $n_{1}+n_{2}=n_{3}$.
(h) shows the how the crossing of interaction lines turns the RPA type ladder into the irreducible pair propagator.
In this figure,
the dashed black lines stands the interactions.}
\end{figure}

\begin{figure}
\centering
\begin{subfigure}
  \centering
  \includegraphics[width=0.4\linewidth]{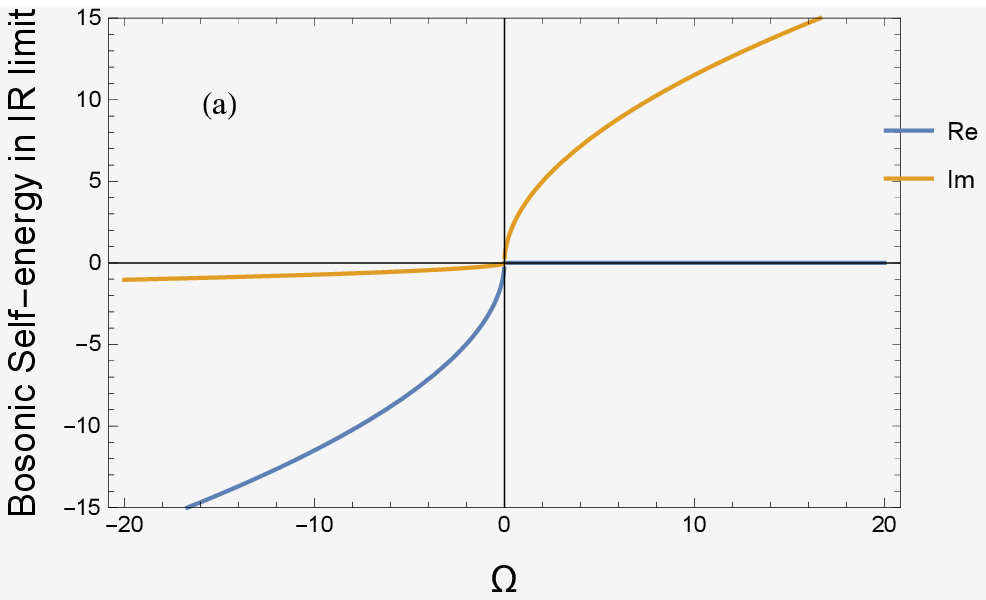}
\end{subfigure}
\begin{subfigure}
  \centering
  \includegraphics[width=0.4\linewidth]{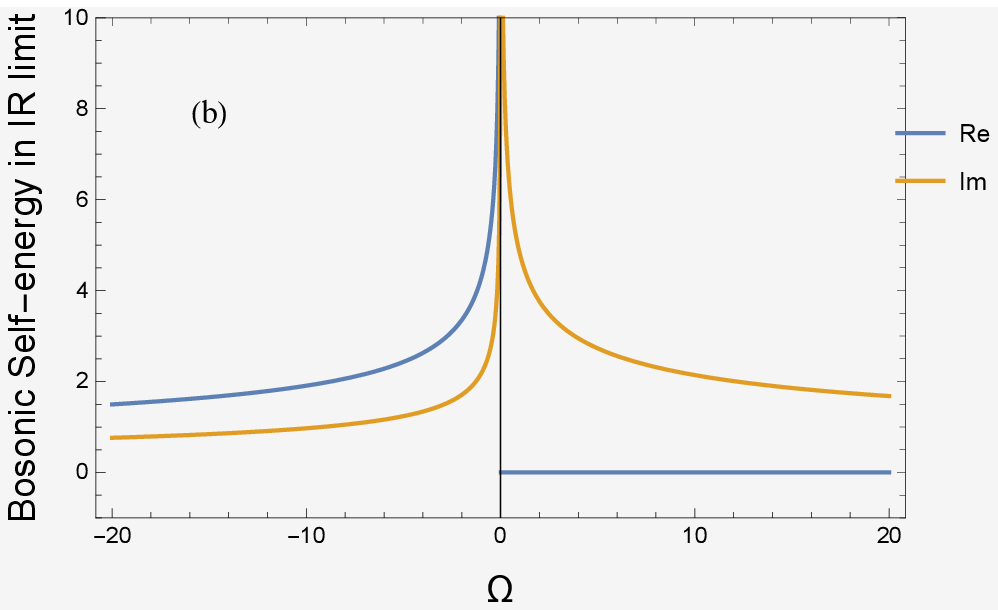}
\end{subfigure}\\
\begin{subfigure}
  \centering
  \includegraphics[width=0.4\linewidth]{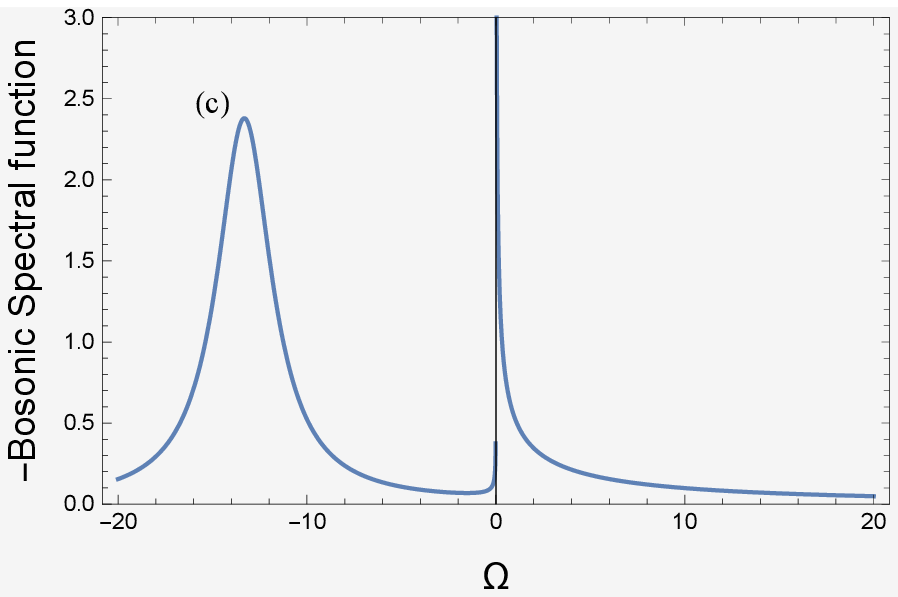}
\end{subfigure}
\begin{subfigure}
  \centering
  \includegraphics[width=0.4\linewidth]{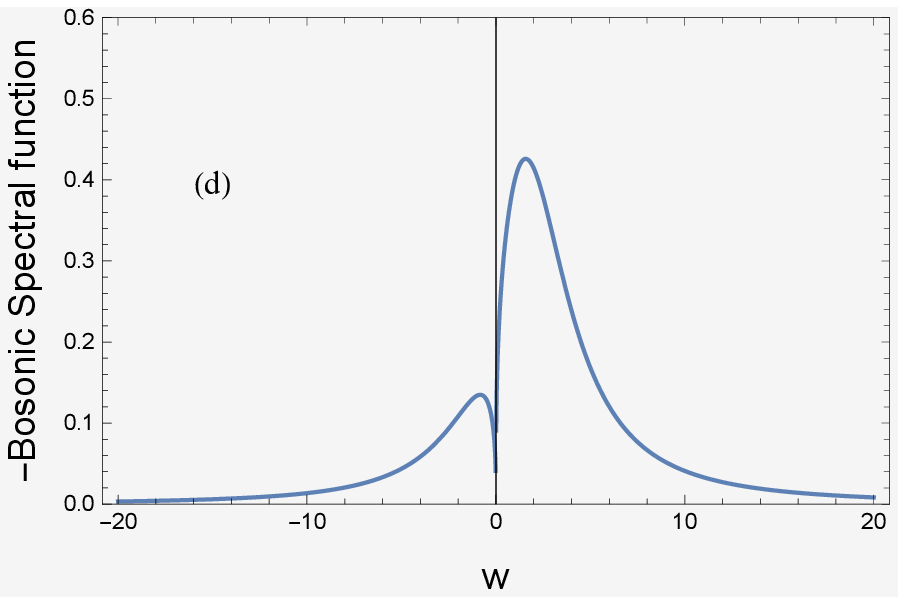}
\end{subfigure}
\caption{
There-point boson mode self-energies in IR limit.
(a) is for half-filling where we set the asmmetry parameter $\theta=0$, and $\alpha=0.16$,
(b) is for the case away from half filling, where we set $\theta=\pi/4$, and $\alpha=0.5$.
The full filling case corresponds to $\theta=\pi/2$, and in which case the boson self-energy is nearly a small constant and thus we donot show it here.
(c) and (d) show the boson spectral function (which is negative since $\rho_{B}={\rm Im}G_{B}<0$) whose parameter setting are the same as (a) and (b), respectively.
}
\end{figure}

\clearpage

\begin{figure}
\centering
\begin{subfigure}
  \centering
  \includegraphics[width=0.4\linewidth]{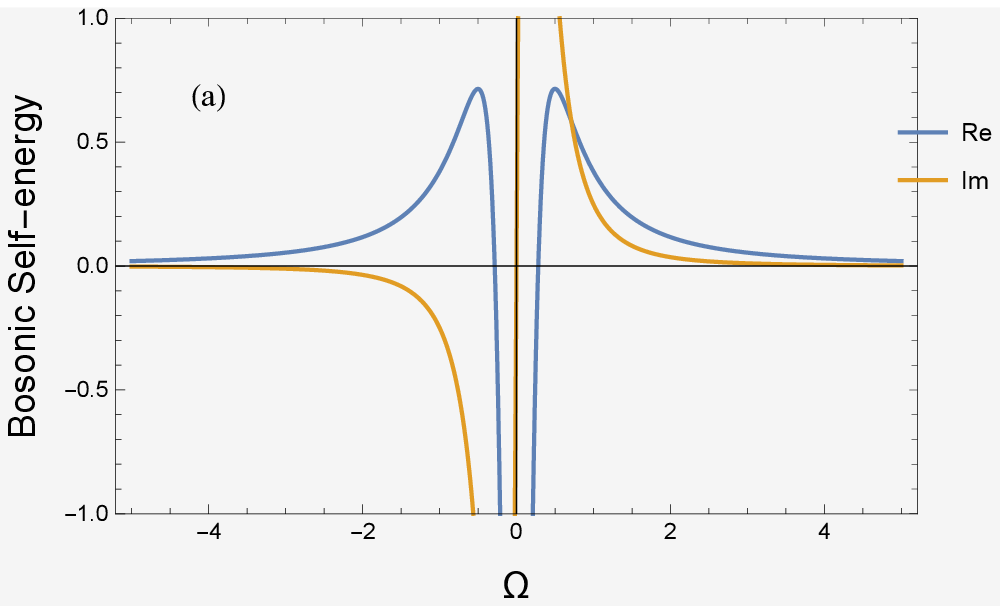}
\end{subfigure}
\begin{subfigure}
  \centering
  \includegraphics[width=0.4\linewidth]{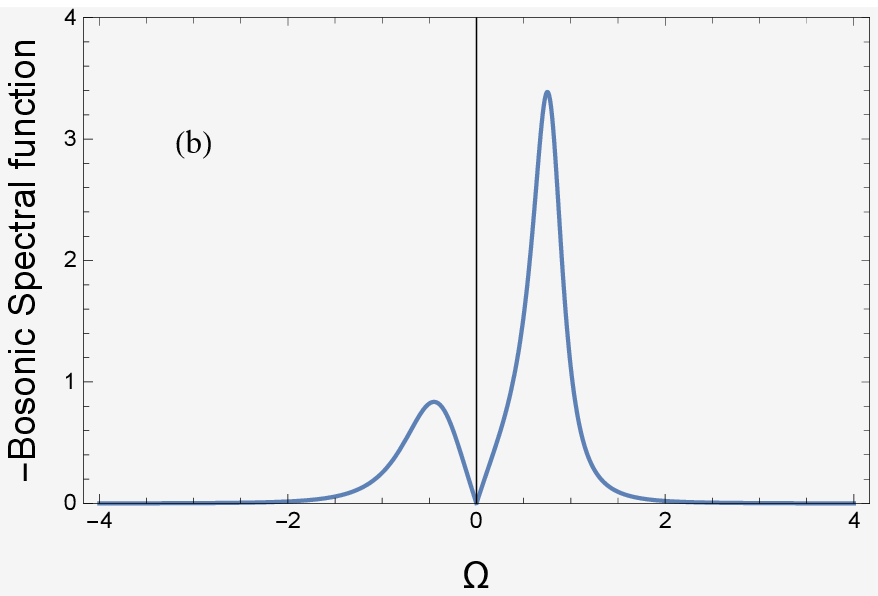}
\end{subfigure}\\
\begin{subfigure}
  \centering
  \includegraphics[width=0.4\linewidth]{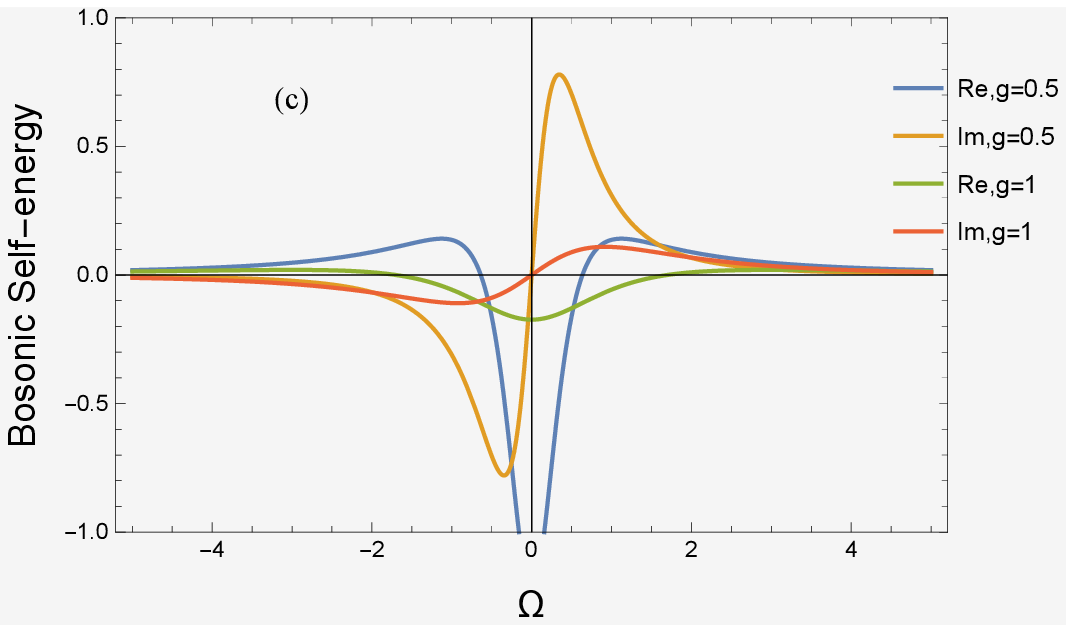}
\end{subfigure}
\begin{subfigure}
  \centering
  \includegraphics[width=0.4\linewidth]{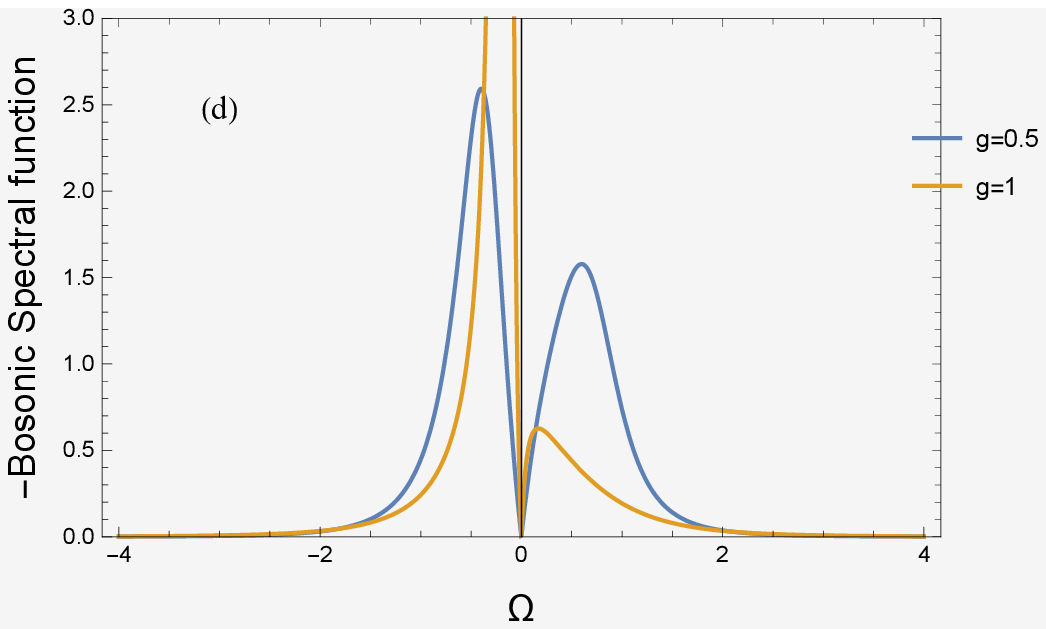}
\end{subfigure}
\caption{
There-point boson mode self-energies and the corresponding spectral functions 
calculated by using Eq.(128) and Eq.(130), respectively.
(a) and (b) donot consider the Hatree-Fock type self-energy while the (c) and (d) do.
Here we set $\varepsilon=0.2$ and $\eta=0.0001$.
}
\end{figure}

\end{document}